\pdfoutput=1
\documentclass[12pt,a4paper]{article}

\usepackage{ifthen} 
\newboolean{pdflatex}
\setboolean{pdflatex}{true} 

\newboolean{articletitles}
\setboolean{articletitles}{true} 

\newboolean{uprightparticles}
\setboolean{uprightparticles}{false} 

\def\paperauthors{LHCb collaboration} 
\def\paperasciititle{Amplitude analysis of the Lambda^+_c  pK^-pi^+ decay and Lambda^+_c baryon polarization measurement in semileptonic beauty hadron decays} 
\def\papertitle{Amplitude analysis of the $\Lambda^+_c\to pK^-\pi^+$ decay and\\ $\Lambda^+_c$ baryon polarization measurement\\ in semileptonic beauty hadron decays} 
\def\paperkeywords{{High Energy Physics}, {LHCb}} 
\def\papercopyright{\the\year\ CERN for the benefit of the LHCb collaboration} 
\def\paperlicence{CC BY 4.0 licence}
\def\paperlicenceurl{https://creativecommons.org/licenses/by/4.0/}

\usepackage[top=1in, bottom=1.25in, left=1in, right=1in]{geometry}

\usepackage{microtype}
\usepackage{lineno}  
\usepackage{xspace} 
\usepackage{caption} 

\usepackage{graphicx}  
\usepackage{color}
\usepackage{colortbl}
\graphicspath{{./figs/}} 

\usepackage{amsmath} 
\usepackage{amssymb}
\usepackage{amsfonts}
\usepackage{upgreek} 

\newcommand*\patchAmsMathEnvironmentForLineno[1]{%
\expandafter\let\csname old#1\expandafter\endcsname\csname #1\endcsname
\expandafter\let\csname oldend#1\expandafter\endcsname\csname
end#1\endcsname
 \renewenvironment{#1}%
   {\linenomath\csname old#1\endcsname}%
   {\csname oldend#1\endcsname\endlinenomath}%
}
\newcommand*\patchBothAmsMathEnvironmentsForLineno[1]{%
  \patchAmsMathEnvironmentForLineno{#1}%
  \patchAmsMathEnvironmentForLineno{#1*}%
}
\AtBeginDocument{%
\patchBothAmsMathEnvironmentsForLineno{equation}%
\patchBothAmsMathEnvironmentsForLineno{align}%
\patchBothAmsMathEnvironmentsForLineno{flalign}%
\patchBothAmsMathEnvironmentsForLineno{alignat}%
\patchBothAmsMathEnvironmentsForLineno{gather}%
\patchBothAmsMathEnvironmentsForLineno{multline}%
\patchBothAmsMathEnvironmentsForLineno{eqnarray}%
}

\usepackage{hyperxmp}

\usepackage[pdftex,
            pdfauthor={\paperauthors},
            pdftitle={\paperasciititle},
            pdfkeywords={\paperkeywords},
            pdfcopyright={Copyright (C) \papercopyright},
            pdflicenseurl={\paperlicenceurl}]{hyperref}

\usepackage[colorinlistoftodos,textsize=scriptsize]{todonotes}

\usepackage[bottom,flushmargin,hang,multiple]{footmisc}

\usepackage[all]{hypcap} 

\usepackage{xspace} 
\usepackage{upgreek}

\def\lhcb   {\mbox{LHCb}\xspace}

\def\MagUp {\mbox{\em Mag\kern -0.05em Up}\xspace}

\ifthenelse{\boolean{uprightparticles}}%
{

 \def\Pmu         {\ensuremath{\upmu}\xspace}

 \def\Ppi         {\ensuremath{\uppi}\xspace}

 \def\Ppsi        {\ensuremath{\uppsi}\xspace}

 \def\PDelta      {\ensuremath{\Delta}\xspace}                 
 \def\PXi         {\ensuremath{\Xi}\xspace}                 
 \def\PLambda     {\ensuremath{\Lambda}\xspace}                 
 \def\PSigma      {\ensuremath{\Sigma}\xspace}                 
 \def\POmega      {\ensuremath{\Omega}\xspace}                 
 \def\PUpsilon    {\ensuremath{\Upsilon}\xspace}
 \let\oldPi\Pi
 \def\PPi         {\ensuremath{\oldPi}\xspace}

 \def\PB      {\ensuremath{\mathrm{B}}\xspace}                 
                  
 \def\PD      {\ensuremath{\mathrm{D}}\xspace}

 \def\PJ      {\ensuremath{\mathrm{J}}\xspace}                 
 \def\PK      {\ensuremath{\mathrm{K}}\xspace}

 \def\Pb      {\ensuremath{\mathrm{b}}\xspace}                 
 \def\Pc      {\ensuremath{\mathrm{c}}\xspace}

 \def\Pi      {\ensuremath{\mathrm{i}}\xspace}

 \def\Ps      {\ensuremath{\mathrm{s}}\xspace}

 \def\thebaroffset{0.0em}
}
{

 \def\Pmu         {\ensuremath{\mu}\xspace}

 \def\Ppi         {\ensuremath{\pi}\xspace}

 \def\Ppsi        {\ensuremath{\psi}\xspace}                 
                  
 \mathchardef\PDelta="7101
 \mathchardef\PXi="7104
 \mathchardef\PLambda="7103
 \mathchardef\PSigma="7106
 \mathchardef\POmega="710A
 \mathchardef\PUpsilon="7107
 \mathchardef\PPi="7105
                  
 \def\PB      {\ensuremath{B}\xspace}                 
                  
 \def\PD      {\ensuremath{D}\xspace}

 \def\PJ      {\ensuremath{J}\xspace}                 
 \def\PK      {\ensuremath{K}\xspace}

 \def\Pb      {\ensuremath{b}\xspace}                 
 \def\Pc      {\ensuremath{c}\xspace}

 \def\Pi      {\ensuremath{i}\xspace}

 \def\Ps      {\ensuremath{s}\xspace}

 \def\thebaroffset{0.18em}
}
\newcommand{\offsetoverline}[2][\thebaroffset]{\kern #1\overline{\kern -#1 #2}}%

\makeatletter
\ifcase \@ptsize \relax
  \newcommand{\miniscule}{\@setfontsize\miniscule{4}{5}}
\or
  \newcommand{\miniscule}{\@setfontsize\miniscule{5}{6}}
\or
  \newcommand{\miniscule}{\@setfontsize\miniscule{5}{6}}
\fi
\makeatother

\DeclareRobustCommand{\optbar}[1]{\shortstack{{\miniscule (\rule[.5ex]{1.25em}{.18mm})}
  \\ [-.7ex] $#1$}}

\def\mun        {{\ensuremath{\Pmu^-}}\xspace} 

\def\mumu       {{\ensuremath{\Pmu^+\Pmu^-}}\xspace}

\def\squark    {{\ensuremath{\Ps}}\xspace}

\def\cquark    {{\ensuremath{\Pc}}\xspace}

\def\bquark    {{\ensuremath{\Pb}}\xspace}

\def\pion   {{\ensuremath{\Ppi}}\xspace}

\def\pip    {{\ensuremath{\pion^+}}\xspace}
\def\pim    {{\ensuremath{\pion^-}}\xspace}

\def\kaon    {{\ensuremath{\PK}}\xspace}

\def\KorKbar {\kern \thebaroffset\optbar{\kern -\thebaroffset \PK}{}\xspace}

\def\Kp      {{\ensuremath{\kaon^+}}\xspace}
\def\Km      {{\ensuremath{\kaon^-}}\xspace}

\def\KS      {{\ensuremath{\kaon^0_{\mathrm{S}}}}\xspace}

\def\D       {{\ensuremath{\PD}}\xspace}

\def\DorDbar {\kern \thebaroffset\optbar{\kern -\thebaroffset \PD}\xspace}
\def\Dz      {{\ensuremath{\D^0}}\xspace}

\def\Dp      {{\ensuremath{\D^+}}\xspace}
\def\Dm      {{\ensuremath{\D^-}}\xspace}

\def\DpDm    {\ensuremath{\Dp {\kern -0.16em \Dm}}\xspace}

\def\Dstarp  {{\ensuremath{\D^{*+}}}\xspace}

\def\Ds      {{\ensuremath{\D^+_\squark}}\xspace}

\def\B       {{\ensuremath{\PB}}\xspace}

\def\BorBbar {\kern \thebaroffset\optbar{\kern -\thebaroffset \PB}\xspace}

\def\Bd      {{\ensuremath{\B^0}}\xspace}

\def\BdorBdbar {\kern \thebaroffset\optbar{\kern -\thebaroffset \Bd}\xspace}
\def\Bu      {{\ensuremath{\B^+}}\xspace}

\def\Bs      {{\ensuremath{\B^0_\squark}}\xspace}

\def\BsorBsbar {\kern \thebaroffset\optbar{\kern -\thebaroffset \Bs}\xspace}

\def\jpsi     {{\ensuremath{{\PJ\mskip -3mu/\mskip -2mu\Ppsi}}}\xspace}

\def\Y#1S{\ensuremath{\PUpsilon{(#1S)}}\xspace}

\def\Deltares    {{\ensuremath{\PDelta}}\xspace}

\def\Lz          {{\ensuremath{\PLambda}}\xspace}

\def\LorLbar     {\kern \thebaroffset\optbar{\kern -\thebaroffset \PLambda}\xspace}

\def\Lc          {{\ensuremath{\Lz^+_\cquark}}\xspace}

\def\Lb           {{\ensuremath{\Lz^0_\bquark}}\xspace}

\newcommand{\decay}[2]{\ensuremath{#1\!\to #2}\xspace} 

\def\to                 {\ensuremath{\rightarrow}\xspace}

\def\AT#1     {\ensuremath{A_{\mathrm{T}}^{#1}}\xspace}           

\def\C#1      {\ensuremath{\mathcal{C}_{#1}}\xspace}                       
\def\Cp#1     {\ensuremath{\mathcal{C}_{#1}^{'}}\xspace}                    
\def\Ceff#1   {\ensuremath{\mathcal{C}_{#1}^{\mathrm{(eff)}}}\xspace}        
\def\Cpeff#1  {\ensuremath{\mathcal{C}_{#1}^{'\mathrm{(eff)}}}\xspace}       
\def\Ope#1    {\ensuremath{\mathcal{O}_{#1}}\xspace}                       
\def\Opep#1   {\ensuremath{\mathcal{O}_{#1}^{'}}\xspace}                    

\newcommand{\ket}[1]{\ensuremath{|#1\rangle}}              
\newcommand{\braket}[2]{\ensuremath{\langle #1|#2\rangle}} 

\newcommand{\nospaceunit}[1]{\ensuremath{\text{#1}}}       
\newcommand{\aunit}[1]{\ensuremath{\text{\,#1}}}       

\newcommand{\tev}{\aunit{Te\kern -0.1em V}\xspace}
\newcommand{\gev}{\aunit{Ge\kern -0.1em V}\xspace}
\newcommand{\mev}{\aunit{Me\kern -0.1em V}\xspace}
\newcommand{\kev}{\aunit{ke\kern -0.1em V}\xspace}
\newcommand{\ev}{\aunit{e\kern -0.1em V}\xspace}
 
\newcommand{\mevc}{\ensuremath{\aunit{Me\kern -0.1em V\!/}c}\xspace}
\newcommand{\gevc}{\ensuremath{\aunit{Ge\kern -0.1em V\!/}c}\xspace}
\newcommand{\mevcc}{\ensuremath{\aunit{Me\kern -0.1em V\!/}c^2}\xspace}
\newcommand{\gevcc}{\ensuremath{\aunit{Ge\kern -0.1em V\!/}c^2}\xspace}

\def\mm   {\aunit{mm}\xspace}

\def\mum  {\ensuremath{\,\upmu\nospaceunit{m}}\xspace}

\def\fb   {\ensuremath{\aunit{fb}}\xspace}
\def\invfb   {\ensuremath{\fb^{-1}}\xspace}

\newcommand{\chisq}{\ensuremath{\chi^2}\xspace}

\newcommand{\chisqip}{\ensuremath{\chi^2_{\text{IP}}}\xspace}

\def\gsim{{~\raise.15em\hbox{$>$}\kern-.85em
          \lower.35em\hbox{$\sim$}~}\xspace}
\def\lsim{{~\raise.15em\hbox{$<$}\kern-.85em
          \lower.35em\hbox{$\sim$}~}\xspace}

\def\pt         {\ensuremath{p_{\mathrm{T}}}\xspace}

\def\ptot       {\ensuremath{p}\xspace}

\def\evtgen     {\mbox{\textsc{EvtGen}}\xspace}

\def\geant      {\mbox{\textsc{Geant4}}\xspace}

\def\photos     {\mbox{\textsc{Photos}}\xspace}

\def\pythia     {\mbox{\textsc{Pythia}}\xspace}

\def\root       {\mbox{\textsc{Root}}\xspace}

\def\tensorflow {\mbox{\textsc{TensorFlow}}\xspace}

\def\tell1  {TELL1\xspace}
\def\ukl1   {UKL1\xspace}

\newcommand{\lhcborcid}[1]{\href{https://orcid.org/#1}{\hspace*{0.1em}\raisebox{-0.45ex}{\includegraphics[width=1em]{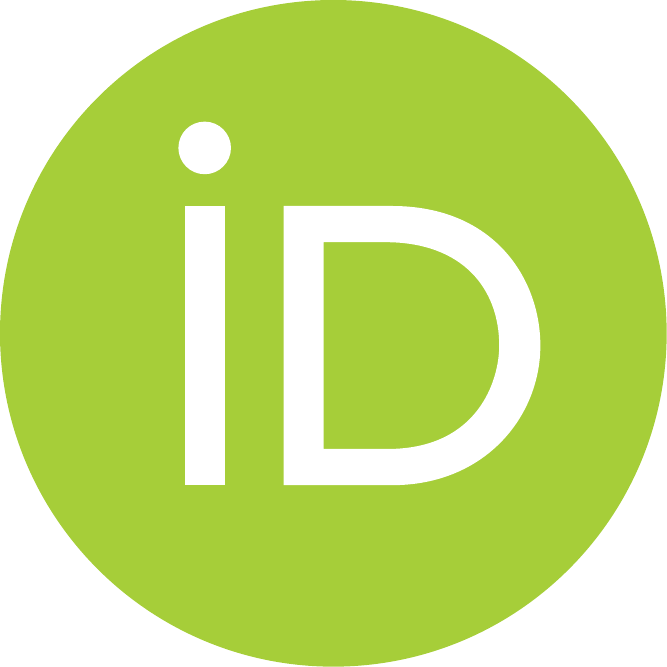}}}}

\usepackage{cite} 
\usepackage{mciteplus}

\usepackage{longtable} 

\usepackage{amsmath}
\usepackage{booktabs,url}
\usepackage{braket}
\usepackage[utf8]{inputenc}
\usepackage{bm} 
\usepackage{bigints} 
\definecolor{darkgreen}{RGB}{0,102,0}
\usepackage{tikz}
\usetikzlibrary{arrows,matrix}

\newcommand{\Lcpkpi}{\ensuremath{\Lc\to pK^-\pi^+}\xspace}

\newcommand{\Dkpipi}{\ensuremath{\Dp\to K^-\pi^+\pi^+}\xspace}

\newcommand{\Dskkpi}{\ensuremath{\Ds\to K^+K^-\pi^+}\xspace}
\newcommand{\mqpk}{\ensuremath{m^2_{pK^-}}\xspace}
\newcommand{\mqkpi}{\ensuremath{m^2_{K^-\pi^+}}\xspace}
\newcommand{\mqppi}{\ensuremath{m^2_{p\pi^+}}\xspace}
\newcommand{\mpk}{\ensuremath{m_{pK^-}}\xspace}
\newcommand{\mppk}{\ensuremath{m'_{pK^-}}\xspace}
\newcommand{\chl}{\ensuremath{\cos\theta^{\Lambda^*}_{K\pi}}\xspace}

\graphicspath{{figs/}}

\begin{document}

\renewcommand{\thefootnote}{\fnsymbol{footnote}}
\setcounter{footnote}{1}


\begin{titlepage}
\pagenumbering{roman}

\vspace*{-1.5cm}
\centerline{\large EUROPEAN ORGANIZATION FOR NUCLEAR RESEARCH (CERN)}
\vspace*{1.5cm}
\noindent
\begin{tabular*}{\linewidth}{lc@{\extracolsep{\fill}}r@{\extracolsep{0pt}}}
\ifthenelse{\boolean{pdflatex}}
{\vspace*{-1.5cm}\mbox{\!\!\!\includegraphics[width=.14\textwidth]{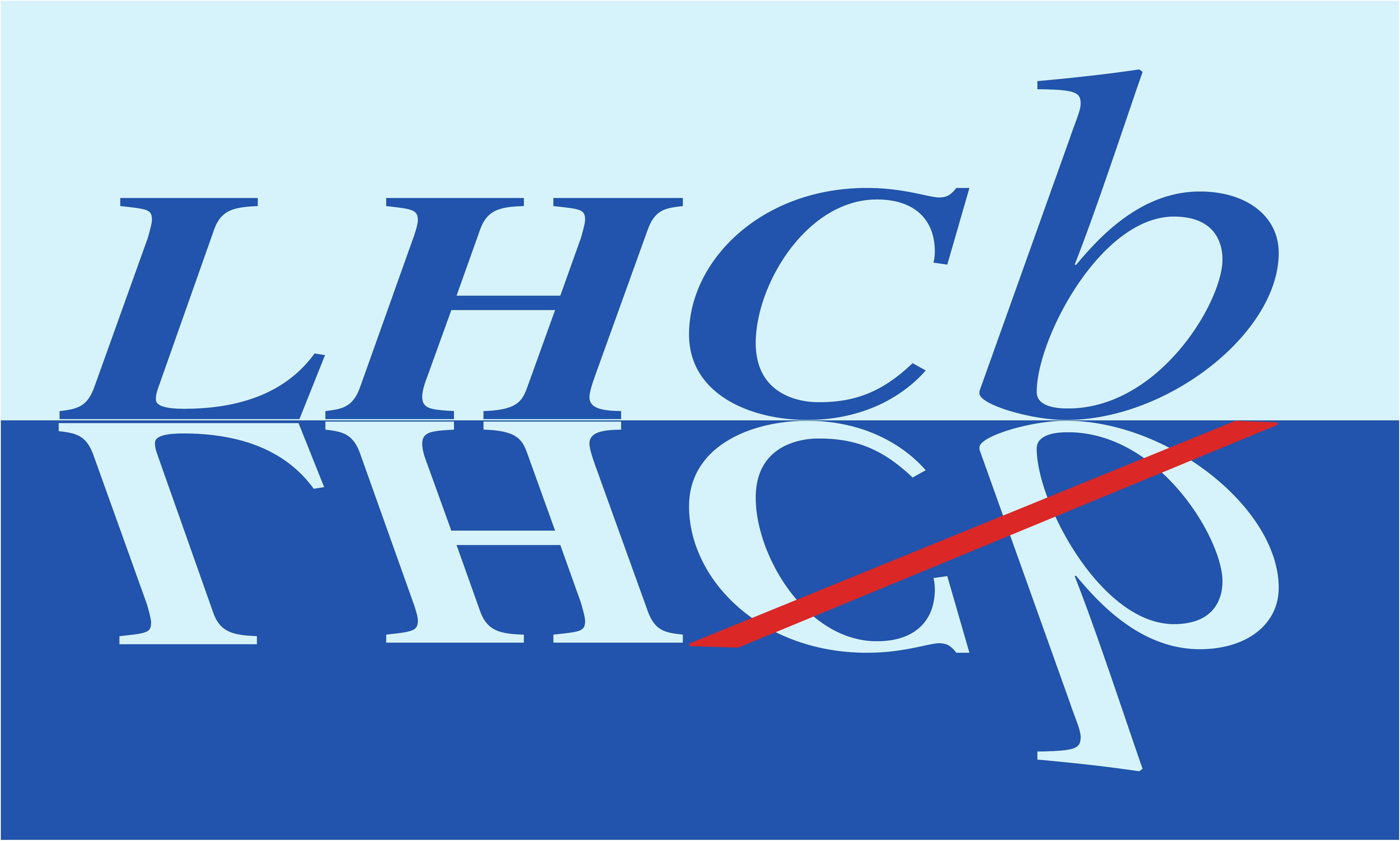}} & &}%
{\vspace*{-1.2cm}\mbox{\!\!\!\includegraphics[width=.12\textwidth]{figs/lhcb-logo.eps}} & &}%
\\
 & & CERN-EP-2022-124 \\  
 & & LHCb-PAPER-2022-002 \\  
 & & Jul 31, 2023\\ 
 & & \\
\end{tabular*}

\vspace*{3.5cm} 

{\normalfont\bfseries\boldmath\huge
\begin{center}
  \papertitle 
\end{center}
}

\vspace*{1.5cm} 

\begin{center}
\paperauthors\footnote{Authors are listed at the end of this paper.}
\end{center}

\vspace{\fill}

\begin{abstract}
  \noindent
An amplitude analysis of \Lcpkpi decays together with a measurement of the \Lc polarization vector in semileptonic beauty hadron decays is presented. A sample of $400\,000$ candidates is selected from proton-proton collisions recorded by the LHCb detector at a center-of-mass energy of 13 \tev. An amplitude model is developed and the resonance fractions as well as two- and three-body decay parameters are reported. The mass and width of the $\Lz(2000)$ state are also determined. A significant \Lc polarization is found. A large sensitivity of the \Lcpkpi decay to the polarization is seen, making the amplitude model suitable for \Lc polarization measurements in other systems. 

\end{abstract}

\vspace*{2.0cm}

\begin{center}
  Published in
  Phys.~Rev.~D108 (2023) 012023
\end{center}

\vspace{\fill}

{\footnotesize 
\centerline{\copyright~\papercopyright. \href{\paperlicenceurl}{\paperlicence}.}}
\vspace*{2mm}

\end{titlepage}


\newpage
\setcounter{page}{2}
\mbox{~}
%
%
%
%

\renewcommand{\thefootnote}{\arabic{footnote}}
\setcounter{footnote}{0}

\cleardoublepage


\pagestyle{plain} 
\setcounter{page}{1}
\pagenumbering{arabic}


\section{Introduction}
\label{sec:Introduction}

The \Lcpkpi decay has a complex resonant structure, with multiple overlapping states in the $\Km\pip$, $p\Km$ and $p\pip$ systems. A full amplitude analysis is therefore necessary to determine the composition of the decay amplitude, allowing the determination of the \Lc polarization, as described in Ref.~\cite{Marangotto:2020ead}.
The developed \Lcpkpi amplitude model therefore has multiple applications ranging from new physics searches to low-energy QCD studies.

The helicity amplitudes characterize each contribution to the \Lcpkpi decay, including intermediate state polarization. In general, the knowledge of the resonant structure is useful in searches for $C\!P$ symmetry violation, which are still unobserved in baryon decays, and can be related to specific contributions or localized in phase-space.
Parity violation is characterized by the decay asymmetry $\alpha$ parameters associated to the two-body resonant contributions, which are determined from the helicity couplings. In this respect, the prediction of the parity-conserving nature of \mbox{$1/2^+ \to 3/2^+ \,0^-$} decays~\cite{Korner:1992wi} can be tested. Parity violation is also studied for the entire three-body \Lcpkpi process, considering a quantity called average event information~\cite{Davier:1992nw}, which represents the sensitivity of the decay to the baryon polarization.

The \Lcpkpi decay amplitude can be exploited in studies of $\Lb\to\Lc l^-\bar{\nu}_l$ decays, in which the possibility to analyze the spin of the \Lc baryon increases the sensitivity to possible contributions beyond the Standard Model~\cite{Konig:1993ze, Pervin:2005ve, Gutsche:2015mxa, Shivashankara:2015cta, Dutta:2015ueb, Faustov:2016pal, Li:2016pdv, Celis:2016azn, Datta:2017aue, Zhu:2018zxb, DiSalvo:2018ngq, Ray:2018hrx, Bernlochner:2018bfn,Penalva:2019rgt, Hu:2020axt}. Such analyses are motivated by the anomalies seen in $b\to c l\nu$ processes~\cite{Lees:2013uzd,Huschle:2015rga,Sato:2016svk,Hirose:2016wfn,LHCb-PAPER-2017-027,LHCb-PAPER-2015-025} but are also interesting for Cabibbo-Kobayashi-Maskawa matrix angle measurements and $C\!P$-violation searches.

The \Lcpkpi amplitude model can be employed to measure the polarization of the \Lc baryon. This \Lc decay mode is particularly suited thanks to its good reconstruction efficiency, through the vertexing of three charged, displaced tracks. For comparison, the two-body decay $\Lc\to \Lz\pi^+$, usable for polarization measurements since its decay asymmetry parameter is known~\cite{PDG2020}, has a lower branching fraction by a factor of about five~\cite{PDG2020} and a reduced detector reconstruction efficiency because of the large $\Lz$ baryon flight distance.

The measurement of the \Lc polarization has been advocated for many production processes: \Lb semileptonic decays, for which several theoretical predictions are available~\cite{Konig:1993wz, Gutsche:2015mxa, Li:2016pdv, Faustov:2016pal, Ray:2018hrx, Hu:2020axt, Mu:2019bin}; weak interaction vector boson decays~\cite{Falk:1993rf,Galanti:2015pqa} and strong interactions~\cite{Falk:1993rf, Goldstein:1999jr, Goldstein:2015aqa}.
Polarization measurements of \Lc baryons are a fundamental probe of their spin structure and formation process via hadronization of heavy charm quarks. According to the heavy quark effective theory, most of the $c$-quark polarization is expected to be retained by the charm baryon~\cite{Mannel:1991bs, Falk:1993rf, Galanti:2015pqa}. In the case of strong force production, the baryon polarization is difficult to predict in the nonperturbative regime of QCD, its measurement discriminates among different low-energy QCD approaches. The \Lc polarization also constitutes an additional probe for new physics effects. Its measurement can complement the aforementioned $b\to c l\nu$ angular analyses providing an additional observable to test. Moreover it enables the  measurement of charm baryon electric and magnetic dipole moments via spin precession~\cite{Botella:2016ksl,Bagli:2017foe,Marangotto:2713231,Aiola:2020yam}.

In this paper, an amplitude analysis of \Lcpkpi decays recorded by the LHCb detector is presented. Charge conjugate states are implied. The analysis is based on a data sample of semileptonic decays of beauty hadrons produced in proton-proton ($pp$) collisions at a center-of-mass energy of $13\tev$, recorded in 2016, corresponding to an integrated luminosity of $1.7 \invfb$. A high purity sample of $400\,000$ candidates is selected for the amplitude fit.  This \Lcpkpi amplitude analysis improves the measurement performed by the E791 experiment at Fermilab~\cite{Aitala:1999uq} by increasing the size data sample by a factor $\approx 400$ and addressing the recently realized issue of the matching of proton spin states among different decay chains~\cite{Marangotto:2019ucc}.

This paper is organized as follows. The description of the LHCb detector and the simulation sample employed is given in Sec.~\ref{sec:detector}. The selection of \Lcpkpi candidates and the invariant mass fit used to determine signal and background yields are described in Sec.~\ref{sec:selection}. The amplitude analysis framework and the development of the \Lcpkpi default amplitude model are described in Sec.~\ref{sec:amplitude_fit}. The evaluation of statistical and systematic uncertainties is covered by Sec.~\ref{sec:systematic}, along with the consistency checks performed. The results of the amplitude fit are reported in Sec.~\ref{sec:results}, and a brief summary of the amplitude analysis is provided in Sec.~\ref{sec:summary}. Appendix~\ref{sec:amplitude_model} presents the building of the \Lcpkpi amplitude model in the helicity formalism following the method for the matching of the proton spin of Ref.~\cite{Marangotto:2019ucc}, along with the definition of the \Lc polarization systems employed and the description of the \Lcpkpi phase-space.

\section{Detector and simulation}
\label{sec:detector}
The \lhcb detector~\cite{LHCb-DP-2008-001,LHCb-DP-2014-002} is a single-arm forward
spectrometer covering the \mbox{pseudorapidity} range $2<\eta <5$,
designed for the study of particles containing \bquark or \cquark
quarks. The detector includes a high-precision tracking system
consisting of a silicon-strip vertex detector surrounding the $pp$
interaction region, a large-area silicon-strip detector located
upstream of a dipole magnet with a bending power of about
$4{\mathrm{\,Tm}}$, and three stations of silicon-strip detectors and straw
drift tubes placed downstream of the magnet.
The tracking system provides a measurement of the momentum, \ptot, of charged particles with
a relative uncertainty that varies from 0.5\% at low momentum to 1.0\% at 200\gev. Natural units with $c = 1$ are used throughout.
The minimum distance of a track to a primary $pp$ collision vertex (PV), the impact parameter, 
is measured with a resolution of $(15+29/\pt)\mum$,
where \pt is the component of the momentum transverse to the beam, in\,\gev.
Different types of charged hadrons are distinguished using information
from two ring-imaging Cherenkov detectors. 
Photons, electrons and hadrons are identified by a calorimeter system consisting of
scintillating-pad and preshower detectors, an electromagnetic
and a hadronic calorimeter. Muons are identified by a
system composed of alternating layers of iron and multiwire
proportional chambers.
The online event selection is performed by a trigger, 
which consists of a hardware stage, based on information from the calorimeter and muon
systems, followed by a software stage, which applies a full event
reconstruction.

At the hardware trigger stage, events are required to have at least one muon with high \pt.
  The software trigger requires a two-, three- or four-track
  secondary vertex with a significant displacement from any primary
  $pp$ interaction vertex. At least one charged particle
  must have a transverse momentum $\pt > 1.6\gev$ and be
  inconsistent with originating from a PV.
  A multivariate algorithm~\cite{BBDT,LHCb-PROC-2015-018} is used for
  the identification of secondary vertices consistent with the decay
  of a \bquark hadron.

The momentum scale is calibrated using samples of $\decay{\jpsi}{\mumu}$ 
and $\decay{\Bu}{\jpsi\Kp}$~decays collected concurrently
with the~data sample used for this analysis~\cite{LHCb-PAPER-2012-048,LHCb-PAPER-2013-011}.
The~relative accuracy of this
procedure is estimated to be $3 \times 10^{-4}$ using samples of other $\bquark$~hadrons, $\PUpsilon$~and
$\KS$~mesons.

Simulation is required to model the effects of the detector acceptance and the
  imposed selection requirements.
  In the simulation, $pp$ collisions are generated using
  \pythia~\cite{Sjostrand:2007gs,*Sjostrand:2006za}
  with a specific \lhcb configuration~\cite{LHCb-PROC-2010-056}.
  Decays of unstable particles
  are described by \evtgen~\cite{Lange:2001uf}, in which final-state
  radiation is generated using \photos~\cite{davidson2015photos}.
  The interaction of the generated particles with the detector, and its response,
  are implemented using the \geant
  toolkit~\cite{Allison:2006ve, *Agostinelli:2002hh} as described in
  Ref.~\cite{LHCb-PROC-2011-006}. 
  The underlying $pp$ interaction is reused multiple times, with an independently generated signal decay for each~\cite{LHCb-DP-2018-004}.

Simulated $\Lb\to\Lc\mun\bar{\nu}_{\mu}$ decays are produced with the \Lcpkpi decay uniformly generated over the phase-space. This sample consists of about $450\,000$ events passing the selection criteria.
  
The particle identification (PID) response in the simulated samples is calibrated by sampling from distributions of $\Dstarp \to \Dz\pip, \Dz \to \Km\pip$ and $\Lb \to \Lc \pim, \Lcpkpi$ decays, considering their kinematics and the detector occupancy. An unbinned method is employed, where the probability density functions are modeled using kernel density estimation~\cite{Poluektov:2014rxa}.
The \Lc momentum and transverse momentum distributions in the simulated samples are corrected in order to reproduce the data distribution using the gradient boost weighting technique of the \texttt{hep\_ml} package\cite{Rogozhnikov:2016bdp}.

\section{Selection and invariant mass fit}
\label{sec:selection}

The data sample used in this analysis corresponds to an integrated luminosity of \mbox{1.7 \invfb}.
The \Lc candidates are reconstructed from combinations of three charged hadron tracks forming a vertex separated from the PV. Semileptonic beauty hadron decay candidates are then reconstructed inclusively requiring that the \Lc candidate and a muon track originate from a common vertex. Small \chisq values are required for all track and vertex fits.
Each track is required to have a good particle identification response, transverse momentum $\pt>250\mev$ for hadrons and $\pt>1\gev$ for muons and momentum $p>2\gev$ for mesons, $p>8\gev$ for protons and $p>6\gev$ for muons.
Final-state particles are ensured to be well displaced from the interaction point by requiring a large track \chisqip with respect to any PV, where \chisqip is defined as the difference in the vertex fit \chisq of a given PV reconstructed with and without the particle under consideration. The associated PV is the one featuring the smallest \chisqip.

Further selection criteria are used to reduce the combinatorial background contamination to roughly $1\%$. Strict requirements are imposed on proton and kaon particle identification probabilities, obtained from neural network classifiers, which combine information from the different subdetectors. Probabilities larger than 0.95 and 0.7 are selected for proton and kaon hypotheses, respectively. The $\Lc\mun$ vertex is required to be significantly displaced from the PV, displacement measured in terms of a \chisq test. A maximum distance along the beam axis between $\Lc\mun$ and $p\Km\pip$ vertexes of $6\mm$ is also imposed.
The only physical sources of background identified before applying the aforementioned selection criteria, \Dkpipi and \Dskkpi decays, are effectively rejected by the tight proton identification requirement. No specific vetoes are therefore applied.

\begin{figure}
\centering
\includegraphics[scale=0.4]{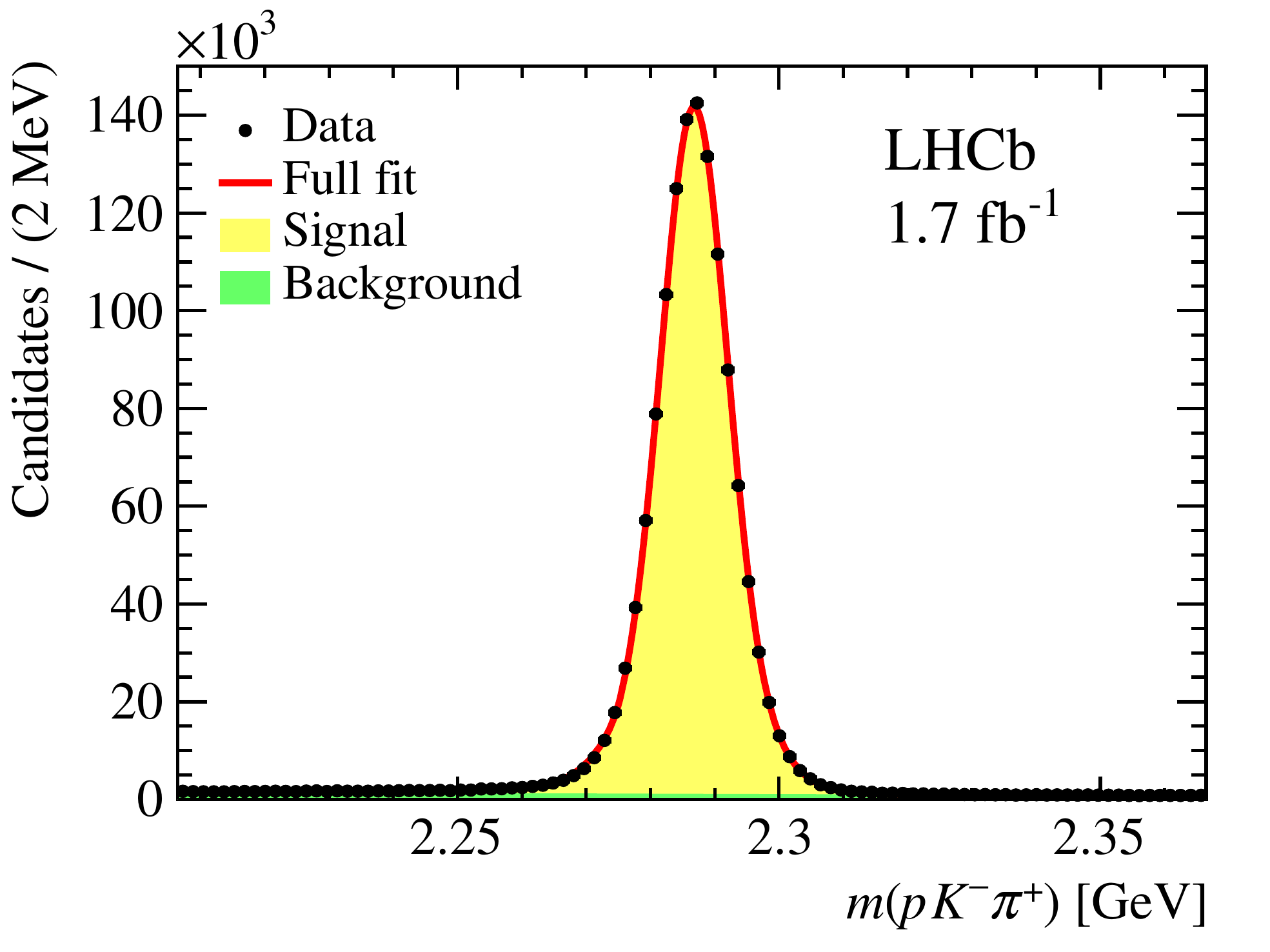}
\caption{Invariant mass distribution of selected $pK^-\pi^+$ candidates from the total dataset. 
The results from the fit described in the text are also shown.
\label{fig:fit}}
\end{figure}

The distribution of the invariant mass $m(pK^-\pi^+)$ of selected $pK^-\pi^+$ candidates from the total dataset is shown in Fig.~\ref{fig:fit}.
An extended unbinned maximum-likelihood fit is performed to the $m(p\Km\pip)$ distribution in the mass range within $80 \mev$ of the known  \Lc mass~\cite{PDG2020}.
The \Lcpkpi signal component is modeled with a double-sided Crystal Ball function~\cite{Skwarnicki:1986xj}, having asymmetric power-law tails on both sides of the peak to describe detector resolution and final-state radiation effects.
An exponential function describes the combinatorial background contribution.

The signal $m(pK^-\pi^+)$ region chosen for the amplitude analysis is within $15 \mev$ of the known \Lc mass~\cite{PDG2020}, containing about $95\%$ of the signal candidates.
The fraction of background in the signal region is $f_b = 1.69\%$.

A subset of $400\,000$ \Lcpkpi candidates is employed, corresponding to roughly 30\% of the analyzed data set, since increasing the dataset does not reduce the leading source of systematic uncertainty (see Sec.~\ref{sec:systematic}).
 The subset is chosen selecting an equal number of candidates per each category of \Lc baryon charge and LHCb magnet polarity.

\section{Amplitude fit}
\label{sec:amplitude_fit}
The amplitude model for the \Lcpkpi decay is written in the helicity formalism~\cite{JacobWick}, following the method and conventions of Ref.~\cite{Marangotto:2019ucc}. In particular, the method for the matching of the proton spin among different decay chains is employed. 
The \Lc polarization is measured in the \Lc rest frame in two different helicity systems defined by a boost along the \Lc momentum from the laboratory frame (\textit{lab}), and from the approximate beauty hadron rest frame ($\tilde{B}$).
The five variables describing the \Lcpkpi decay are chosen to be
\begin{equation}
\Omega = (\mqpk,\mqkpi,\cos\theta_p,\phi_p,\chi),
\label{eq:phase_space_vars}
\end{equation}
where \mqpk and \mqkpi are the squared invariant masses and $\theta_p,\phi_p$ and $\chi$ are the angles describing the decay orientation with respect to the \Lc polarization system. The angles $\theta_p$ and $\phi_p$ are the polar and azimuthal angles of the proton momentum, while $\chi$ is the angle between the plane formed by the proton momentum and the \Lc quantization axis and the plane formed by the kaon and pion momenta, where momenta are expressed in the \Lc rest frame. 
The detailed definition of the amplitude model, \Lc helicity systems and phase-space variables is given in Appendix~\ref{sec:amplitude_model}.

The free parameters of the amplitude model, denoted $\omega$, are determined by an unbinned maximum-likelihood fit to the five phase-space observables $\Omega$, defined in Eq.~\eqref{eq:phase_space_vars}, with $\omega$ including helicity couplings, lineshape parameters and polarization components. The negative logarithmic likelihood (NLL) minimized in the fit is
\begin{equation}
-\log\mathcal{L}(\omega) = - \sum_{i=1}^N \log p_{\text{tot}}(\Omega_i|\omega),
\label{eq:log_likelihhod_def}
\end{equation}
where $N$ is the number of events, with the probability density function (PDF)
\begin{equation}
p_{\text{tot}}(\Omega_i|\omega) = \frac{p(\Omega_i|\omega) \epsilon(\Omega_i)}{I(\omega)} (1-f_b) + p_{\text{bkg}}(\Omega_i) f_b,
\end{equation}
which includes the \Lcpkpi decay rate $p(\Omega|\omega)$, the detector efficiency $\epsilon(\Omega)$ and the background contribution $p_{\text{bkg}}(\Omega)$. The background fraction $f_b$ is fixed to that determined from the $m(p\Km\pip)$ invariant mass fit, while $I(\omega)$ is the normalization of the signal part of the total PDF,
\begin{equation}
I(\omega) = \int p(\Omega|\omega) \epsilon(\Omega) d\Omega,
\label{eq:model_sig_normalization}
\end{equation}
computed numerically
using simulated events reconstructed through the detector.
The phase-space variables $\Omega$ are computed after constraining the mass of the $p\Km\pip$ candidate to the known \Lc mass~\cite{PDG2020}.

The Equation~\eqref{eq:log_likelihhod_def} can be rewritten as
\begin{align}
-\log\mathcal{L}(\omega) &= - \sum_{i=1}^N \log \left[ p(\Omega_i|\omega) + \frac{f_b}{1-f_b} \frac{p_{\text{bkg}}(\Omega_i)I(\omega)}{\epsilon(\Omega_i)}\right]\nonumber\\
&+ N\log I(\omega) + \mathrm{constant}.
\label{eq:log_likelihood}
\end{align}
Efficiency and background parametrizations are expressed in terms of factorized Legendre polynomial expansions, derived from simulation and data sidebands, respectively. They are described in Appendix~\ref{sec:parametrizations}.

The amplitude fitting code is based on a version of the \mbox{\textsc{TensorFlowAnalysis}} package~\cite{TFA} adapted to three-body amplitudes in five-dimensional phase-space fits~\cite{Marangotto:2713231}; this package depends on the machine-learning framework \tensorflow~\cite{tensorflow2015-whitepaper} interfaced with \mbox{\textsc{MINUIT}} minimization~\cite{James:1975dr} via the \root package~\cite{Brun:1997pa}.
The fit is performed using multiple gradient-descent minimization with different, randomized, initial values of the parameters, choosing the result with minimum NLL. This ensures that the global NLL minimum is reached.

The fit quality is measured by a \chisq test performed over a five-dimensional binning of the phase-space; an adaptive binning technique is employed to guarantee a similar number of candidates in each bin. Probability values are obtained assuming a number of degrees of freedom equal to the number of bins employed subtracted by the number of free parameters in the amplitude fit, including normalization.

The normalization of the model, Eq.~\eqref{eq:model_sig_normalization}, is determined by setting one of the helicity coupling values, defined in Eq.~\eqref{eq:helicity_couplings} of Appendix~\ref{sec:amplitude_model}, to unity, with the value of the other couplings expressed relative to this reference. The $\mathcal{H}^{K^*(892)}_{1/2,0}$ helicity coupling is chosen as reference.
The default model is obtained from the amplitude fit in which the \Lc polarization is expressed in the helicity system reached from the laboratory frame.

The fit fraction, $FF$, for each resonance $R$ is obtained by computing the integral of the amplitude model over the phase-space where only the $R$ contribution is left. Fit fractions are normalized to the complete amplitude model integral, see Appendix~\ref{sec:amplitude_model} for a detailed definition.
For each two-body contribution, the associated decay asymmetry parameter $\alpha$ is measured. It characterizes the two-body decay angular distribution,
\begin{equation}
\frac{d\Gamma}{d\cos\theta_R} \propto \frac{1}{2} \left( 1 + \alpha P \cos\theta_R \right),
\label{eq:two_body_decay_dist}
\end{equation}
in which $\cos\theta_R = \hat{\bm{P}} \cdot \hat{\bm{p}}(R)$ is the cosine of the angle between the polarization vector $\bm{P}$, its modulus denoted by $P$, and the direction of the intermediate resonance in the \Lc rest frame $\hat{\bm{p}}(R)$. The $\alpha$ parameters can be expressed as a combination of helicity coupling squared moduli: for a \Lc weak decay to a pair of states with spin $J$ and $0$ (for $\Lz$, $\Deltares$, $K^*$ spin zero intermediate states) it is
\begin{equation}
\alpha = \frac{|\mathcal{H}_{1/2,0}|^2-|\mathcal{H}_{-1/2,0}|^2}{|\mathcal{H}_{1/2,0}|^2+|\mathcal{H}_{-1/2,0}|^2}.
\label{eq:alpha_par}
\end{equation}

In light of the application of the \Lcpkpi amplitude model as a \Lc polarization analyser, it is important to give a quantitative estimation of the sensitivity to the baryon polarization given by the decay. The sensitivity is related to the amount of parity violation in the decay; for instance, a parity conserving decay has no sensitivity to the decaying baryon polarization. The sensitivity is studied in comparison to that given by two-body decays, for which the sensitivity is given by the decay asymmetry parameter.

The sensitivity to polarization of the decay can be measured by the average event information, $S^2$, evaluated at zero polarization, which, for a given number of events $N$, is inversely proportional to the variance $\sigma^2$ of the polarization measurement,
\begin{align}
S^2 = \frac{1}{N\sigma^2} = \left\langle \frac{g^2}{f^2}(\Omega) \right\rangle = \int \frac{g^2}{f} (\Omega) d\Omega,
\label{eq:fisher_info}
\end{align}
 (see Refs.~\cite{Davier:1992nw,Marangotto:2713231,Aiola:2020yam}) with $f(\Omega)$, $g(\Omega)$ obtained expressing the decay rate PDF Eq.~\eqref{eq:decay_rate_Lc} of Appendix~\ref{sec:amplitude_model} as
\begin{equation}
p_{\text{PDF}}(\Omega|P_z) = f(\Omega) + P_z g(\Omega),
\end{equation}
where the choice of the polarization direction is arbitrary, thanks to rotational invariance. Indeed, $S^2$ does not depend on the polarization direction. In practice, direction-dependent sensitivities could be induced by experimental effects, since detector efficiencies are usually not isotropic.
The average event information for a two-body decay is related to the decay asymmetry parameter $\alpha$ as, applying Eqs.~\eqref{eq:two_body_decay_dist} to~\eqref{eq:fisher_info},
\begin{equation}
S = \frac{|\alpha|}{\sqrt{3}},
\label{eq:S_alpha_relation}
\end{equation}
so that the modulus of $\alpha$ can be seen as a measure of the polarization sensitivity.
The sensitivity to the polarization of the \Lcpkpi three-body decay is obtained from Eq.~\eqref{eq:fisher_info} using its amplitude model. Like two-body decay asymmetry parameters, $S^2$ is a measure of parity violation, which ranges between zero (parity-conservation) and $1/3$ (maximum parity-violation).
For an easier comparison to two-body decays, the sensitivity to polarization is expressed as $\sqrt{3}S$, which ranges between zero and unity.

For \Lc quasi two-body decays to a pair of baryon and pseudoscalar, the $\sqrt{3}S$ quantity is equal to the absolute value of the $\alpha$ decay parameter computed via Eq.~\eqref{eq:alpha_par}. For the $K^*(892)$ contribution, characterized by a different spin structure, the $\sqrt{3}S$ quantity is considered.

\begin{figure}
\centering
\includegraphics[width=0.75\textwidth]{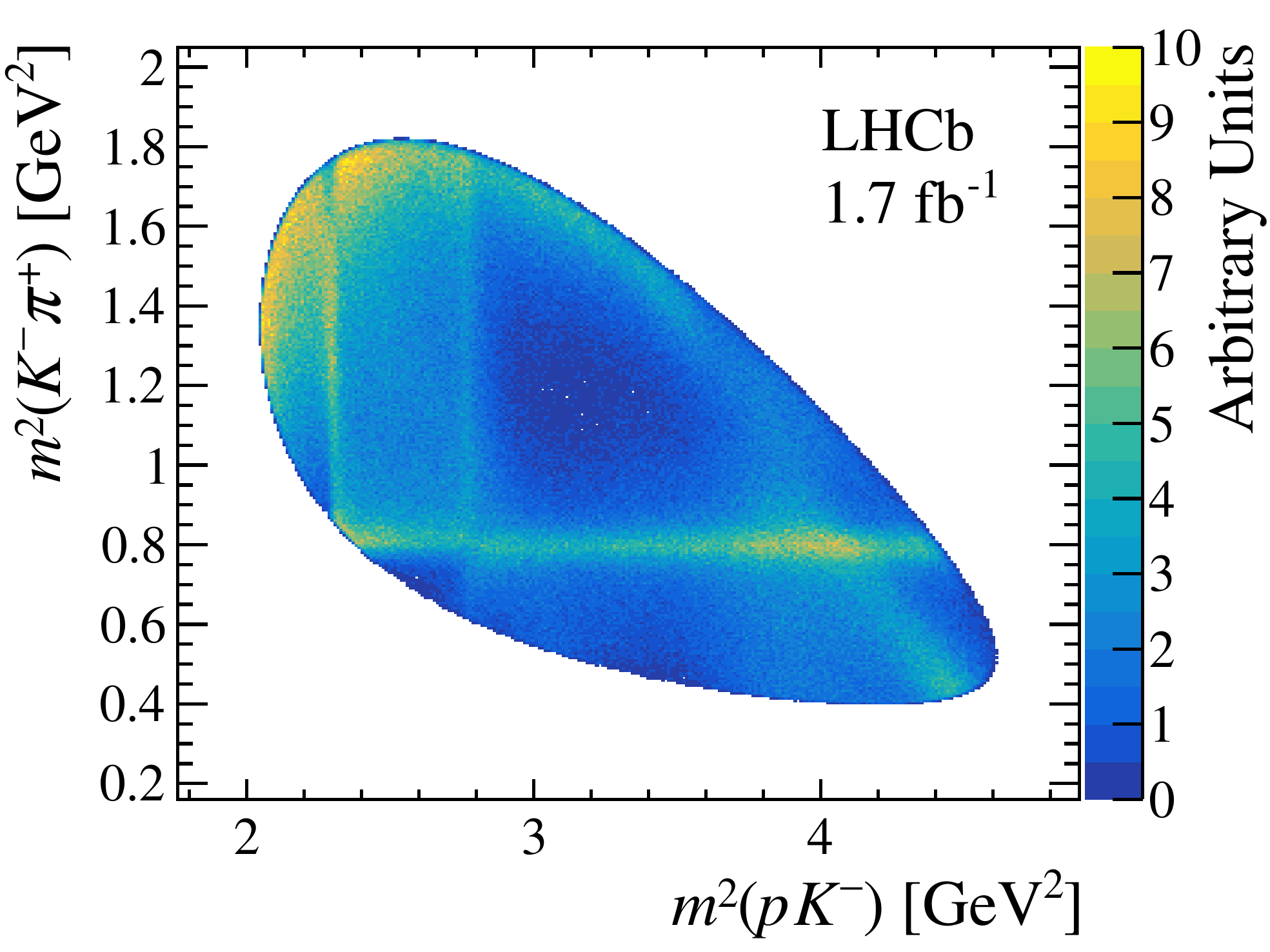}
\caption{Dalitz plot for the total sample of \Lcpkpi candidates. \label{fig:data_dalitz_plot}}
\end{figure}

The Dalitz plot of the total reconstructed  \Lcpkpi sample is presented in Fig.~\ref{fig:data_dalitz_plot}. The plot contains candidates from the signal region prior to efficiency correction. It displays a rich structure with resonant contributions from all three possible pairs of final state particles: $\Lz\to pK^-$ resonances are visible as vertical bands, $K^*\to \Km\pip$ as horizontal bands and $\Deltares^{++}\to p\pip$ as diagonal bands. The different intensity patterns can be explained by the spin of the resonance, by interference patterns or nonuniform detector efficiency.
Regarding the $\Lz$ pattern, there are two narrow structures corresponding to $\Lz(1520)$ and $\Lz(1670)$ resonances, plus broader bands indicating possible $\Lz(1405)$ and $\Lz(1600)$ contributions. The only other vertical band is in the $\mqpk$ region $3.8 - 4.0 \gev$, where no clear resonances are reported according to Ref.~\cite{PDG2020}. Regarding $K^*$ resonances, the clear band is due to the $K^{*}(892)$ meson. Higher-mass resonances, having pole masses outside the allowed phase-space, may contribute thanks to the lower-mass tail of their broad distribution, possibly explaining the presence of an enhanced number of events at high \mqkpi. Even if a spin-zero nonresonant component cannot be excluded visually, the center of the Dalitz plot is almost empty of events. Finally, besides the apparent diagonal band due to the $\Deltares(1232)^{++}$ resonance, there is a slightly enhanced diagonal band at higher \mqppi mass, a possible sign of additional $\Deltares^{++}$ resonances. The Dalitz plot shows interference effects among resonances belonging to different decay chains, which are needed for the simultaneous measurement of helicity amplitudes and \Lc polarization vector~\cite{Marangotto:2020ead}. It can be noted that the $K^*(892)$ band gets shifted when crossing the $\Lz(1670)$ contribution, while the $\Lz(1520)$ band shows a destructive interference pattern with high mass $K^*$ contributions at the upper corner of the Dalitz plot.

The default amplitude model is built starting from the contributions visible in Fig.~\ref{fig:data_dalitz_plot} and adding resonant states according to those listed in Ref.~\cite{PDG2020}. Contributions which significantly improve the fit quality are added to the default model; those giving similar qualities are considered as alternative models for systematic uncertainty evaluation. The same criterion is employed for choosing among different descriptions of the same contribution.

\begin{table}
\centering
\caption{Resonant composition of the default \Lcpkpi amplitude model, with  spin-parity $J^P$, and the Breit--Wigner mass and width parameters, which, in the amplitude fit, are left free within the reported range or fixed to the given value if no interval is quoted. \label{tab:nominal_model}}
\begin{tabular}{lccc}
\toprule
Resonance & $J^P$ & Mass (\mev) & Width (\mev)\\
\midrule
$\Lz(1405)$ & $1/2^-$ & $1405.1$ & $50.5$\\
$\Lz(1520)$ & $3/2^-$ & $1515-1523$ & $10-20$\\
$\Lz(1600)$ & $1/2^+$ & $1630$ & $250$\\
$\Lz(1670)$ & $1/2^-$ & $1670$ & $30$\\
$\Lz(1690)$ & $3/2^-$ & $1690$ & $70$\\
$\Lz(2000)$ & $1/2^-$ & $1900-2100$ & $20-400$\\
\midrule
$\Deltares(1232)^{++}$ & $3/2^+$ & $1232$ & $117$\\
$\Deltares(1600)^{++}$ & $3/2^+$ & $1640$ & $300$\\
$\Deltares(1700)^{++}$ & $3/2^-$ & $1690$ & $380$\\
\midrule
$K_0^{*}(700)$ & $0^+$ & $824$ & $478$\\
$K^{*}(892)$ & $1^-$ & $895.5$ & $47.3$\\
$K_0^{*}(1430)$ & $0^+$ & $1375$ & $190$\\
\bottomrule
\end{tabular}
\end{table}

The resonances included in the default model are listed in Table~\ref{tab:nominal_model}. Most resonance parameters are fixed to the mean values reported in Ref.~\cite{PDG2020}. The parameters of the broad $\Lz(1600)$, $\Deltares(1600)^{++}$, $\Deltares(1700)^{++}$ and $K_0^{*}(1430)$ contributions are set to the edges of the intervals quoted by Ref.~\cite{PDG2020} giving better fit quality. These are the upper values of the $\Lz(1600)$ and $\Deltares(1600)^{++}$ masses, the lower values of the $\Deltares(1700)^{++}$ and $K_0^{*}(1430)$ masses, the upper values of the $\Lz(1600)$, $\Deltares(1600)^{++}$, $\Deltares(1700)^{++}$ widths and the lower value of the $K_0^{*}(1430)$ width. The mass and width of the narrow $\Lz(1520)$ state are left free to absorb resolution effects. A large contribution from a resonant state in the $m(pK^-) \approx 2 \gev$ region is observed and it is well described as a single $J^P = 1/2^-$ state, indicated as $\Lz(2000)$, with Breit--Wigner parameters determined from the fit. Different spin-parity assignments are rejected by the fit.

The invariant mass dependence (lineshape) of resonant contributions is parametrized by default with relativistic Breit--Wigner functions, whose implementation is described in Appendix~\ref{sec:BW}. 
Some of the resonances employed in the amplitude model cannot be parametrized by relativistic Breit--Wigner lineshapes. The spin-zero contribution in the $K^*$ decay channel is modeled as $K^*_0(700)$ and $K^*_0(1430)$ resonant states, each described by a simplified version of the parametrization proposed in Ref.~\cite{Bugg:2005xx}.
It consists of a Breit--Wigner lineshape, Eq.~\eqref{eq:breit_wigner_lineshape}, in which the mass-dependent width is given by
\begin{equation}
\Gamma(m) = \frac{m^2 - s_A}{m^2_0 - s_A} \Gamma_0 e^{-\gamma m^2},
\end{equation}
which features a singularity (Adler zero) at $s_A = m^2_K - 0.5 m^2_\pi$ and an exponential form factor on the $K\pi$ width driven by the parameter $\gamma$. The $\gamma$ parameter is left free separately for each contribution.
Comparing this lineshape to the parametrization presented in Ref.~\cite{Bugg:2005xx}, the additional overall exponential form factor $\exp\left[-\alpha q^2(m)\right]$, with $q(m)$ the momentum of the decay products in the $\Lc\to K^*p$ decay, and the opening of $K\eta$, $K\eta'$ decay channels, are neglected in the default fit.

The $\Lz(1405)$ resonance, having its pole mass below the $pK^-$ threshold, is parametrized by a Flatt\'{e} lineshape~\cite{Flatte:1976xu} describing the opening of the $pK$ decay channel in addition to the $\Sigma\pi$ decay.
It is composed of a Breit--Wigner lineshape, Eq.~\eqref{eq:breit_wigner_lineshape}, with a total width being the sum of the widths associated to the two decay channels,
\begin{equation}
\Gamma(m) = \Gamma_{pK}(m) + \Gamma_{\Sigma\pi}(m),
\end{equation}
which are calculated considering the decay to $\Sigma\pi$ (see Appendix~\ref{sec:BW}). Assuming that both channels are dynamically equally likely and differ only by the phase-space factors, the Breit--Wigner width $\Gamma_0$ is set to the total width of the $\Lz(1405)$ resonance in both terms.

\section{Uncertainties and consistency checks}
\label{sec:systematic}

Statistical and systematic uncertainties are computed for amplitude model parameters, polarization components, fit fractions, decay asymmetry parameters and sensitivity to the polarization.
Statistical uncertainties are obtained fitting the default model to 1000 pseudoexperiments sampled with Monte Carlo methods from the same model with parameters fixed to the central values obtained in the real data fit.
For each pseudoexperiment the simulation sample used to compute the model normalization is also
generated to reproduce finite size effects. Pseudoexperiments and simulation samples are generated with the same size as for the amplitude fit. Statistical uncertainties for each parameter are determined as the standard deviation
of the distribution of the results from the pseudoexperiments.
Statistical uncertainties are reported along with the final results in Sec.~\ref{sec:results}; they are found to be larger than the Hessian uncertainties computed in the fit but significantly smaller than the main source of systematic uncertainty.

Different sources of systematic uncertainties are considered. These are grouped into contributions coming from the model choice, the background determination, the kinematics of the decay, the PID, and the fit bias. Since the model choice is the largest contribution for most of the measured quantities, it is quoted separately from the other systematic contributions, which are combined in the final results.

The systematic uncertainty associated to the amplitude model choice is estimated by determining the measured parameters employing alternative models, with fit quality similar to the default fit. They include: leaving as free fit parameters the Breit--Wigner parameters of resonances having fixed mass and width in the default model; addition of the $\Lz(1800)$, $\Lz(1810)$ and $\Deltares(1620)^{++}$ states; use of a relativistic Breit--Wigner function for describing the $K^*_0(700)$ and $K^*_0(1430)$ lineshapes; varying the \Lc radius in the lineshape (see Appendix~\ref{sec:BW}); and using spin-orbit instead of helicity couplings. Each modification to the default amplitude model described in Table~\ref{tab:nominal_model} is considered separately. The maximum absolute variation among the alternative models, with respect to the default model parameters, is assigned as systematic uncertainty.

The uncertainty associated to the background description includes the uncertainty on the background fraction $f_b$, estimated using an alternative model for the $m(p\Km\pip)$ mass shape, and that on the Legendre polynomial parametrization, estimated using an alternative parametrization determined from the upper mass sideband only. The uncertainty on $f_b$ is at the $10^{-3}$ level.
The uncertainties related to the corrections applied to the simulation are estimated varying the calibration samples employed and the functional form of the corrections, separately for the baryon kinematics and particle identification. A possible bias in the determination of fit parameters is considered by assigning the mean deviation of 1000 pseudoexperiments from the default result as systematic uncertainty. 
Systematic uncertainties for contributions from fit parameters, polarization components, fit fractions, and decay asymmetries are reported in Tables~\ref{tab:syst_fit_pars_LD}--\ref{tab:syst_alphapars}.
\begin{table}
\centering
\caption{Systematic uncertainty contributions on fit parameters describing the $\Lz$ contributions. Total* includes all contributions except for the choice of the amplitude model.\label{tab:syst_fit_pars_LD}}
\begin{tabular}{lcccccc}
\toprule
Parameter & Model Choice & Total* & Background & Kinematics & PID & Fit Bias \\
\midrule 
Re$\mathcal{H}^{\Lz(1405)}_{1/2,0}$ & 3.3 & 0.1 & 0.0 & 0.0 & 0.0 & 0.0 \\
Im$\mathcal{H}^{\Lz(1405)}_{1/2,0}$ & 3.2 & 0.1 & 0.1 & 0.0 & 0.1 & 0.0 \\
Re$\mathcal{H}^{\Lz(1405)}_{-1/2,0}$ & 12 & 0.2 & 0.1 & 0.1 & 0.0 & 0.1 \\
Im$\mathcal{H}^{\Lz(1405)}_{-1/2,0}$ & 3.7 & 0.3 & 0.2 & 0.1 & 0.2 & 0.0 \\
\midrule
Re$\mathcal{H}^{\Lz(1520)}_{1/2,0}$ & 0.12 & 0.01 & 0.00 & 0.00 & 0.00 & 0.00 \\
Im$\mathcal{H}^{\Lz(1520)}_{1/2,0}$ & 0.12 & 0.02 & 0.01 & 0.00 & 0.01 & 0.00 \\
Re$\mathcal{H}^{\Lz(1520)}_{-1/2,0}$ & 0.69 & 0.03 & 0.01 & 0.02 & 0.02 & 0.00 \\
Im$\mathcal{H}^{\Lz(1520)}_{-1/2,0}$ & 1.3 & 0.0 & 0.0 & 0.0 & 0.0 & 0.0 \\
$m^{\Lz(1520)} \left[\mev\right]$ & 0.65 & 0.03 & 0.03 & 0.01 & 0.01 & 0.01 \\
$\Gamma^{\Lz(1520)} \left[\mev\right]$ & 1.3 & 0.1 & 0.1 & 0.1 & 0.1 & 0.0 \\
\midrule
Re$\mathcal{H}^{\Lz(1600)}_{1/2,0}$ & 5.0 & 0.1 & 0.1 & 0.0 & 0.0 & 0.0 \\
Im$\mathcal{H}^{\Lz(1600)}_{1/2,0}$ & 3.7 & 0.1 & 0.1 & 0.0 & 0.1 & 0.0 \\
Re$\mathcal{H}^{\Lz(1600)}_{-1/2,0}$ & 8.7 & 0.1 & 0.0 & 0.0 & 0.0 & 0.1 \\
Im$\mathcal{H}^{\Lz(1600)}_{-1/2,0}$ & 2.0 & 0.2 & 0.0 & 0.1 & 0.2 & 0.0 \\
\midrule
Re$\mathcal{H}^{\Lz(1670)}_{1/2,0}$ & 0.35 & 0.01 & 0.00 & 0.00 & 0.01 & 0.00 \\
Im$\mathcal{H}^{\Lz(1670)}_{1/2,0}$ & 0.22 & 0.02 & 0.01 & 0.00 & 0.02 & 0.00 \\
Re$\mathcal{H}^{\Lz(1670)}_{-1/2,0}$ & 0.46 & 0.02 & 0.01 & 0.01 & 0.02 & 0.00 \\
Im$\mathcal{H}^{\Lz(1670)}_{-1/2,0}$ & 1.2 & 0.0 & 0.0 & 0.0 & 0.0 & 0.0 \\
\midrule
Re$\mathcal{H}^{\Lz(1690)}_{1/2,0}$ & 0.23 & 0.02 & 0.01 & 0.01 & 0.01 & 0.00 \\
Im$\mathcal{H}^{\Lz(1690)}_{1/2,0}$ & 0.44 & 0.02 & 0.02 & 0.00 & 0.01 & 0.00 \\
Re$\mathcal{H}^{\Lz(1690)}_{-1/2,0}$ & 2.4 & 0.0 & 0.0 & 0.0 & 0.0 & 0.0 \\
Im$\mathcal{H}^{\Lz(1690)}_{-1/2,0}$ & 0.60 & 0.06 & 0.04 & 0.03 & 0.03 & 0.00 \\
\midrule
Re$\mathcal{H}^{\Lz(2000)}_{1/2,0}$ & 11 & 0 & 0 & 0 & 0 & 0 \\
Im$\mathcal{H}^{\Lz(2000)}_{1/2,0}$ & 7.7 & 0.2 & 0.2 & 0.0 & 0.1 & 0.0 \\
Re$\mathcal{H}^{\Lz(2000)}_{-1/2,0}$ & 3.4 & 0.2 & 0.1 & 0.0 & 0.1 & 0.0 \\
Im$\mathcal{H}^{\Lz(2000)}_{-1/2,0}$ & 3.7 & 0.1 & 0.1 & 0.0 & 0.0 & 0.0 \\
$m^{\Lz(2000)} \left[\mev\right]$ & 21 & 1 & 1 & 0 & 0 & 0 \\
$\Gamma^{\Lz(2000)} \left[\mev\right]$ & 16 & 3 & 3 & 1 & 0 & 0 \\
\bottomrule
\end{tabular}
\end{table}
\begin{table}
\centering
\caption{Systematic uncertainty contributions on fit parameters describing the $K^*$ and $\Deltares^{++}$ contributions. Total* includes all contributions except for the choice of the amplitude model. \label{tab:syst_fit_pars_K}}
\begin{tabular}{lcccccc}
\toprule
Parameter & Model Choice & Total* & Background & Kinematics & PID & Fit Bias \\
\midrule 
Re$\mathcal{H}^{K^*_0(700)}_{1/2,0}$ & 2.1 & 0.2 & 0.2 & 0.1 & 0.0 & 0.0 \\
Im$\mathcal{H}^{K^*_0(700)}_{1/2,0}$ & 1.3 & 0.1 & 0.0 & 0.0 & 0.0 & 0.0 \\
Re$\mathcal{H}^{K^*_0(700)}_{-1/2,0}$ & 0.93 & 0.09 & 0.07 & 0.00 & 0.05 & 0.01 \\
Im$\mathcal{H}^{K^*_0(700)}_{-1/2,0}$ & 3.2 & 0.1 & 0.1 & 0.0 & 0.0 & 0.0 \\
$\gamma^{K^*_0(700)} \left[\gev^{-2}\right]$ & 0.35 & 0.04 & 0.03 & 0.02 & 0.00 & 0.00 \\
\midrule
Re$\mathcal{H}^{K^*(892)}_{1/2,0}$ & \multicolumn{6}{c}{0 (fixed)}\\
Im$\mathcal{H}^{K^*(892)}_{1/2,0}$ & \multicolumn{6}{c}{0 (fixed)}\\
Re$\mathcal{H}^{K^*(892)}_{1/2,-1}$ & 0.85 & 0.05 & 0.04 & 0.01 & 0.02 & 0.01 \\
Im$\mathcal{H}^{K^*(892)}_{1/2,-1}$ & 0.96 & 0.02 & 0.02 & 0.00 & 0.00 & 0.01 \\
Re$\mathcal{H}^{K^*(892)}_{-1/2,1}$ & 2.4 & 0.1 & 0.0 & 0.0 & 0.0 & 0.0 \\
Im$\mathcal{H}^{K^*(892)}_{-1/2,1}$ & 2.7 & 0.1 & 0.1 & 0.0 & 0.0 & 0.0 \\
Re$\mathcal{H}^{K^*(892)}_{-1/2,0}$ & 1.5 & 0.1 & 0.0 & 0.0 & 0.1 & 0.0 \\
Im$\mathcal{H}^{K^*(892)}_{-1/2,0}$ & 4.5 & 0.1 & 0.1 & 0.0 & 0.0 & 0.0 \\
\midrule
Re$\mathcal{H}^{K^*_0(1430)}_{1/2,0}$ & 2.4 & 0.3 & 0.2 & 0.1 & 0.1 & 0.0 \\
Im$\mathcal{H}^{K^*_0(1430)}_{1/2,0}$ & 9.7 & 0.1 & 0.1 & 0.0 & 0.0 & 0.1 \\
Re$\mathcal{H}^{K^*_0(1430)}_{-1/2,0}$ & 5.4 & 0.4 & 0.3 & 0.1 & 0.2 & 0.0 \\
Im$\mathcal{H}^{K^*_0(1430)}_{-1/2,0}$ & 14 & 0 & 0 & 0 & 0 & 0 \\
$\gamma^{K^*_0(1430)} \left[\gev^{-2}\right]$ & 0.33 & 0.01 & 0.00 & 0.00 & 0.00 & 0.00 \\
\midrule
Re$\mathcal{H}^{\Deltares(1232)^{++}}_{1/2,0}$ & 5.3 & 0.2 & 0.2 & 0.0 & 0.0 & 0.0 \\
Im$\mathcal{H}^{\Deltares(1232)^{++}}_{1/2,0}$ & 1.4 & 0.1 & 0.1 & 0.0 & 0.1 & 0.0 \\
Re$\mathcal{H}^{\Deltares(1232)^{++}}_{-1/2,0}$ & 12 & 0 & 0 & 0 & 0 & 0 \\
Im$\mathcal{H}^{\Deltares(1232)^{++}}_{-1/2,0}$ & 3.7 & 0.3 & 0.3 & 0.1 & 0.2 & 0.0 \\
\midrule
Re$\mathcal{H}^{\Deltares(1600)^{++}}_{1/2,0}$ & 9.8 & 0.2 & 0.2 & 0.1 & 0.1 & 0.1 \\
Im$\mathcal{H}^{\Deltares(1600)^{++}}_{1/2,0}$ & 1.4 & 0.2 & 0.0 & 0.1 & 0.2 & 0.0 \\
Re$\mathcal{H}^{\Deltares(1600)^{++}}_{-1/2,0}$ & 7.2 & 0.2 & 0.2 & 0.1 & 0.1 & 0.0 \\
Im$\mathcal{H}^{\Deltares(1600)^{++}}_{-1/2,0}$ & 2.1 & 0.1 & 0.0 & 0.0 & 0.1 & 0.0 \\
\midrule
Re$\mathcal{H}^{\Deltares(1700)^{++}}_{1/2,0}$ & 7.2 & 0.3 & 0.3 & 0.1 & 0.0 & 0.1 \\
Im$\mathcal{H}^{\Deltares(1700)^{++}}_{1/2,0}$ & 5.2 & 0.2 & 0.1 & 0.1 & 0.1 & 0.0 \\
Re$\mathcal{H}^{\Deltares(1700)^{++}}_{-1/2,0}$ & 13 & 0 & 0 & 0 & 0 & 0 \\
Im$\mathcal{H}^{\Deltares(1700)^{++}}_{-1/2,0}$ & 6.0 & 0.3 & 0.2 & 0.1 & 0.1 & 0.0 \\
\bottomrule
\end{tabular}
\end{table}
\begin{table}
\centering
\caption{Systematic uncertainty contributions on polarization components in percentage. Total* includes all contributions except for the choice of the amplitude model. \label{tab:syst_pol}}
\begin{tabular}{lcccccc}
\toprule
Parameter & Model Choice & Total* & Background & Kinematics & PID & Fit Bias \\
\midrule 
$P_x$ (\textit{lab}) & 0.98 & 0.21 & 0.10 & 0.17 & 0.05 & 0.05 \\
$P_y$ (\textit{lab}) & 0.16 & 0.07 & 0.03 & 0.04 & 0.05 & 0.00 \\
$P_z$ (\textit{lab}) & 0.3\phantom{0} & 1.1\phantom{0} & 0.0\phantom{0} & 1.0\phantom{0} & 0.2\phantom{0} & 0.0\phantom{0} \\
\midrule
$P_x$ ($\tilde{B}$) & 0.36\phantom{0} & 0.15\phantom{0} & 0.02\phantom{0} & 0.15\phantom{0} & 0.01\phantom{0} & 0.05\phantom{0} \\
$P_y$ ($\tilde{B}$) & 0.087 & 0.081 & 0.019 & 0.057 & 0.054 & 0.001 \\
$P_z$ ($\tilde{B}$) & 1.1\phantom{00} & 0.1\phantom{00} & 0.1\phantom{00} & 0.1\phantom{00} & 0.1\phantom{00} & 0.0\phantom{00} \\
\bottomrule
\end{tabular}
\end{table}
\begin{table}
\centering
\caption{Systematic uncertainty contributions on fit fractions. Total* includes all contributions except for the choice of the amplitude model. \label{tab:syst_fitfraction}}
\begin{tabular}{lcccccc}
\toprule
Resonance FF & Model Choice & Total* & Background & Kinematics & PID & Fit Bias \\
\midrule 
$\Lz(1405)$ & 0.030\phantom{0} & 0.002\phantom{0} & 0.002\phantom{0} & 0.001\phantom{0} & 0.000\phantom{0} & 0.000\phantom{0} \\
$\Lz(1520)$ & 0.0023 & 0.0003 & 0.0000 & 0.0002 & 0.0001 & 0.0002 \\
$\Lz(1600)$ & 0.019\phantom{0} & 0.001\phantom{0} & 0.001\phantom{0} & 0.000\phantom{0} & 0.000\phantom{0} & 0.000\phantom{0} \\
$\Lz(1670)$ & 0.0032 & 0.0001 & 0.0001 & 0.0000 & 0.0001 & 0.0000 \\
$\Lz(1690)$ & 0.0034 & 0.0001 & 0.0001 & 0.0001 & 0.0001 & 0.0000 \\
$\Lz(2000)$ & 0.0093 & 0.0023 & 0.0020 & 0.0010 & 0.0004 & 0.0001 \\
$\Deltares(1232)^{++}$ & 0.0076 & 0.0016 & 0.0014 & 0.0003 & 0.0006 & 0.0002 \\
$\Deltares(1600)^{++}$ & 0.015\phantom{0} & 0.001\phantom{0} & 0.000\phantom{0} & 0.000\phantom{0} & 0.000\phantom{0} & 0.000\phantom{0} \\
$\Deltares(1700)^{++}$ & 0.0094 & 0.0007 & 0.0005 & 0.0004 & 0.0001 & 0.0001 \\
$K^*_0(700)$ & 0.0092 & 0.0018 & 0.0018 & 0.0005 & 0.0002 & 0.0001 \\
$K^*(892)$ & 0.0064 & 0.0004 & 0.0000 & 0.0003 & 0.0002 & 0.0002 \\
$K^*_0(1430)$ & 0.027\phantom{0} & 0.001\phantom{0} & 0.001\phantom{0} & 0.001\phantom{0} & 0.000\phantom{0} & 0.000\phantom{0} \\
\bottomrule
\end{tabular}
\end{table}
\begin{table}
\centering
\caption{Systematic uncertainties on $\sqrt{3}S$ and decay asymmetry parameters. Total* includes all contributions except for the choice of the amplitude model. \label{tab:syst_alphapars}}
\begin{tabular}{lcccccc}
\toprule
$\alpha$ & Model Choice & Total* & Background & Kinematics & PID & Fit Bias \\
\midrule 
Model $\sqrt{3}S$ & 0.010 & 0.007 & 0.002 & 0.002 & 0.001 & 0.007 \\
$K^*(892)$ $\sqrt{3}S$ & 0.023 & 0.003 & 0.001 & 0.000 & 0.002 & 0.002 \\
\midrule
$\Lz(1405)$ & 0.28\phantom{0} & 0.01\phantom{0} & 0.01\phantom{0} & 0.00\phantom{0} & 0.00\phantom{0} & 0.00\phantom{0} \\
$\Lz(1520)$ & 0.084 & 0.005 & 0.000 & 0.001 & 0.003 & 0.004 \\
$\Lz(1600)$ & 0.50\phantom{0} & 0.03\phantom{0} & 0.02\phantom{0} & 0.01\phantom{0} & 0.00\phantom{0} & 0.01\phantom{0} \\
$\Lz(1670)$ & 0.073 & 0.006 & 0.001 & 0.002 & 0.004 & 0.003 \\
$\Lz(1690)$ & 0.027 & 0.006 & 0.004 & 0.002 & 0.002 & 0.004 \\
$\Lz(2000)$ & 0.19\phantom{0} & 0.01\phantom{0} & 0.01\phantom{0} & 0.00\phantom{0} & 0.00\phantom{0} & 0.00\phantom{0} \\
$\Deltares(1232)^{++}$ & 0.036 & 0.004 & 0.003 & 0.000 & 0.002 & 0.000 \\
$\Deltares(1600)^{++}$ & 0.17\phantom{0} & 0.01\phantom{0} & 0.01\phantom{0} & 0.00\phantom{0} & 0.01\phantom{0} & 0.00\phantom{0} \\
$\Deltares(1700)^{++}$ & 0.075 & 0.011 & 0.007 & 0.005 & 0.003 & 0.006 \\
$K^*_0(700)$ & 0.24\phantom{0} & 0.23\phantom{0} & 0.02\phantom{0} & 0.01\phantom{0} & 0.00\phantom{0} & 0.23\phantom{0} \\
$K^*_0(1430)$ & 0.14\phantom{0} & 0.01\phantom{0} & 0.01\phantom{0} & 0.00\phantom{0} & 0.01\phantom{0} & 0.00\phantom{0} \\
\bottomrule
\end{tabular}
\end{table}

The stability of the default amplitude model is checked by repeating the fit splitting the data set for different data-taking magnet polarities, \Lc charge, \Lc transverse momentum, \Lc lifetime, and polarization systems. All the amplitude models obtained are compatible within uncertainties with the default one.
Detector resolution effects on invariant masses are found to be one order of magnitude smaller than the width of the two narrowest structures, $\Lz(1520)$ and $\Lz(1670)$, and are not considered.

\clearpage

\section{Results}
\label{sec:results}

\begin{figure}
\centering
\includegraphics[width=\textwidth]{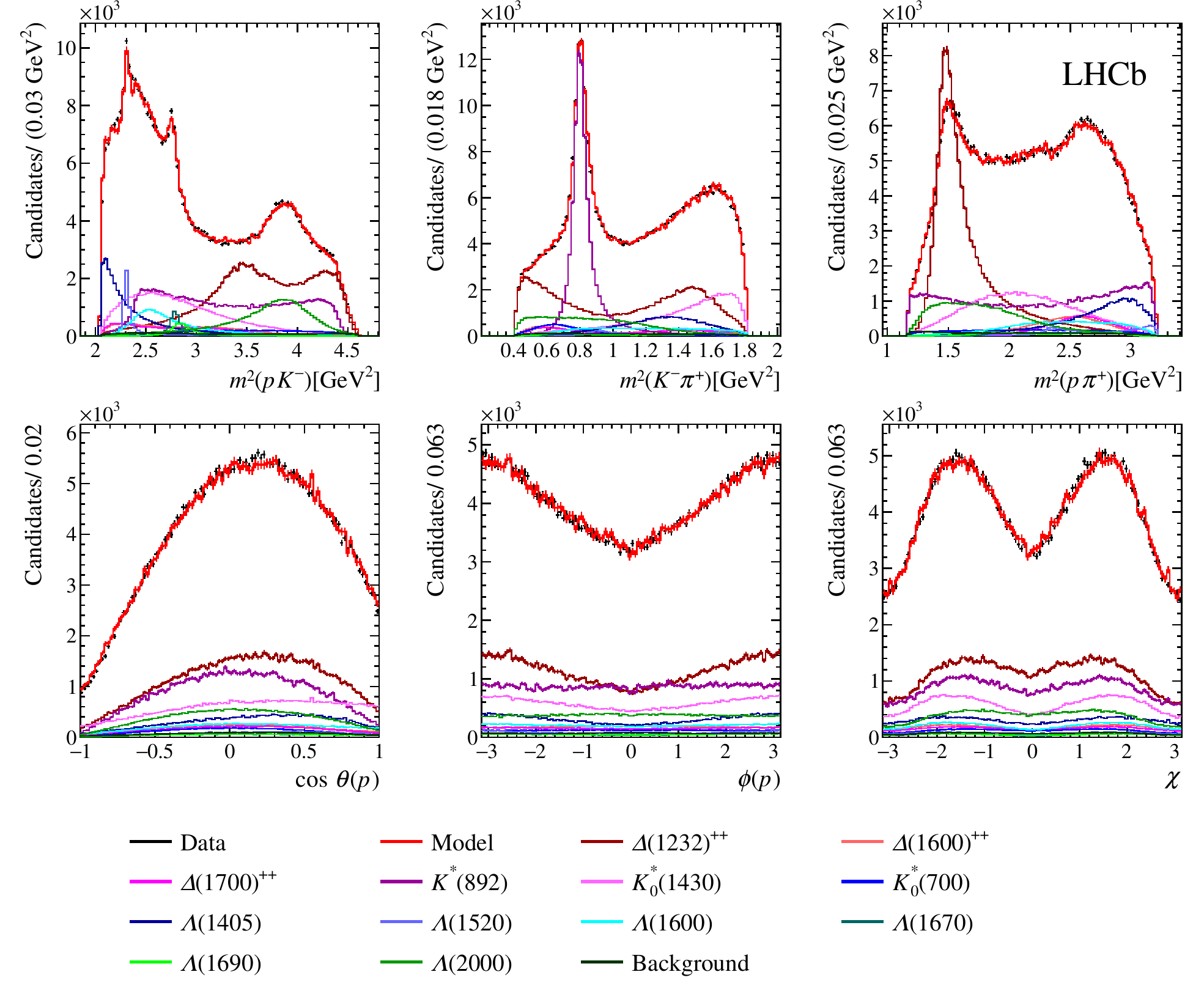}
\caption{Distributions for selected candidates together with amplitude fit projections in the \textit{lab} system for (top row) invariant mass squared projections; (bottom row) decay orientation angle projections.\label{fig:data_100k_nominal}}
\end{figure}

\begin{figure}
\centering
\includegraphics[width=\textwidth]{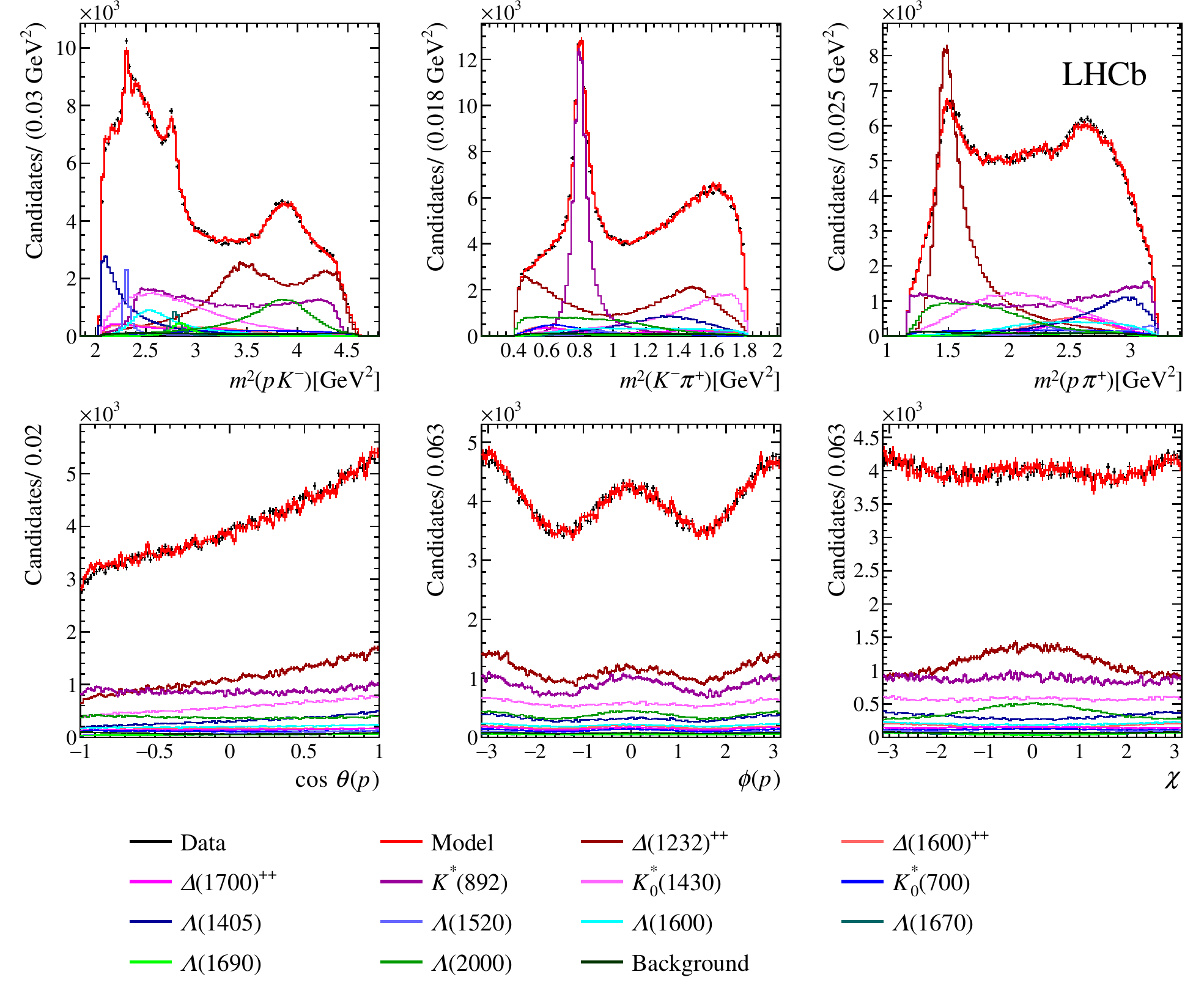}
\caption{Distributions for selected candidates together with amplitude fit projections in the $\tilde{B}$ system for (top row) invariant mass squared projections; (bottom row) decay orientation angle projections.\label{fig:data_100k_nominal_boost}}
\end{figure}

The comparison between \Lcpkpi data and default amplitude fit projections is displayed in Figs.~\ref{fig:data_100k_nominal} and~\ref{fig:data_100k_nominal_boost} for \Lc polarization in the laboratory or approximate $B$ systems, respectively. The amplitude model distributions are obtained from the \Lcpkpi simulation sample which reproduces detector efficiency effects. The fit qualities are good given the large number of events.

The polarization components in the laboratory and approximate $B$ systems are reported in Table~\ref{tab:results_final_pol}.
This analysis demonstrates the possibility of a precision measurement of the full \Lc polarization vector, with absolute uncertainties of order $1\%$ on each component. A large polarization is measured in both  \Lc helicity frames considered. In the one reached from the laboratory it has a modulus $P \approx 65\%$, with a dominating positive transverse component $P_x \approx 60\%$ and a smaller negative longitudinal component $P_z \approx -25\%$. In the system reached from the approximate beauty hadron rest frame it has a modulus $P \approx 70\%$, with a dominating negative longitudinal component $P_z \approx -66\%$ and a smaller positive transverse component $P_x \approx 22\%$. The latter polarization components follow the theoretical predictions~\cite{Konig:1993wz, Gutsche:2015mxa, Li:2016pdv, Faustov:2016pal, Ray:2018hrx, Hu:2020axt, Mu:2019bin}. A precise comparison of the measured \Lc polarization values with theoretical predictions is beyond the scope of the present analysis. This would require an exclusive selection of $\Lb\to\Lc \mun \nu$ decays from other contributions, like those from excited charm resonances $\Lb\to\Lz_c^{*+}(\to \Lc \pip\pim) \mun \nu$, and an understanding of the effect of partial semileptonic decay reconstruction on the polarization components obtained from the true beauty hadron rest frame.

The relation between the two measured polarization vectors can be explained in terms of the different orientation in space of the two polarization systems considered, with the rotation from mostly longitudinal to mostly transverse polarization reproduced by a study performed on simulated events.
The leading uncertainty for transverse polarization in both systems and longitudinal polarization in the $\tilde{B}$ frame is that associated to the amplitude model choice, while the main uncertainty on longitudinal polarization in laboratory frame is due to baryon kinematic corrections.
The normal polarization $P_y$, sensitive to time-reversal violation effects and final-state interactions as explained in Appendix~\ref{sec:amplitude_model}, is compatible with zero at the $1\%$ level, for both  polarization systems considered.
The dominant uncertainty on $P_y$ is statistical. The reduced impact of systematic contributions on $P_y$ is explained by the fact that this component is compatible with zero: indeed the presence of a polarization component is model-independent (can be seen directly from decay orientation angle distribution), while the determination of the polarization magnitude depends on the resonance interference pattern described by the amplitude model~\cite{Marangotto:2020ead}.

The measured parameters of the default amplitude model, for \Lc polarization measured in the \textit{lab} system, are reported in Tables~\ref{tab:results_final},\ref{tab:results_final_K}, fit fractions for each resonant contribution in Table~\ref{tab:fitfractions_final}, $\sqrt{3}S$ and two-body decay asymmetry parameters in Table~\ref{tab:alphapars_final}.
As studied in Ref.~\cite{Marangotto:2020ead}, the \Lcpkpi amplitude analysis is sensitive to all the parameters describing the \Lcpkpi amplitude model, thanks to the significant \Lc polarization and interference effects among different decay chains. The leading uncertainty comes from the amplitude model choice for all parameters.

A significant contribution from a resonant state in the $m(pK^-) \approx 2 \gev$ region, where clear resonances have not previously been reported by Ref.~\cite{PDG2020}, has been established. This contribution is well described as a single $J^P = 1/2^-$ state, with Breit--Wigner parameters $m = 1988 \pm 2 \pm 21\mev$ and $\Gamma = 179 \pm 4 \pm 16 \mev$, where the first uncertainty is statistical the second combines all the systematic sources. The closest resonance reported by Ref.~\cite{PDG2020} is the $\Lz(2000)$: the measured values are fairly compatible with those quoted by Ref.~\cite{Zhang:2013sva}, $m = 2020 \pm 16 \mev$ and $\Gamma = 255 \pm 63 \mev$, and Ref.~\cite{Cameron:1978qi}, $m = 2030 \pm 30 \mev$ and $\Gamma = 125 \pm 25 \mev$. This structure can be therefore interpreted as a $\Lz(2000)$ state contribution.

The largest contributions to the amplitude model, measured from fit fractions, come from the $\Deltares(1232)^{++}$, $K^*(892)$ and $K^*_0(1430)$ resonances. Among the $\Lz$ resonances the largest contributions are from the $\Lz(1405)$ and $\Lz(2000)$ states. The $\Lz(1520)$ parameters are compatible with those reported by Ref.~\cite{PDG2020}.

A large sensitivity of the \Lcpkpi decay to the polarization is measured, \mbox{$\sqrt{3}S = 0.662 \pm 0.005 \pm 0.015$}, which is also an observation of parity-violation in the decay. The large sensitivity, combined with the large value of the branching fraction, makes the \Lcpkpi decay the best probe for \Lc polarization and spin precession measurements, providing the smallest uncertainties. 
Many two-body decay asymmetry parameters are significantly different from zero, indicating parity-violation. In particular, the $\alpha$ parameter associated to the $3/2^+$ $\Deltares(1232)^{++}$ contribution is nonzero, in contrast with the prediction of Ref.~\cite{Korner:1992wi}.

\begin{table}
\centering
\caption{Measured polarization components. The first uncertainty is statistical, the second is the amplitude model choice systematic contribution and the third is the combination of the other systematic uncertainties.\label{tab:results_final_pol}}
\begin{tabular}{lc}
\toprule
Component & Value (\%)\\
\midrule
$P_x$ (\textit{lab}) & $60.32 \pm 0.68 \pm 0.98 \pm 0.21$ \\
$P_y$ (\textit{lab}) & $-0.41 \pm 0.61 \pm 0.16 \pm 0.07$ \\
$P_z$ (\textit{lab}) & $-24.7 \pm 0.6 \pm 0.3 \pm 1.1$ \\
\midrule
$P_x$ ($\tilde{B}$) & $21.65 \pm 0.68 \pm 0.36 \pm 0.15$ \\
$P_y$ ($\tilde{B}$) & $1.08 \pm 0.61 \pm 0.09 \pm 0.08$ \\
$P_z$ ($\tilde{B}$) & $-66.5 \pm 0.6 \pm 1.1 \pm 0.1$ \\
\bottomrule
\end{tabular}
\end{table}

\begin{table}
\centering
\caption{Default amplitude model measured fit parameters describing the $\Lz$ contributions. \label{tab:results_final}}
\begin{tabular}{lcccc}
\toprule
Parameter & Central Value & Stat. Unc. & Model Unc. & Syst. Unc. \\
\midrule 
Re$\mathcal{H}^{\Lz(1405)}_{1/2,0}$ & $-$4.6  & 0.5  & 3.3  & 0.1 \\
Im$\mathcal{H}^{\Lz(1405)}_{1/2,0}$ & 3.2  & 0.5  & 3.2  & 0.1 \\
Re$\mathcal{H}^{\Lz(1405)}_{-1/2,0}$ & 10  & 1  & 12  & 0 \\
Im$\mathcal{H}^{\Lz(1405)}_{-1/2,0}$ & 2.8  & 1.1  & 3.7  & 0.3 \\
\midrule
Re$\mathcal{H}^{\Lz(1520)}_{1/2,0}$ & 0.29  & 0.05  & 0.12  & 0.01 \\
Im$\mathcal{H}^{\Lz(1520)}_{1/2,0}$ & 0.04  & 0.05  & 0.12  & 0.02 \\
Re$\mathcal{H}^{\Lz(1520)}_{-1/2,0}$ & $-$0.16  & 0.14  & 0.69  & 0.03 \\
Im$\mathcal{H}^{\Lz(1520)}_{-1/2,0}$ & 1.5  & 0.1  & 1.3  & 0.0 \\
$m^{\Lz(1520)} \left[\mev\right]$ & 1518.47  & 0.36  & 0.65  & 0.03 \\
$\Gamma^{\Lz(1520)} \left[\mev\right]$ & 15.2  & 0.8  & 1.3  & 0.1 \\
\midrule
Re$\mathcal{H}^{\Lz(1600)}_{1/2,0}$ & 4.8  & 0.5  & 5.0  & 0.1 \\
Im$\mathcal{H}^{\Lz(1600)}_{1/2,0}$ & 3.1  & 0.5  & 3.7  & 0.1 \\
Re$\mathcal{H}^{\Lz(1600)}_{-1/2,0}$ & $-$7.0  & 0.5  & 8.7  & 0.1 \\
Im$\mathcal{H}^{\Lz(1600)}_{-1/2,0}$ & 0.8  & 0.6  & 2.0  & 0.2 \\
\midrule
Re$\mathcal{H}^{\Lz(1670)}_{1/2,0}$ & $-$0.34  & 0.05  & 0.35  & 0.01 \\
Im$\mathcal{H}^{\Lz(1670)}_{1/2,0}$ & $-$0.14  & 0.05  & 0.22  & 0.02 \\
Re$\mathcal{H}^{\Lz(1670)}_{-1/2,0}$ & $-$0.57  & 0.10  & 0.46  & 0.02 \\
Im$\mathcal{H}^{\Lz(1670)}_{-1/2,0}$ & 1.0  & 0.1  & 1.2  & 0.0 \\
\midrule
Re$\mathcal{H}^{\Lz(1690)}_{1/2,0}$ & $-$0.39  & 0.10  & 0.23  & 0.02 \\
Im$\mathcal{H}^{\Lz(1690)}_{1/2,0}$ & $-$0.11  & 0.09  & 0.44  & 0.02 \\
Re$\mathcal{H}^{\Lz(1690)}_{-1/2,0}$ & $-$2.7  & 0.2  & 2.4  & 0.0 \\
Im$\mathcal{H}^{\Lz(1690)}_{-1/2,0}$ & $-$0.35  & 0.23  & 0.60  & 0.06 \\
\midrule
Re$\mathcal{H}^{\Lz(2000)}_{1/2,0}$ & $-$8  & 1  & 11  & 0 \\
Im$\mathcal{H}^{\Lz(2000)}_{1/2,0}$ & $-$7.6  & 0.8  & 7.7  & 0.2 \\
Re$\mathcal{H}^{\Lz(2000)}_{-1/2,0}$ & $-$4.3  & 0.5  & 3.4  & 0.2 \\
Im$\mathcal{H}^{\Lz(2000)}_{-1/2,0}$ & $-$3.8  & 0.4  & 3.7  & 0.1 \\
$m^{\Lz(2000)} \left[\mev\right]$ & 1988  & 2  & 21  & 1 \\
$\Gamma^{\Lz(2000)} \left[\mev\right]$ & 179  & 4  & 16  & 3 \\
\bottomrule
\end{tabular}
\end{table}

\begin{table}
\centering
\caption{Default amplitude model measured fit parameters describing the $K^*$ and $\Deltares^{++}$ contributions. \label{tab:results_final_K}}
\begin{tabular}{lcccc}
\toprule
Parameter & Central Value & Stat. Unc. & Model Unc. & Syst. Unc. \\
\midrule 
Re$\mathcal{H}^{K^*_0(700)}_{1/2,0}$ & $-$2.7  & 0.2  & 2.1  & 0.2 \\
Im$\mathcal{H}^{K^*_0(700)}_{1/2,0}$ & 0.0  & 0.3  & 1.3  & 0.1 \\
Re$\mathcal{H}^{K^*_0(700)}_{-1/2,0}$ & 0.07  & 0.23  & 0.93  & 0.09 \\
Im$\mathcal{H}^{K^*_0(700)}_{-1/2,0}$ & 2.5  & 0.2  & 3.2  & 0.1 \\
$\gamma^{K^*_0(700)} \left[\gev^{-2}\right]$ & 0.94  & 0.07  & 0.35  & 0.04 \\
\midrule
Re$\mathcal{H}^{K^*(892)}_{1/2,0}$ & 1 (fixed)\\
Im$\mathcal{H}^{K^*(892)}_{1/2,0}$ & 0 (fixed)\\
Re$\mathcal{H}^{K^*(892)}_{1/2,-1}$ & 1.19  & 0.13  & 0.85  & 0.05 \\
Im$\mathcal{H}^{K^*(892)}_{1/2,-1}$ & $-$1.03  & 0.12  & 0.96  & 0.02 \\
Re$\mathcal{H}^{K^*(892)}_{-1/2,1}$ & $-$3.1  & 0.4  & 2.4  & 0.1 \\
Im$\mathcal{H}^{K^*(892)}_{-1/2,1}$ & $-$3.3  & 0.3  & 2.7  & 0.1 \\
Re$\mathcal{H}^{K^*(892)}_{-1/2,0}$ & $-$0.7  & 0.3  & 1.5  & 0.1 \\
Im$\mathcal{H}^{K^*(892)}_{-1/2,0}$ & $-$4.2  & 0.3  & 4.5  & 0.1 \\
\midrule
Re$\mathcal{H}^{K^*_0(1430)}_{1/2,0}$ & 0.2  & 0.8  & 2.4  & 0.3 \\
Im$\mathcal{H}^{K^*_0(1430)}_{1/2,0}$ & 8.7  & 0.7  & 9.7  & 0.1 \\
Re$\mathcal{H}^{K^*_0(1430)}_{-1/2,0}$ & $-$6.7  & 1.0  & 5.4  & 0.4 \\
Im$\mathcal{H}^{K^*_0(1430)}_{-1/2,0}$ & 10  & 1  & 14  & 0 \\
$\gamma^{K^*_0(1430)} \left[\gev^{-2}\right]$ & 0.02  & 0.02  & 0.33  & 0.01 \\
\midrule
Re$\mathcal{H}^{\Deltares(1232)^{++}}_{1/2,0}$ & $-$6.8  & 0.5  & 5.3  & 0.2 \\
Im$\mathcal{H}^{\Deltares(1232)^{++}}_{1/2,0}$ & 3.1  & 0.6  & 1.4  & 0.1 \\
Re$\mathcal{H}^{\Deltares(1232)^{++}}_{-1/2,0}$ & $-$13  & 1  & 12  & 0 \\
Im$\mathcal{H}^{\Deltares(1232)^{++}}_{-1/2,0}$ & 4.5  & 1.1  & 3.7  & 0.3 \\
\midrule
Re$\mathcal{H}^{\Deltares(1600)^{++}}_{1/2,0}$ & 11.4  & 0.9  & 9.8  & 0.2 \\
Im$\mathcal{H}^{\Deltares(1600)^{++}}_{1/2,0}$ & $-$3.1  & 1.0  & 1.4  & 0.2 \\
Re$\mathcal{H}^{\Deltares(1600)^{++}}_{-1/2,0}$ & 6.7  & 0.7  & 7.2  & 0.2 \\
Im$\mathcal{H}^{\Deltares(1600)^{++}}_{-1/2,0}$ & $-$1.0  & 0.7  & 2.1  & 0.1 \\
\midrule
Re$\mathcal{H}^{\Deltares(1700)^{++}}_{1/2,0}$ & 10.4  & 0.9  & 7.2  & 0.3 \\
Im$\mathcal{H}^{\Deltares(1700)^{++}}_{1/2,0}$ & 1.4  & 0.9  & 5.2  & 0.2 \\
Re$\mathcal{H}^{\Deltares(1700)^{++}}_{-1/2,0}$ & 13  & 1  & 13  & 0 \\
Im$\mathcal{H}^{\Deltares(1700)^{++}}_{-1/2,0}$ & 2.1  & 1.1  & 6.0  & 0.3 \\
\bottomrule
\end{tabular}
\end{table}

\begin{table}
\centering
\caption{Fit fractions of the resonant contributions included in the default amplitude model. The first uncertainty is statistical, the second is amplitude model choice systematic contribution, the third is the combination of the other systematic uncertainties. \label{tab:fitfractions_final}}
\begin{tabular}{lcccc}
\toprule
Resonance & Fit Fraction (\%) & Stat. Unc. & Model Unc. & Syst. Unc. \\
\midrule 
$\Lz(1405)$ & \phantom{0}7.7\phantom{0}  & \phantom{0}0.2\phantom{0}  & \phantom{0}3.0\phantom{0}  & \phantom{0}0.2\phantom{0} \\
$\Lz(1520)$ & \phantom{0}1.86  & \phantom{0}0.09  & \phantom{0}0.23  & \phantom{0}0.03 \\
$\Lz(1600)$ & \phantom{0}5.2\phantom{0}  & \phantom{0}0.2\phantom{0}  & \phantom{0}1.9\phantom{0}  & \phantom{0}0.1\phantom{0} \\
$\Lz(1670)$ & \phantom{0}1.18  & \phantom{0}0.06  & \phantom{0}0.32  & \phantom{0}0.01 \\
$\Lz(1690)$ & \phantom{0}1.19  & \phantom{0}0.09  & \phantom{0}0.34  & \phantom{0}0.01 \\
$\Lz(2000)$ & \phantom{0}9.58  & \phantom{0}0.27  & \phantom{0}0.93  & \phantom{0}0.23 \\
$\Deltares(1232)^{++}$ & 28.60  & \phantom{0}0.29  & \phantom{0}0.76  & \phantom{0}0.16 \\
$\Deltares(1600)^{++}$ & \phantom{0}4.5\phantom{0}  & \phantom{0}0.3\phantom{0}  & \phantom{0}1.5\phantom{0}  & \phantom{0}0.1\phantom{0} \\
$\Deltares(1700)^{++}$ & \phantom{0}3.90  & \phantom{0}0.20  & \phantom{0}0.94  & \phantom{0}0.07 \\
$K^*_0(700)$ & \phantom{0}3.02  & \phantom{0}0.16  & \phantom{0}0.92  & \phantom{0}0.18 \\
$K^*(892)$ & 22.14  & \phantom{0}0.23  & \phantom{0}0.64  & \phantom{0}0.04 \\
$K^*_0(1430)$ & 14.7\phantom{0}  & \phantom{0}0.6\phantom{0}  & \phantom{0}2.7\phantom{0}  & \phantom{0}0.1\phantom{0} \\
\bottomrule
\end{tabular}
\end{table}

\begin{table}
\centering
\caption{Sensitivity to polarization $\sqrt{3}S$ of the default amplitude model and decay asymmetry $\alpha$ parameters of single resonant contributions. The first uncertainty is statistical, the second is the amplitude model choice systematic contribution, the third is the combination of the other systematic uncertainties. \label{tab:alphapars_final}}
\begin{tabular}{lcccc}
\toprule
Resonance & $\alpha$ & Stat. Unc. & Model Unc. & Syst. Unc. \\
\midrule 
Model $\sqrt{3}S$ & 0.662  & 0.005  & 0.010  & 0.007 \\
$K^*(892)$ $\sqrt{3}S$ & 0.873  & 0.010  & 0.023  & 0.003\\
\midrule
$\Lz(1405)$ & $-$0.58\phantom{0}  & 0.05\phantom{0}  & 0.28\phantom{0}  & 0.01\phantom{0} \\
$\Lz(1520)$ & $-$0.925  & 0.025  & 0.084  & 0.005 \\
$\Lz(1600)$ & $-$0.20\phantom{0}  & 0.06\phantom{0}  & 0.50\phantom{0}  & 0.03\phantom{0} \\
$\Lz(1670)$ & $-$0.817  & 0.042  & 0.073  & 0.006 \\
$\Lz(1690)$ & $-$0.958  & 0.020  & 0.027  & 0.006 \\
$\Lz(2000)$ & \phantom{$-$}0.57\phantom{0}  & 0.03\phantom{0}  & 0.19\phantom{0}  & 0.01\phantom{0} \\
$\Deltares(1232)^{++}$ & $-$0.548  & 0.014  & 0.036  & 0.004 \\
$\Deltares(1600)^{++}$ & \phantom{$-$}0.50\phantom{0}  & 0.05\phantom{0}  & 0.17\phantom{0}  & 0.01\phantom{0} \\
$\Deltares(1700)^{++}$ & $-$0.216  & 0.036  & 0.075  & 0.011 \\
$K^*_0(700)$ & \phantom{$-$}0.06\phantom{0}  & 0.66\phantom{0}  & 0.24\phantom{0}  & 0.23\phantom{0} \\
$K^*_0(1430)$ & $-$0.34\phantom{0}  & 0.03\phantom{0}  & 0.14\phantom{0}  & 0.01\phantom{0} \\
\bottomrule
\end{tabular}
\end{table}

\section{Summary}
\label{sec:summary}

A full amplitude analysis of \Lcpkpi decays with measurement of the \Lc polarization vector in semileptonic beauty hadron decays is presented. A sample of $400\,000$ candidates is selected from proton-proton collisions recorded by the LHCb detector at a center-of-mass energy of 13 \tev, featuring a very small residual background contribution. The amplitude model is written in the helicity formalism with a general method to deal with the matching of final particle spin states in different decay chains and includes a generic \Lc polarization vector. A maximum-likelihood fit is performed to determine the amplitude model parameters. All the parameters of the amplitude model and the baryon polarization have been measured; fit fractions and decay asymmetry parameters for each two-body resonant contributions are also reported together with the effective three-body decay asymmetry parameter of the \Lcpkpi decay.
The most important resonances contributing to the \Lcpkpi decay are from $\Deltares(1232)^{++}$, $K^*(892)$ and spin-zero $K^*$ states. A significant enhancement in the \mqpk spectrum, in a region where no clear $\Lz$ resonances have been reported in Ref.~\cite{PDG2020}, is well described by a spin $1/2^-$ state, identified as a $\Lz(2000)$ resonance. Its mass and width parameters are determined to be $1988 \pm 2 \pm 21\mev$ and $179 \pm 4 \pm 16 \mev$, respectively. A large \Lc polarization is found, of order $65-70\%$, measured with absolute uncertainties of order $1\%$. The normal polarization, sensitive to time-reversal violation effects and final-state interactions, is compatible with zero.
A large sensitivity to the polarization is measured, showing the \Lcpkpi decay to be the best probe for \Lc polarization. The amplitude model obtained provides a complete description of the \Lcpkpi decay, with applications ranging from new physics searches to low-energy QCD. Such applications include an increased sensitivity to angular analyses of semileptonic baryon decays, and the most precise measurements of the $\Lc$ polarization and electromagnetic dipole moments via spin precession.


\clearpage
\section*{Acknowledgements}
%
%
\noindent We express our gratitude to our colleagues in the CERN
accelerator departments for the excellent performance of the LHC. We
thank the technical and administrative staff at the LHCb
institutes.
We acknowledge support from CERN and from the national agencies:
CAPES, CNPq, FAPERJ and FINEP (Brazil); 
MOST and NSFC (China); 
CNRS/IN2P3 (France); 
BMBF, DFG and MPG (Germany); 
INFN (Italy); 
NWO (Netherlands); 
MNiSW and NCN (Poland); 
MEN/IFA (Romania); 
MICINN (Spain); 
SNSF and SER (Switzerland); 
NASU (Ukraine); 
STFC (United Kingdom); 
DOE NP and NSF (USA).
We acknowledge the computing resources that are provided by CERN, IN2P3
(France), KIT and DESY (Germany), INFN (Italy), SURF (Netherlands),
PIC (Spain), GridPP (United Kingdom), 
CSCS (Switzerland), IFIN-HH (Romania), CBPF (Brazil),
Polish WLCG  (Poland) and NERSC (USA).
We are indebted to the communities behind the multiple open-source
software packages on which we depend.
Individual groups or members have received support from
ARC and ARDC (Australia);
Minciencias (Colombia);
AvH Foundation (Germany);
EPLANET, Marie Sk\l{}odowska-Curie Actions and ERC (European Union);
A*MIDEX, ANR, IPhU and Labex P2IO, and R\'{e}gion Auvergne-Rh\^{o}ne-Alpes (France);
Key Research Program of Frontier Sciences of CAS, CAS PIFI, CAS CCEPP, 
Fundamental Research Funds for the Central Universities, 
and Sci. \& Tech. Program of Guangzhou (China);
GVA, XuntaGal, GENCAT and Prog.~Atracci\'on Talento, CM (Spain);
SRC (Sweden);
the Leverhulme Trust, the Royal Society
 and UKRI (United Kingdom).

\clearpage

\appendix

\section{Amplitude model}
\label{sec:amplitude_model}

The helicity formalism allows the association of a decay amplitude to a two-body decay, dependent on the spin projection of the decaying particle and the helicities of the final-state particles. The amplitude is obtained establishing a relation between two sets of two-particle states: plane-wave helicity states, describing propagating particles with well defined momentum; and spherical-wave helicity states, describing states of definite total angular momentum.
In Ref.~\cite{Marangotto:2019ucc} the plane-wave two-particle state is defined by the product of a pair of helicity and opposite-helicity states,
\begin{equation}
\Ket{p,\theta_1,\phi_1, \lambda_1,\bar{\lambda}_2} \equiv \Ket{\bm{p}_1, s_1, \lambda_1} \otimes \Ket{\bm{p}_2, s_2, \bar{\lambda}_2},
\end{equation}
instead of the product of helicity states for both final particles as in Ref.~\cite{JacobWick}. Helicity (opposite-helicity) states are those in which the spin is quantized along (opposite to) the momentum direction. Helicity values are indicated with $\lambda$ ($\bar{\lambda}$); while $p$, $\theta_1$ and $\phi_1$ are the spherical coordinates of the  momentum $\bm{p}_1$ of particle 1 in the two-particle center-of-mass frame,
\begin{align}
p &= |\bm{p}_1|, \nonumber\\
\theta_1 &= \arccos\left(\hat{\bm{z}}\cdot\hat{\bm{p}}_1\right), \nonumber\\
\phi_1 &= \mathrm{arctan}\left( \hat{\bm{y}} \cdot \hat{\bm{p}}_1, \hat{\bm{x}} \cdot \hat{\bm{p}}_1\right),
\label{eq:helicity_angles}
\end{align}
where the coordinate axes are those representing the total angular momentum components. The function $\mathrm{arctan}(y,x) \in [-\pi,\pi]$ computes the signed angle between the $x$ axis and the vector having components $(x,y)$.
The relation expressing plane-wave states in terms of spherical-wave states is
\begin{align}
\Ket{p,\theta_1,\phi_1, \lambda_1,\bar{\lambda}_2} &= \sum_{J,M} \sqrt{\frac{2J+1}{4\pi}} D^J_{M,\lambda_1+\bar{\lambda}_2}(\phi_1,\theta_1,0)\nonumber\\
&\times \Ket{p,J,M,\lambda_1,\bar{\lambda}_2},
\label{eq:plane_spherical_wave_relation_corrected}
\end{align}
where $D^J_{M,\lambda_1+\bar{\lambda}_2}(\phi_1,\theta_1,0)$ is the Wigner $D$-matrix representing the Euler rotation from the total momentum spin system to the particle 1 helicity system. A basic introduction to Euler rotations and their representation on spin states is given in Appendix~\ref{sec:euler_rotations}.

With the present definition of plane-wave states, both final-state particle spin states are defined by the same rotation $R(\phi_1,\theta_1,0)$ (details are given in Ref.~\cite{Marangotto:2019ucc}), without the need to invert the particle 2 helicity states as in Ref.~\cite{JacobWick}. This choice of the two-particle state eases the handling of phase differences and the matching of proton spin states, addressed later in the construction of the \Lcpkpi amplitude model.

The amplitude associated to a two-body decay $A\to 1 \, 2$, with $\hat{T}$ the relevant transition operator, is obtained by introducing the $A$ particle spin states $\Ket{s_A,m_A}$,
\begin{align}
\mathcal{A}_{m_A,\lambda_1,\bar{\lambda}_2}(\theta_1,\phi_1) &\equiv\Braket{p,\theta_1,\phi_1, \lambda_1,\bar{\lambda}_2|\hat{T}|s_A,m_A} \nonumber\\
&= \mathcal{H}_{\lambda_1,\bar{\lambda}_2} D^{*s_A}_{m_A,\lambda_1+\bar{\lambda}_2}(\phi_1,\theta_1,0)\mathcal{R}(m^2_{12}),
\label{eq:two_body_amplitude_corrected}
\end{align}
in which the Wigner $D$-matrix contains the dependence of the decay amplitude on the helicity angles. The decay dynamics is described by complex numbers called helicity couplings,
\begin{equation}
\mathcal{H}_{\lambda_1,\bar{\lambda}_2} \equiv \Braket{s_A,m_A,\lambda_1,\bar{\lambda}_2|\hat{T}|s_A,m_A},
\label{eq:helicity_couplings}
\end{equation}
and a (lineshape) function $\mathcal{R}(m^2_{12})$, modeling the dependence on the decaying $A$ particle invariant mass, in case it has a nonnegligible width.
Helicity couplings cannot depend on $s_A,m_A$ by rotational invariance. The helicity values allowed by angular momentum conservation are
\begin{equation}
|\lambda_1|\leq s_1,\hspace{1cm} |\bar{\lambda}_2|\leq s_2,\hspace{1cm} |\lambda_1+\bar{\lambda}_2|\leq s_A.
\label{eq:allowed_helicity_couplings}
\end{equation}
If the decay respects parity symmetry, the helicity couplings for opposite helicities are constrained by the relation
\begin{align}
\mathcal{H}_{\lambda_1,\bar{\lambda}_2} &\equiv \eta_A \eta_1 \eta_2 (-1)^{s_1+s_2-s_A}\mathcal{H}_{-\lambda_1,-\bar{\lambda}_2},
\label{eq:couplings_parity_conservation}
\end{align}
where $\eta$ are the parity eigenvalues of the particle states.
\subsection{\boldmath\Lc baryon polarization and decay kinematics}
\label{sec:Lc_pol_frame}
Charm baryons in semileptonic beauty decays are polarized due to the parity-violating nature of the weak interaction.
The \Lc polarization is measured using helicity coordinate systems defined from \Lc and muon momenta in two reference frames: the laboratory frame (\textit{lab}), precisely determined, and an approximate beauty hadron rest frame ($\tilde{B}$). The latter is obtained assuming the proper velocity along the beam axis, $\gamma\beta_z$, of the reconstructed $\Lc\mun$ pair is equal to that of the beauty baryon.

\begin{figure}
\centering
\begin{tikzpicture}[scale=3.5]
\draw [->] (0,0) -- (1,0);
\draw [->] (0,0) -- (0,1);
\draw [dashed,->] (0,0) -- (0.91651514,0.4);
\node [below] at (1,0) {$\bm{\hat{z}}_{\Lc}\equiv \bm{\hat{p}}(\Lc)$};
\node [below,left] at (0,0) {$\bm{\hat{y}}_{\Lc}\equiv \bm{\hat{z}}_{\Lc} \times \bm{\hat{x}}_{\Lc}$};
\node [right] at (0,1) {$\bm{\hat{x}}_{\Lc}$};
\node [right] at (0.91651514,0.4) {$\bm{\hat{p}}(\mun)$};
\draw (0,0) circle (1pt);
\end{tikzpicture}
\caption{Definition of the \Lc polarization system. The $\bm{\hat{y}}_{\Lc}$ axis is orthogonal to the page, towards the reader.\label{fig:polarization_frame}}
\end{figure}
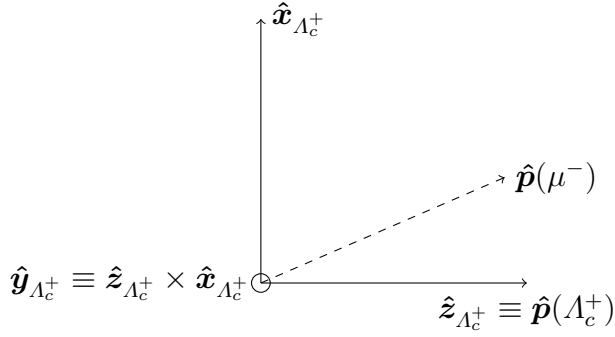

The \Lc momentum defines the longitudinal $P_z$ polarization component, the muon momentum component orthogonal to the \Lc momentum defines the orthogonal $P_x$ component, with the normal $P_y$ component defined through a right-handed system, as shown in Fig.~\ref{fig:polarization_frame},
\begin{align}
P_z &= \bm{P} \cdot \bm{\hat{z}}_{\Lc} = \bm{P} \cdot \bm{\hat{p}}(\Lc) , \nonumber\\
P_x &= \bm{P} \cdot \bm{\hat{x}}_{\Lc} = \bm{P} \cdot \left[ \frac{\bm{p}(\Lc) \times \bm{p}(\mu^-)}{\left|\bm{p}(\Lc) \times \bm{p}(\mu^-)\right|} \times \bm{\hat{p}}(\Lc) \right] , \nonumber\\
P_y &= \bm{P} \cdot \bm{\hat{y}}_{\Lc} = \bm{P} \cdot \frac{\bm{p}(\Lc) \times \bm{p}(\mu^-)}{\left|\bm{p}(\Lc) \times \bm{p}(\mu^-)\right|}.
\label{eq:polarization_frame_definition}
\end{align}
Considering the effect of the time-reversal operation $T$, longitudinal and transverse components are $T$-even quantities, while normal polarization is $T$-odd, implying the latter to be produced only by time-reversal violating effects or final-state interactions between particles produced in the semileptonic decay~\cite{Sozzi:1087897}. Final-state interactions between the \Lc and \mun particles can occur due to electromagnetic interactions only, expected to be $\mathcal{O}(1\%)$.
The $T$-odd property of normal polarization does not depend on the reference system used for its definition, as studied with $\Lb\to\Lc\mun\bar{\nu}_{\mu}$ simulated decays. It has been checked that a null $P_y$ in the helicity system reached from the true \Lb rest frame is zero also when rotated to the helicity systems reached from the laboratory or the approximate $B$ systems; the polarization rotation is confined to the $zx$ plane.

The five degrees of freedom of a baryon three-body decay can be described by the following quantities: two two-body invariant mass squared (``Dalitz'' variables) modeling the decay dynamics and three decay orientation angles describing the decay plane orientation with respect to the baryon polarization system; the latter are important when the baryon polarization is considered.
For the \Lcpkpi decay, \mqpk and \mqkpi are selected as Dalitz variables. The decay orientation angles are defined to be the Euler angles describing the rotation from the polarization system in Eq.~\eqref{eq:polarization_frame_definition} to a decay plane (DP) reference system constructed as follows: the proton momentum defines the $z$ axis, while the component of the kaon momentum orthogonal to the proton momentum defines the $x$ axis (see Fig.~\ref{fig:polarization_frame} with $\Lc\leftrightarrow p$, $\mu^- \leftrightarrow K^-$ substitutions),
\begin{align}
\bm{\hat{z}}_{\rm DP} &= \bm{\hat{p}}(p) , \nonumber\\
\bm{\hat{x}}_{\rm DP} &= \frac{\bm{p}(p) \times \bm{p}(K^-)}{\left|\bm{p}(p) \times \bm{p}(K^-)\right|} \times \bm{\hat{p}}(p) , \nonumber\\
\bm{\hat{y}}_{\rm DP} &= \bm{\hat{z}}_{\rm DP} \times \bm{\hat{x}}_{\rm DP} ,
\label{eq:decay_plane_definition}
\end{align}
where momenta are expressed in the \Lc rest frame.

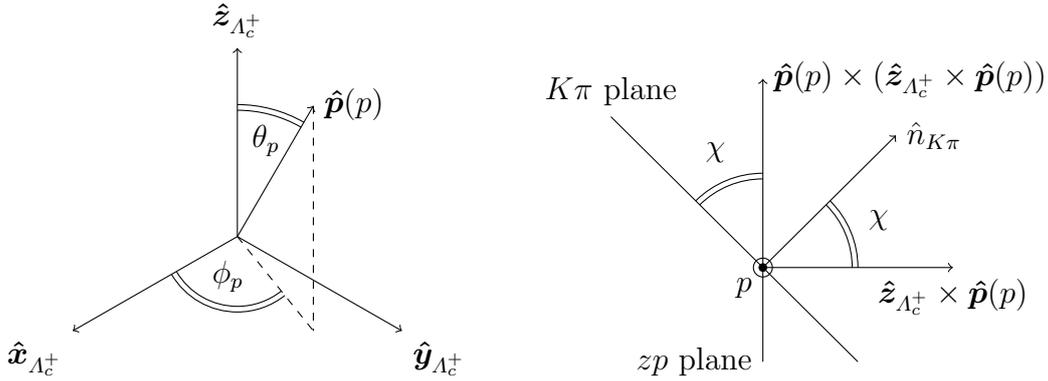
\begin{figure}
\centering
\begin{tikzpicture}[scale=2.5]
\draw [->] (0,0) -- (0,1);
\draw [->] (0,0) -- (-0.866025,-0.5);
\draw [->] (0,0) -- (0.866025,-0.5);
\node [above] at (0,1) {$\bm{\hat{z}}_{\Lc}$};
\node [below left] at (-0.866025,-0.5) {$\bm{\hat{x}}_{\Lc}$};
\node [below right] at (0.866025,-0.5) {$\bm{\hat{y}}_{\Lc}$};
\draw [->] (0,0) -- (0.4,0.69282);
\draw [dashed] (0.4,0.69282) -- (0.4,-0.5);
\draw [dashed] (0,0) -- (0.4,-0.5);
\node [right] at (0.4,0.7) {$\bm{\hat{p}}(p)$};
\draw (0,0.7) arc [radius=0.7,start angle=90,end angle=60];
\draw (0,0.67) arc [radius=0.67,start angle=90,end angle=60];
\node [below] at (0.15,0.65) {$\theta_p$};
\draw (-0.34641,-0.2) arc [radius=0.4,start angle=210,end angle=308.65980825];
\draw (-0.32042925,-0.185) arc [radius=0.37,start angle=210,end angle=308.65980825];
\node [above] at (-0.05,-0.35) {$\phi_p$};
\end{tikzpicture}
\hspace{0.5cm}
\begin{tikzpicture}[scale=2.5]
\draw [->] (0,0) -- (1,0);
\draw [->] (0,-0.5) -- (0,1);
\draw (0,0) circle [radius=0.05];
\draw [fill] (0,0) circle [radius=0.02];
\node [below left] at (0,0) {$p$};
\node [below] at (1,0) {$\bm{\hat{z}}_{\Lc}\times\bm{\hat{p}}(p)$};
\node [right] at (0,1) {$\bm{\hat{p}}(p)\times(\bm{\hat{z}}_{\Lc}\times\bm{\hat{p}}(p))$};
\draw [->] (0,0) -- (0.7,0.7);
\node [right] at (0.7,0.7) {$\hat{n}_{K\pi}$};
\draw (0.5,-0.5) -- (-0.8,0.8);
\node [above] at (-0.8,0.8) {$K\pi$ plane};
\node [left] at (0,-0.5) {$zp$ plane};
\draw (0.5,0) arc [radius=0.5,start angle=0,end angle=45];
\draw (0.47,0) arc [radius=0.47,start angle=0,end angle=45];
\node [right] at (0.5,0.25) {$\chi$};
\draw (0,0.5) arc [radius=0.5,start angle=90,end angle=135];
\draw (0,0.47) arc [radius=0.47,start angle=90,end angle=135];
\node [above] at (-0.25,0.5) {$\chi$};
\end{tikzpicture}
\caption{Definition of the Euler angles describing the rotation from the \Lc polarization system to the decay plane reference system: (left) proton polar and azimuthal angles and (right) $\chi$ angle.\label{fig:orientation_angles}}
\end{figure}

With this definition the Euler angles $\alpha$ and $\beta$  are the azimuthal and polar angles of the proton in the \Lc polarization system, $\phi_p$ and $\theta_p$, respectively, and the $\gamma$ angle is the signed angle named $\chi$ formed by the proton and the \Lc quantization axis $\bm{\hat{z}}_{\Lc}$ and the plane formed by the kaon and the pion. Following Eq.~\eqref{eq:euler_angles} and the definition in Fig.~\ref{fig:orientation_angles} one has,
\begin{align}
\phi_p &= \text{arctan}(\bm{\hat{p}}(p)\cdot\bm{\hat{y}}_{\Lc},\bm{\hat{p}}(p)\cdot\bm{\hat{x}}_{\Lc}) , \nonumber\\
\theta_p &= \arccos\left(\bm{\hat{p}}(p)\cdot\bm{\hat{z}}_{\Lc}\right) , \nonumber\\
\chi &= \text{arctan}\left\lbrace\bm{\hat{z}}_{\Lc}\cdot \frac{\bm{p}(p) \times \bm{p}(K^-)}{\left| \bm{p}(p) \times \bm{p}(K^-) \right|},-\bm{\hat{z}}_{\Lc}\cdot \left[\frac{\bm{p}(p) \times \bm{p}(K^-)}{\left|\bm{p}(p) \times \bm{p}(K^-)\right|} \times \bm{\hat{p}}(p) \right] \right\rbrace.
\label{eq:orientation_angles}
\end{align}

The five phase-space variables are therefore chosen to be
\begin{equation}
\Omega = (\mqpk,\mqkpi,\cos\theta_p,\phi_p,\chi),
\label{eq:phase_space_vars_app}
\end{equation}
with the phase-space density being uniform over them. Their allowed range is \mbox{$\cos\theta_p \in \left[-1,1\right]$} and $\phi_p,\chi \in \left[-\pi,\pi\right]$, while the mass distributions are constrained to a rounded-triangle shape in the $(\mqpk,\mqkpi)$ plane (Dalitz plot, see Fig.~\ref{fig:data_dalitz_plot}).

\subsection{Amplitude model for the \boldmath\Lcpkpi decay}
The amplitude model for the \Lcpkpi decay is written by decomposing the three-body decay into two sequential two-body decays by introducing intermediate resonant states. Three decay channels are possible: $K^*\to \Km\pip$, $\Lz\to p\Km$ and $\Deltares^{++}\to p\pip$.
Helicity amplitudes are obtained for each channel and summed coherently using the method for matching proton spin states among different decay chains of Ref.~\cite{Marangotto:2019ucc}. The matching step allows the transformation of proton helicity states, different for each decay channel, to a common definition of proton spin states.

To simplify the amplitude model expression, the invariant mass and decay orientation angle degrees of freedom can be separated applying the Dalitz plot decomposition~\cite{Mikhasenko:2019rjf},
\begin{equation}
\mathcal{A}_{m_{\Lc},m_p}(\Omega) = \sum_{\nu_{\Lc}} D^{*1/2}_{m_{\Lc},\nu_{\Lc}}(\phi_p,\theta_p,\chi) \mathcal{O}_{\nu_{\Lc},m_p}(\mqpk,\mqkpi),
\label{eq:amplitude_decomposition}
\end{equation}
in which the Wigner $D$-matrix describes the rotation of \Lc spin states from the \Lc polarization system in Eq.~\eqref{eq:polarization_frame_definition}, $|1/2,m_{\Lc}\rangle$, to the decay plane system in Eq.~\eqref{eq:decay_plane_definition}, $|1/2,\nu_{\Lc}\rangle$. The term $\mathcal{O}_{\nu_{\Lc},m_p}(\mqpk,\mqkpi)$ describes the \Lc decay amplitudes in terms of \Lc spin states $|1/2,\nu_{\Lc}\rangle$ and proton states defined in the canonical spin system reached from the \Lc decay plane system, $\Ket{1/2,m_p}$. These proton states are later employed for the matching of proton spin states among different decay chains.
The decomposition allows the definition of helicity and canonical spin states using polar rotations only, around the $y$ axis of the decay plane system. Therefore, the angular part of the decay amplitude simplifies, depending solely on polar helicity angles via reduced Wigner $d$-matrices.

Considering the $\Lc\to pK^*(\to K^-\pi^+)$ decay chain, the weak $\Lc\to pK^*$ decay amplitude is given by Eq.~\eqref{eq:two_body_amplitude_corrected} applied starting from the decay plane coordinate system,
\begin{equation}
\mathcal{A}^{\Lc\to pK^*}_{\nu_{\Lc},m_p,\bar{\lambda}_{K^*}} = \mathcal{H}^{\Lc\to pK^*}_{m_p,\bar{\lambda}_{K^*}} \delta_{\nu_{\Lc},m_p+\bar{\lambda}_{K^*}},
\end{equation}
where spin states are expressed in the DP system, and the amplitude is written in terms of the proton helicity $m_p$ and the $K^*$ opposite helicity $\bar{\lambda}_{K^*}$. The proton helicity states reached from the DP system coincide with the canonical states since in the DP system the proton momentum is already aligned to the quantization axis. Since no rotation of spin states is involved, the $D$-matrix becomes a constraint on the helicity values $m_p+\bar{\lambda}_{K^*}=\nu_{\Lc}$.

For spin-zero $K^*$ resonances the angular momentum conservation relations in Eq.~\eqref{eq:allowed_helicity_couplings} allow two complex couplings corresponding to $m_p=\pm 1/2$. For higher-spin resonances, four couplings are allowed, corresponding to $\lbrace m_p=1/2$; $\bar{\lambda}_{K^*}=0,-1\rbrace$ and $\lbrace m_p=-1/2$; $\bar{\lambda}_{K^*}=0,1 \rbrace$. Couplings are independent of each other because of parity violation in weak decays. The strong $K^*\to K^-\pi^+$ decay amplitude is
\begin{equation}
\mathcal{A}^{K^*\to K^-\pi^+}_{\bar{\lambda}_{K^*}} = \mathcal{H}^{K^*\to K^-\pi^+}_{0,0} d^{*J_{K^*}}_{\bar{\lambda}_{K^*},0}(\bar{\theta}_K)\mathcal{R}(\mqkpi),
\label{eq:amplitude_Kstar_Kpi_DPD}
\end{equation}
where $\mathcal{R}(\mqkpi)$ is the lineshape of the $K^*$ resonance and $\bar{\theta}_{K}$ is the kaon momentum signed polar angle in the $K^*$ opposite-helicity coordinate system,
\begin{equation}
\bar{\theta}_{K} = \mathrm{arctan}\left(p_x^{K^*}(K^-),p_z^{K^*}(K^-)\right).
\end{equation}
Signed polar angles are used as helicity angles in order to have rotations only around the $y$ axis of the decay plane system. Otherwise, the use of positive polar angles would require additional azimuthal rotations around the $z$ axis (to flip the $y$ axis direction) complicating unnecessarily the expression of the helicity amplitudes.
In the amplitude fit the coupling $\mathcal{H}^{K^*\to K^-\pi^+}_{0,0}$ can not be determined independently of $\mathcal{H}^{\Lc\to K^* p}_{m_p,\bar{\lambda}_{K^*}}$ couplings, therefore it is set equal to 1 and absorbed into the latter.

Considering the $\Lc\to\Lz(\to pK^-)\pi^+$ decay chain, the $\Lc\to\Lz\pi^+$ decay amplitude is
\begin{equation}
\mathcal{A}^{\Lc\to\Lz\pi^+}_{\nu_{\Lc},\lambda_{\Lz}} = \mathcal{H}^{\Lc\to\Lz\pi^+}_{\lambda_{\Lz},0} d^{1/2}_{\nu_{\Lc},\lambda_{\Lz}}(\theta_{\Lz}),
\end{equation}
in which $\lambda_{\Lz}$ is the $\Lz$ helicity reached from the \Lc system and $\theta_{\Lz}$ is the signed polar angle of the $\Lz$ momentum in the DP system,
\begin{equation}
\theta_{\Lz} = \mathrm{arctan}\left(p_x^{\Lc}(\Lz),p_z^{\Lc}(\Lz)\right).
\end{equation}
The angular momentum conservation relations in Eq.~\eqref{eq:allowed_helicity_couplings} allow two helicity couplings, $\lambda_{\Lz}=\pm 1/2$, to fit for each resonance whatever $J_{\Lz}$ is.

The $\Lz\to pK^-$ decay amplitude is
\begin{equation}
\mathcal{A}^{\Lz\to pK^-}_{\lambda_{\Lz},\lambda^{\Lz}_p} = \mathcal{H}^{\Lz\to pK^-}_{\lambda^{\Lz}_p,0} d^{J_{\Lz}}_{\lambda_{\Lz},\lambda^{\Lz}_p}(\theta^{\Lz}_p)\mathcal{R}(\mqpk),
\end{equation}
where $\lambda^{\Lz}_p$ is the proton helicity, $\theta^{\Lz}_p$ the proton signed polar angle, both defined in the helicity coordinate system reached from the $\Lz$ resonance frame, while $\mathcal{R}(\mqpk)$ is the lineshape of the $\Lz$ resonance. The two helicity couplings corresponding to $\lambda^{\Lz}_p=\pm 1/2$ are related by Eq.~\eqref{eq:couplings_parity_conservation} thanks to parity conservation in strong interactions,
\begin{equation}
\mathcal{H}^{\Lz\to pK^-}_{-\lambda^{\Lz}_p,0} = -P_{\Lz} (-1)^{J_{\Lz}-1/2} \mathcal{H}^{\Lz\to pK^-}_{\lambda^{\Lz}_p,0},
\end{equation}
with $P_{\Lz}$ the parity of the $\Lz$ resonance, and the proton and kaon parities $P_p=1$, $P_K=-1$ have been inserted. In the amplitude fit these couplings can not be determined independently of $\mathcal{H}^{\Lc\to\Lz\pi^+}_{\lambda_{\Lz},0}$ couplings, so they are absorbed into the latter setting them as $\mathcal{H}^{\Lz\to pK^-}_{+1/2,0}=1$ and $\mathcal{H}^{\Lz\to pK^-}_{-1/2,0}=-P_{\Lz} (-1)^{J_{\Lz}-1/2}$, with zero imaginary parts.

The matching of proton spin states from the $\Lz$ helicity system to the canonical system is performed applying the method of Ref.~\cite{Marangotto:2019ucc}. The transformation sequence applied to reach the proton helicity system is undone step-by-step in order to ensure a consistent phase definition of fermion spin states. Three rotations must be applied: two of angles $\theta^{\Lz}_p$ and $\theta_{\Lz}$, plus the Wigner rotation accounting for the different boost sequence applied to reach the two systems. The Wigner rotation can be written in angle-axis decomposition~\cite{Gourgoulhon}, obtained in terms of gamma factors $\gamma^{S}_{S'}$ and boost vectors $\bm{u}^S_{S'}$ associated to Lorentz transformations connecting $S$ to $S'$ reference frames.
The Wigner rotation has angle
\begin{equation}
\alpha^W_{\Lz} = \arccos\left[ \frac{\left( 1+ \gamma^{\Lc}_p + \gamma^{\Lc}_{\Lz} + \gamma^{\Lz}_p \right)^2}{(1+\gamma^{\Lc}_p) (1+\gamma^{\Lc}_{\Lz}) (1+\gamma^{\Lz}_p)} -1 \right],
\end{equation}
and axis,
\begin{equation}
\bm{a}^W_{\Lz} = \frac{\bm{u}^{\Lc}_{\Lz} \times \bm{u}^{\Lz}_p}{\left| \bm{u}^{\Lc}_{\Lz} \times \bm{u}^{\Lz}_p \right|} = \bm{\hat{y}}_{\rm DP},
\end{equation}
which coincides with the $y$ axis of the decay plane coordinate system. All these rotations are around the same $y$ axis, so they can be combined into just one rotation $R_y(\theta^{\Lz}_p+\theta_{\Lz}+\alpha^W_{\Lz})$, represented by $d^{1/2}_{m_p,\lambda^{\Lz}_p}(\theta^{\Lz}_p+\theta_{\Lz}+\alpha^W_{\Lz})$.

The amplitude associated to the $\Lc\to\Deltares^{++}(\to p\pi^+)K^-$ decay chain is similar to the $\Lz$ one, having the same spin structure. The $\Lc\to\Deltares^{++}K^-$ decay amplitude is
\begin{equation}
\mathcal{A}^{\Lc\to\Deltares^{++}K^-}_{\nu_{\Lc},\lambda_{\Deltares}} = \mathcal{H}^{\Lc\to\Deltares^{++}K^-}_{\lambda_{\Deltares},0} d^{1/2}_{\nu_{\Lc},\lambda_{\Deltares}}(\theta_{\Deltares}),
\end{equation}
with $\lambda_{\Deltares}$ the $\Deltares$ helicity and $\theta_{\Deltares}$ the signed polar angle of the $\Deltares$ momentum in the DP system. Two helicity couplings corresponding to $\lambda_{\Deltares}=\pm 1/2$ are allowed for each resonance. The $\Deltares^{++}\to p\pi^+$ decay amplitude is
\begin{equation}
\mathcal{A}^{\Deltares^{++}\to p\pi^+}_{\lambda_{\Deltares},\lambda^{\Deltares}_p} = \mathcal{H}^{\Deltares^{++}\to p\pi^+}_{\lambda^{\Deltares}_p,0} d^{J_{\Deltares}}_{\lambda_{\Deltares},\lambda^{\Deltares}_p}(\theta^{\Deltares}_p)\mathcal{R}(\mqppi),
\end{equation}
with $\lambda^{\Deltares}_p$ the proton helicity and $\theta^{\Deltares}_p$ the signed polar angle, both defined in the $\Deltares$ helicity system. The strong decay couplings are absorbed into $\mathcal{H}^{\Lc\to\Deltares^{++}K^-}_{\lambda_{\Deltares},0}$ setting them to $\mathcal{H}^{\Deltares^{++}\to p\pi^+}_{+1/2,0}=1$ and $\mathcal{H}^{\Deltares^{++}\to p\pi^+}_{-1/2,0}=-P_{\Deltares} (-1)^{J_{\Deltares}-1/2}$.

The matching of proton spin states from the $\Deltares$ helicity system to the canonical system is performed similarly to the $\Lz$ decay chain. The Wigner rotation angle is
\begin{equation}
\alpha^W_{\Deltares} = \arccos\left[ \frac{\left( 1+ \gamma^{\Lc}_p + \gamma^{\Lc}_{\Deltares} + \gamma^{\Deltares}_p \right)^2}{(1+\gamma^{\Lc}_p) (1+\gamma^{\Lc}_{\Deltares}) (1+\gamma^{\Deltares}_p)} -1 \right],
\end{equation}
around the axis
\begin{equation}
\bm{a}^W_{\Deltares} = \frac{\bm{u}^{\Lc}_{\Deltares} \times \bm{u}^{\Deltares}_p}{\left| \bm{u}^{\Lc}_{\Deltares} \times \bm{u}^{\Deltares}_p \right|} = -\bm{\hat{y}}_{\rm DP},
\end{equation}
which is opposite to the $y$ axis of the decay plane coordinate system. Therefore, the proton spin rotation can be written as $R_y(\theta^{\Deltares}_p+\theta_{\Deltares}-\alpha^W_{\Deltares})$, with reversed Wigner angle sign.

The decay amplitude for each intermediate helicity state is the product of the two two-body decay amplitudes in which the three-body decay is decomposed, summed over the allowed helicity states for each resonant contribution.
For the $\Lz$ and $\Deltares^{++}$ decay chains the amplitude is further multiplied by the Wigner $d$-matrices representing the spin matching rotation and summed over proton helicities,
\begin{align}
\mathcal{A}^{K^*}_{\nu_{\Lc},m_p} &= \sum_{\bar{\lambda}_{K^*}} \mathcal{H}^{\Lc\to pK^*}_{m_p,\bar{\lambda}_{K^*}} \delta_{\nu_{\Lc},m_p+\bar{\lambda}_{K^*}} d^{*J_{K^*}}_{\bar{\lambda}_{K^*},0}(\bar{\theta}_K)\mathcal{R}(\mqkpi),\nonumber\\
\mathcal{A}^{\Lz}_{\nu_{\Lc},m_p} &=  \sum_{\lambda_{\Lz}} \sum_{\lambda^{\Lz}_p} d^{1/2}_{m_p,\lambda^{\Lz}_p}(\theta^{\Lz}_p+\theta_{\Lz}+\alpha^W_{\Lz})\nonumber\\
&\times \mathcal{H}^{\Lc\to\Lz\pi^+}_{\lambda_{\Lz},0} \mathcal{H}^{\Lz\to pK^-}_{\lambda^{\Lz}_p,0} d^{1/2}_{\nu_{\Lc},\lambda_{\Lz}}(\theta_{\Lz}) d^{J_{\Lz}}_{\lambda_{\Lz},\lambda^{\Lz}_p}(\theta^{\Lz}_p)\mathcal{R}(\mqpk), \nonumber\\
\mathcal{A}^{\Deltares^{++}}_{\nu_{\Lc},m_p} &= \sum_{\lambda_{\Deltares^{++}}} \sum_{\lambda^{\Deltares}_p} d^{1/2}_{m_p,\lambda^{\Deltares}_p}(\theta^{\Deltares}_p+\theta_{\Deltares}-\alpha^W_{\Deltares})\nonumber\\
&\times\mathcal{H}^{\Lc\to\Deltares^{++}K^-}_{\lambda_{\Deltares},0} \mathcal{H}^{\Deltares^{++}\to p\pi^+}_{\lambda^{\Deltares}_p,0} d^{1/2}_{\nu_{\Lc},\lambda_{\Deltares}}(\theta_{\Deltares})
d^{J_{\Deltares}}_{\lambda_{\Deltares},\lambda^{\Deltares}_p}(\theta^{\Deltares}_p)\mathcal{R}(\mqppi).
\label{eq:helicity_amplitudes_with_DPD}
\end{align}

The total amplitude for the \Lcpkpi decay is obtained summing the amplitudes for the different intermediate states,
\begin{align}
\mathcal{O}_{\nu_{\Lc},m_p}(\mqpk,\mqkpi) &= \sum_{i=1}^{N_{K^*}} \mathcal{A}^{K^*_i}_{\nu_{\Lc},m_p} \nonumber\\
&+ \sum_{j=1}^{N_{\Lz}} \mathcal{A}^{\Lz_i^*}_{\nu_{\Lc},m_p} \nonumber\\
&+ \sum_{k=1}^{N_{\Deltares^{++}}}  \mathcal{A}^{\Deltares_k^{*++}}_{\nu_{\Lc},m_p}.
\label{eq:helicity_amplitudes_Lc}
\end{align}

Fit fractions, $FF$, for each resonance $R$ are obtained by computing the integral of the amplitude model over the phase-space where only the $R$ contribution is left. Fit fractions are normalized to the complete amplitude model integral,
\begin{equation}
FF = \frac{\bigints d\Omega \sum_{m_{\Lc},m_p}  \left|\mathcal{A}^R_{m_{\Lc},m_p}(\Omega)\right|^2}
{\bigintss d\Omega \sum_{m_{\Lc},m_p} \left|\mathcal{A}_{m_{\Lc},m_p}(\Omega)\right|^2}. 
\end{equation}

\subsection{Polarized decay rate}
\label{sec:polarised_decay_rate}

A generic polarization state of the \Lc baryon is described by the spin-1/2 density matrix
\begin{equation}
\rho^{\Lc} = \frac{1}{2}\left(\mathbb{I} + \bm{P}\cdot\bm{\sigma}\right) = \frac{1}{2}
\left(
\begin{array}{cc}
1+P_z & P_x-iP_y\\
P_x+iP_y & 1-P_z\\
\end{array}
\right),
\end{equation}
where $P_x$, $P_y$ and $P_z$ are the components of the polarization vector $\bm{P}$ in the polarization system (see definition in Sec.~\ref{sec:Lc_pol_frame}) and $\bm{\sigma}$ are the three Pauli matrices.
The differential decay rate for a \Lcpkpi decay with a generic \Lc polarization is
\begin{equation}
p(\Omega,\bm{P}) = \sum_{m_\Lc=-1/2}^{1/2}
\sum_{m'_\Lc=-1/2}^{1/2}
\sum_{m_p=-1/2}^{1/2}
\left(\rho^{\Lc}\right)_{m_{\Lc},m'_{\Lc}} \mathcal{A}_{m'_{\Lc},m_p}(\Omega)
\mathcal{A}^*_{m_{\Lc},m_p}(\Omega),
\label{eq:decay_rate_polarization}
\end{equation}
in which proton spin states are summed over since the proton polarization is not measured. Substituting the \Lc density matrix expression, the differential rate takes the form
\begin{align}
p(\Omega,\bm{P}) = \frac{1}{\mathcal{N}} \sum_{m_p=\pm 1/2} &\left\lbrace (1+P_z) |\mathcal{A}^{\phantom{*}}_{1/2,m_p}(\Omega)|^2 + (1-P_z) |\mathcal{A}_{-1/2,m_p}(\Omega)|^2\right. \nonumber\\
&+ \left. 2\text{Re} \left[ (P_x-iP_y) \mathcal{A}^*_{1/2,m_p}(\Omega)\mathcal{A}_{-1/2,m_p}(\Omega) \right] \right\rbrace,
\label{eq:decay_rate_Lc}
\end{align}
where $\mathcal{N}$ is a normalization constant introduced to make $p$ a probability density function.

\section{Rotation of reference systems and spin states}
\label{sec:euler_rotations}
The rotation of an initial Cartesian reference frame $(x,y,z)$ into a final one $(X,Y,Z)$ can be described by an Euler rotation parametrized by three angles $\alpha$, $\beta$, $\gamma$. Taking the $z$-$y$-$z$ convention for the rotation axes, the Euler rotation is composed by a first rotation of angle $\alpha$ around the $z$ axis, a second rotation of angle $\beta$ around the rotated $y'$ axis and a third one of angle $\gamma$ around the twice rotated $z''$ axis,
\begin{equation}
\mathcal{R}(\alpha,\beta,\gamma) = R_{z''}(\gamma) R_{y'}(\beta) R_{z}(\alpha) = e^{-i\gamma \hat{J}_{z''}} e^{-i\beta \hat{J}_{y'}} e^{-i\alpha \hat{J}_{z}}.
\label{eq:euler_rotation_intrinsic}
\end{equation}
The latter equality expresses rotations in terms of the generating angular momentum operators. Here, active rotations are considered, in which the normalized vectors $\hat{\bm{x}},\hat{\bm{y}},\hat{\bm{z}}$ defining the initial reference frame are actively rotated to those describing the final reference frame $\hat{\bm{X}},\hat{\bm{Y}},\hat{\bm{Z}}$,
\begin{equation}
\hat{\bm{X}}^i = \mathcal{R}(\alpha,\beta,\gamma) \hat{\bm{x}}^i.
\end{equation}
The three Euler angles can be computed as follows: given the vector $\bm{N} = \hat{\bm{z}}\times \hat{\bm{Z}}$, $\alpha$ is the angle between $\hat{\bm{y}}$ and $\hat{\bm{N}}$, $\beta$ is the angle between $\hat{\bm{z}}$ and $\hat{\bm{Z}}$ axes, and $\gamma$ is the angle between $\hat{\bm{N}}$ and $\hat{\bm{Y}}$ axes. In formulae,
\begin{align}
\alpha &= \mathrm{arctan} \left( \hat{\bm{y}} \cdot \hat{\bm{Z}}, \hat{\bm{x}} \cdot \hat{\bm{Z}} \right) \in [-\pi,\pi], \nonumber\\
\beta &= \arccos\left(  \hat{\bm{z}} \cdot \hat{\bm{Z}} \right) \in [0,\pi], \nonumber\\
\gamma &= \mathrm{arctan} \left( \hat{\bm{z}} \cdot \hat{\bm{Y}}, - \hat{\bm{z}} \cdot \hat{\bm{X}} \right) \in [-\pi,\pi].
\label{eq:euler_angles}
\end{align}
The Euler rotation can be also expressed in terms of rotations around the initial reference frame axes only, by means of the equality
\begin{equation}
\mathcal{R}(\alpha,\beta,\gamma) = R_{z}(\alpha) R_{y}(\beta) R_{z}(\gamma) = e^{-i\alpha \hat{J}_{z}} e^{-i\beta \hat{J}_{y}} e^{-i\gamma \hat{J}_{z}}.
\label{eq:euler_rotation_extrinsic}
\end{equation}

The action of the rotation operators $\mathcal{R}(\alpha,\beta,\gamma)$ on angular momentum eigenstates $\ket{J,m}$ can be written as
\begin{equation}
\mathcal{R}(\alpha,\beta,\gamma)\ket{J,m} = \sum_{m'=-J}^{J} D^J_{m',m}(\alpha,\beta,\gamma)\ket{J,m'},
\label{eq:euler_rotation_spin_states}
\end{equation}
where the Wigner $D$-matrices $D^J_{m'm}(\alpha,\beta,\gamma)$ are the matrix elements of the rotation operator in the given spin reference frame,
\begin{equation}
D^{J}_{m',m}(\alpha,\beta,\gamma) = \braket{J,m'|\mathcal{R}(\alpha,\beta,\gamma)|J,m}.
\label{eq:Wigner_D_matrix_definition}
\end{equation}
From the latter equality of Eq.~\eqref{eq:euler_rotation_extrinsic}, the Wigner $D$-matrices can be factorized as
\begin{align}
D^{J}_{m',m}(\alpha,\beta,\gamma) &= \braket{J,m'|e^{-i\alpha \hat{J}_{z}} e^{-i\beta \hat{J}_{y}} e^{-i\gamma \hat{J}_{z}}|J,m}\nonumber\\
& = e^{-im'\alpha}d^J_{m',m}(\beta)e^{-im\gamma},
\label{eq:Wigner_d_matrix_definition}
\end{align}
where the Wigner $d$-matrices elements are known combinations of trigonometric functions of $\beta$ depending on the parameters $J,m$ and $m'$.

\section{Efficiency and background models}
\label{sec:parametrizations}

Efficiency and background models are obtained using factorized Legendre polynomial expansions over the decay phase-space. To this end, the five phase-space variables are transformed into new variables defined in the range $\left[-1,1\right]$. The mass variables \mqpk and \mqkpi are replaced by the so-called square Dalitz plot variables,
\begin{align}
\mppk = 2\text{ }\frac{\mpk-\mpk^{\text{min}}}{\mpk^{\text{max}}-\mpk^{\text{min}}}-1,
\hspace{1cm}
\chl = \frac{m^2_{K^-} + m^2_{\pi^+} + 2E_{K^-} E_{\pi^+} - \mqpk}{2p_{K^-} p_{\pi^+}},
\end{align}
where $\mpk^{\text{min}}=m_p+m_{K^-}$ and $\mpk^{\text{max}}=m_{\Lc}-m_{\pi^+}$ are the minimum and maximum $pK^-$ invariant mass allowed values, and \chl is the cosine of the angle between the kaon and the pion in the \Lz rest frame. The transformation of the decay plane orientation variables is simple, $\cos\theta'_p = \cos\theta_p$, $\phi'_p=\phi_p/\pi$ and $\chi'=\chi/\pi$.

Exploiting the completeness and orthogonality of Legendre polynomials, a generic function $f$ over a phase-space variable $x$ is expanded as
\begin{align}
f(x) &= \sum_i c_i L(x,i) , \nonumber\\
c_i &= \sum_{n=0}^N \frac{w_n}{\rho(x)} \left(2i+1\right) L(x,i),
\end{align}
where $L(x,l)$ is the Legendre polynomial of order $l$, $w_n$ is the weight associated to the phase-space point, for simulated events only, and $\rho$ represents the nonuniformity of the phase-space density over the \mppk variable.

Both parametrizations are obtained by building five one-dimensional polynomial expansions multiplied together,
\begin{equation}
f(\Omega) = f(\mppk) f(\chl) f(\cos\theta'_p) f(\phi'_p) f(\chi'),
\end{equation}
with correlations among phase-space variables found to be negligible.

The efficiency parametrization $\epsilon(\Omega)$ is derived from simulated \Lcpkpi events generated uniformly in phase-space, while the background parametrization $p_{\text{bkg}}(\Omega)$ is obtained from data candidates selected in the invariant mass sidebands within 40 and 80 \mev from the known \Lc mass~\cite{PDG2020}.

\section{Breit--Wigner lineshape}
\label{sec:BW}
To reproduce the typical suppression of transitions involving nonzero orbital angular momentum, a mass-dependent width is introduced in the Breit--Wigner parametrization, the latter multiplied by angular barrier terms involving Blatt--Weisskopf form factors~\cite{Blatt:1952ije,VonHippel:1972fg},
\begin{equation}
\mathcal{R}_{\rm BW}(m^2) = \left[\frac{q(m)}{q_0}\right]^{L_{\Lc}}\left[\frac{p(m)}{p_0}\right]^{L_{R}}
                             \frac{F_{\Lc}(m,L_{\Lc}) F_R(m,L_R)}{m_0^2-m^2 - i m_0\Gamma(m)},
\label{eq:breit_wigner_lineshape}
\end{equation}
where the mass-dependent width
\begin{equation}
\Gamma(m) = \Gamma_0\left[\frac{p(m)}{p_0}\right]^{2L_R+1}\frac{m_0}{m} F_R^2(m,L_R),
\end{equation}
is introduced. The definition of the different quantities entering the above expressions are the following: $m$ is the invariant mass of the resonance, $m_0$ and $\Gamma_0$ are its Breit--Wigner mass and width, $p(m)$ is the momentum of one of the decay products in the resonance two-body decay, $p_0\equiv p(m_0)$, $q(m)$ is the momentum of one of the decay products in the \Lc two-body decay $\Lc\to Rh$, $q_0=q(m_0)$, both defined in the rest frame of the decaying particle, $L_{\Lc}$ and $L_R$ are the orbital angular momenta associated to the \Lc and $R$ decays, respectively. 
The Blatt--Weisskopf form factors for the resonance, $F_R(m,L_R)$, and for the $\Lc$, $F_{\Lc}(m,L_{\Lc})$,
are parametrized as
\begin{equation}
 F_{R,\Lc}(m,L) = \left\{
  \begin{array}{ll}
    1                      & L=0 \\
    \sqrt{\frac{1+z_0^2}{1+z^2(m)}} & L=1 \\
    \sqrt{\frac{9+3z_0^2+z_0^4}{9+3z^2(m)+z^4(m)}} & L=2 \\
  \end{array}
 \right.,
\end{equation}
in which the definitions of the terms $z(m)$ and $z_0$ depend on whether the form factor for the resonance $R$ or for the \Lc is being considered. For $R$ these terms are given by $z(m)=p(m)d$ and $z_0=p_0d$, where $p(m)$ is the momentum of one of the decay products in the resonance two-body decay, $p_0\equiv p(m_0)$, and $d$ is a radial parameter taken to be $1.5\gev^{-1}$. The angular barrier factors arise from the non-relativistic quantization of a particle in a radial potential, and $d$ is often interpreted as the radius of the resonance.
For the \Lc baryon the respective functions are $z(m)=q(m)d$ and $z_0=q_0d$, in which $q(m)$ is the momentum of one of the decay products in the \Lc two-body decay $\Lc\to Rh$, $q_0=q(m_0)$, and $d=5.0\gev^{-1}$.

The mass-dependent width and the form factors depend on the orbital angular momenta of the two-body decays. 
For the \Lc weak decay, the orbital angular momentum is not constrained: the minimum possible value is assumed, since one can expect higher assignments to be energetically disfavored. For half-integer spin $\Lz$ and $\Deltares^{++}$ resonances it is $L_{\Lc}=J_R-1/2$, $J_R$ being the resonance spin, it is $L_{\Lc}=0$ for spin-zero $K^*$ resonances and $L_{\Lc}=J_R-1$ for higher-spin $K^*$ resonances.
A fit performed using $LS$ couplings instead of helicity couplings, including higher orbital angular momentum states is considered as an alternative model (Sec.~\ref{sec:systematic}).
For the strong decay of $\Lz$ and $\Deltares^{++}$ resonances, the orbital angular momentum $L_R$ is determined by the conservation of angular momentum, which requires $L_R=J_R\pm 1/2$, and the parity of the resonance, $P_R=-(-1)^{L_R}$, which chooses one of the $L_R$ values. The additional minus sign is given by the negative parity of the final-state meson. For $K^*$ resonances decaying into two mesons, the orbital angular momentum is $L_R=J_R$.


\addcontentsline{toc}{section}{References}
\bibliographystyle{LHCb}
\bibliography{main,pol,standard,LHCb-PAPER,LHCb-CONF,LHCb-DP,LHCb-TDR}

\ifx\mcitethebibliography\mciteundefinedmacro
\PackageError{LHCb.bst}{mciteplus.sty has not been loaded}
{This bibstyle requires the use of the mciteplus package.}\fi
\providecommand{\href}[2]{#2}
\begin{mcitethebibliography}{10}
\mciteSetBstSublistMode{n}
\mciteSetBstMaxWidthForm{subitem}{\alph{mcitesubitemcount})}
\mciteSetBstSublistLabelBeginEnd{\mcitemaxwidthsubitemform\space}
{\relax}{\relax}

\bibitem{Marangotto:2020ead}
D.~Marangotto, \ifthenelse{\boolean{articletitles}}{\emph{{Extracting maximum
  information from polarised baryon decays via amplitude analysis: The
  $\Lambda^+_c \to pK^-\pi^+$ case}},
  }{}\href{https://doi.org/10.1155/2020/7463073}{Adv.\ High Energy Phys.\
  \textbf{2020} (2020) 7463073},
  \href{http://arxiv.org/abs/2004.12318}{{\normalfont\ttfamily
  arXiv:2004.12318}}\relax
\mciteBstWouldAddEndPuncttrue
\mciteSetBstMidEndSepPunct{\mcitedefaultmidpunct}
{\mcitedefaultendpunct}{\mcitedefaultseppunct}\relax
\EndOfBibitem
\bibitem{Korner:1992wi}
J.~G. K\"orner and M.~Kr\"amer,
  \ifthenelse{\boolean{articletitles}}{\emph{{Exclusive nonleptonic charm
  baryon decays}}, }{}\href{https://doi.org/10.1007/BF01561305}{Z.\ Phys.\
  \textbf{C55} (1992) 659}\relax
\mciteBstWouldAddEndPuncttrue
\mciteSetBstMidEndSepPunct{\mcitedefaultmidpunct}
{\mcitedefaultendpunct}{\mcitedefaultseppunct}\relax
\EndOfBibitem
\bibitem{Davier:1992nw}
M.~Davier, L.~Duflot, F.~Le~Diberder, and A.~Rouge,
  \ifthenelse{\boolean{articletitles}}{\emph{{The optimal method for the
  measurement of tau polarization}},
  }{}\href{https://doi.org/10.1016/0370-2693(93)90101-M}{Phys.\ Lett.\
  \textbf{B306} (1993) 411}\relax
\mciteBstWouldAddEndPuncttrue
\mciteSetBstMidEndSepPunct{\mcitedefaultmidpunct}
{\mcitedefaultendpunct}{\mcitedefaultseppunct}\relax
\EndOfBibitem
\bibitem{Konig:1993ze}
B.~K\"onig, J.~G. K\"orner, M.~Kr\"amer, and P.~Kroll,
  \ifthenelse{\boolean{articletitles}}{\emph{{Infinite momentum frame
  calculation of semileptonic heavy $\Lambda_b \to \Lambda_c$ transitions
  including HQET improvements}},
  }{}\href{https://doi.org/10.1103/PhysRevD.56.4282}{Phys.\ Rev.\  \textbf{D56}
  (1997) 4282},
  \href{http://arxiv.org/abs/hep-ph/9701212}{{\normalfont\ttfamily
  arXiv:hep-ph/9701212}}\relax
\mciteBstWouldAddEndPuncttrue
\mciteSetBstMidEndSepPunct{\mcitedefaultmidpunct}
{\mcitedefaultendpunct}{\mcitedefaultseppunct}\relax
\EndOfBibitem
\bibitem{Pervin:2005ve}
M.~Pervin, W.~Roberts, and S.~Capstick,
  \ifthenelse{\boolean{articletitles}}{\emph{{Semileptonic decays of heavy
  lambda baryons in a quark model}},
  }{}\href{https://doi.org/10.1103/PhysRevC.72.035201}{Phys.\ Rev.\
  \textbf{C72} (2005) 035201},
  \href{http://arxiv.org/abs/nucl-th/0503030}{{\normalfont\ttfamily
  arXiv:nucl-th/0503030}}\relax
\mciteBstWouldAddEndPuncttrue
\mciteSetBstMidEndSepPunct{\mcitedefaultmidpunct}
{\mcitedefaultendpunct}{\mcitedefaultseppunct}\relax
\EndOfBibitem
\bibitem{Gutsche:2015mxa}
T.~Gutsche {\em et~al.},
  \ifthenelse{\boolean{articletitles}}{\emph{{Semileptonic decay $\Lambda_b \to
  \Lambda_c \tau^- \bar{\nu_\tau}$ in the covariant confined quark model}},
  }{}\href{https://doi.org/10.1103/PhysRevD.91.074001}{Phys.\ Rev.\
  \textbf{D91} (2015) 074001},
  \href{http://arxiv.org/abs/1502.04864}{{\normalfont\ttfamily
  arXiv:1502.04864}}, [Erratum: Phys.Rev.D 91, 119907 (2015)]\relax
\mciteBstWouldAddEndPuncttrue
\mciteSetBstMidEndSepPunct{\mcitedefaultmidpunct}
{\mcitedefaultendpunct}{\mcitedefaultseppunct}\relax
\EndOfBibitem
\bibitem{Shivashankara:2015cta}
S.~Shivashankara, W.~Wu, and A.~Datta,
  \ifthenelse{\boolean{articletitles}}{\emph{{$\Lambda_b \to \Lambda_c \tau
  \bar{\nu}_{\tau}$ decay in the standard model and with new physics}},
  }{}\href{https://doi.org/10.1103/PhysRevD.91.115003}{Phys.\ Rev.\
  \textbf{D91} (2015) 115003},
  \href{http://arxiv.org/abs/1502.07230}{{\normalfont\ttfamily
  arXiv:1502.07230}}\relax
\mciteBstWouldAddEndPuncttrue
\mciteSetBstMidEndSepPunct{\mcitedefaultmidpunct}
{\mcitedefaultendpunct}{\mcitedefaultseppunct}\relax
\EndOfBibitem
\bibitem{Dutta:2015ueb}
R.~Dutta, \ifthenelse{\boolean{articletitles}}{\emph{{$\Lambda_b \to
  (\Lambda_c,\,p)\,\tau\,\nu$ decays within standard model and beyond}},
  }{}\href{https://doi.org/10.1103/PhysRevD.93.054003}{Phys.\ Rev.\
  \textbf{D93} (2016) 054003},
  \href{http://arxiv.org/abs/1512.04034}{{\normalfont\ttfamily
  arXiv:1512.04034}}\relax
\mciteBstWouldAddEndPuncttrue
\mciteSetBstMidEndSepPunct{\mcitedefaultmidpunct}
{\mcitedefaultendpunct}{\mcitedefaultseppunct}\relax
\EndOfBibitem
\bibitem{Faustov:2016pal}
R.~N. Faustov and V.~O. Galkin,
  \ifthenelse{\boolean{articletitles}}{\emph{{Semileptonic decays of
  $\Lambda_b$ baryons in the relativistic quark model}},
  }{}\href{https://doi.org/10.1103/PhysRevD.94.073008}{Phys.\ Rev.\
  \textbf{D94} (2016) 073008},
  \href{http://arxiv.org/abs/1609.00199}{{\normalfont\ttfamily
  arXiv:1609.00199}}\relax
\mciteBstWouldAddEndPuncttrue
\mciteSetBstMidEndSepPunct{\mcitedefaultmidpunct}
{\mcitedefaultendpunct}{\mcitedefaultseppunct}\relax
\EndOfBibitem
\bibitem{Li:2016pdv}
X.-Q. Li, Y.-D. Yang, and X.~Zhang,
  \ifthenelse{\boolean{articletitles}}{\emph{{$ {\varLambda}_b\to
  {\varLambda}_c\tau {\overline{\nu}}_{\tau } $ decay in scalar and vector
  leptoquark scenarios}},
  }{}\href{https://doi.org/10.1007/JHEP02(2017)068}{JHEP \textbf{02} (2017)
  068}, \href{http://arxiv.org/abs/1611.01635}{{\normalfont\ttfamily
  arXiv:1611.01635}}\relax
\mciteBstWouldAddEndPuncttrue
\mciteSetBstMidEndSepPunct{\mcitedefaultmidpunct}
{\mcitedefaultendpunct}{\mcitedefaultseppunct}\relax
\EndOfBibitem
\bibitem{Celis:2016azn}
A.~Celis, M.~Jung, X.-Q. Li, and A.~Pich,
  \ifthenelse{\boolean{articletitles}}{\emph{{Scalar contributions to $b\to c
  (u) \tau \nu$ transitions}},
  }{}\href{https://doi.org/10.1016/j.physletb.2017.05.037}{Phys.\ Lett.\
  \textbf{B771} (2017) 168},
  \href{http://arxiv.org/abs/1612.07757}{{\normalfont\ttfamily
  arXiv:1612.07757}}\relax
\mciteBstWouldAddEndPuncttrue
\mciteSetBstMidEndSepPunct{\mcitedefaultmidpunct}
{\mcitedefaultendpunct}{\mcitedefaultseppunct}\relax
\EndOfBibitem
\bibitem{Datta:2017aue}
A.~Datta, S.~Kamali, S.~Meinel, and A.~Rashed,
  \ifthenelse{\boolean{articletitles}}{\emph{{Phenomenology of $ {\Lambda}_b\to
  {\Lambda}_c\tau {\overline{\nu}}_{\tau } $ using lattice QCD calculations}},
  }{}\href{https://doi.org/10.1007/JHEP08(2017)131}{JHEP \textbf{08} (2017)
  131}, \href{http://arxiv.org/abs/1702.02243}{{\normalfont\ttfamily
  arXiv:1702.02243}}\relax
\mciteBstWouldAddEndPuncttrue
\mciteSetBstMidEndSepPunct{\mcitedefaultmidpunct}
{\mcitedefaultendpunct}{\mcitedefaultseppunct}\relax
\EndOfBibitem
\bibitem{Zhu:2018zxb}
J.~Zhu {\em et~al.}, \ifthenelse{\boolean{articletitles}}{\emph{{Probing the
  R-parity violating supersymmetric effects in \mbox{$B_c\to
  J/\psi\ell^-\bar{\nu}_{\ell},\eta_c\ell^-\bar{\nu}_{\ell}$} and
  $\Lambda_b\to\Lambda_c\ell^-\bar{\nu}_{\ell}$ decays}},
  }{}\href{https://doi.org/10.1016/j.nuclphysb.2018.07.011}{Nucl.\ Phys.\
  \textbf{B934} (2018) 380},
  \href{http://arxiv.org/abs/1801.00917}{{\normalfont\ttfamily
  arXiv:1801.00917}}\relax
\mciteBstWouldAddEndPuncttrue
\mciteSetBstMidEndSepPunct{\mcitedefaultmidpunct}
{\mcitedefaultendpunct}{\mcitedefaultseppunct}\relax
\EndOfBibitem
\bibitem{DiSalvo:2018ngq}
E.~Di~Salvo, F.~Fontanelli, and Z.~J. Ajaltouni,
  \ifthenelse{\boolean{articletitles}}{\emph{{Detailed study of the decay
  $\Lambda_b \to \Lambda_c \tau {\bar \nu}_{\tau}$}},
  }{}\href{https://doi.org/10.1142/S0217751X18501695}{Int.\ J.\ Mod.\ Phys.\
  \textbf{A33} (2018) 1850169},
  \href{http://arxiv.org/abs/1804.05592}{{\normalfont\ttfamily
  arXiv:1804.05592}}\relax
\mciteBstWouldAddEndPuncttrue
\mciteSetBstMidEndSepPunct{\mcitedefaultmidpunct}
{\mcitedefaultendpunct}{\mcitedefaultseppunct}\relax
\EndOfBibitem
\bibitem{Ray:2018hrx}
A.~Ray, S.~Sahoo, and R.~Mohanta,
  \ifthenelse{\boolean{articletitles}}{\emph{{Probing new physics in
  semileptonic $\Lambda_b$ decays}},
  }{}\href{https://doi.org/10.1103/PhysRevD.99.015015}{Phys.\ Rev.\
  \textbf{D99} (2019) 015015},
  \href{http://arxiv.org/abs/1812.08314}{{\normalfont\ttfamily
  arXiv:1812.08314}}\relax
\mciteBstWouldAddEndPuncttrue
\mciteSetBstMidEndSepPunct{\mcitedefaultmidpunct}
{\mcitedefaultendpunct}{\mcitedefaultseppunct}\relax
\EndOfBibitem
\bibitem{Bernlochner:2018bfn}
F.~U. Bernlochner, Z.~Ligeti, D.~J. Robinson, and W.~L. Sutcliffe,
  \ifthenelse{\boolean{articletitles}}{\emph{{Precise predictions for
  $\Lambda_b \to \Lambda_c$ semileptonic decays}},
  }{}\href{https://doi.org/10.1103/PhysRevD.99.055008}{Phys.\ Rev.\
  \textbf{D99} (2019) 055008},
  \href{http://arxiv.org/abs/1812.07593}{{\normalfont\ttfamily
  arXiv:1812.07593}}\relax
\mciteBstWouldAddEndPuncttrue
\mciteSetBstMidEndSepPunct{\mcitedefaultmidpunct}
{\mcitedefaultendpunct}{\mcitedefaultseppunct}\relax
\EndOfBibitem
\bibitem{Penalva:2019rgt}
N.~Penalva, E.~Hern\'andez, and J.~Nieves,
  \ifthenelse{\boolean{articletitles}}{\emph{{Further tests of lepton flavour
  universality from the charged lepton energy distribution in $b\to c$
  semileptonic decays: The case of $\Lambda_b\to \Lambda_c \ell
  \bar\nu_\ell$}}, }{}\href{https://doi.org/10.1103/PhysRevD.100.113007}{Phys.\
  Rev.\  \textbf{D100} (2019) 113007},
  \href{http://arxiv.org/abs/1908.02328}{{\normalfont\ttfamily
  arXiv:1908.02328}}\relax
\mciteBstWouldAddEndPuncttrue
\mciteSetBstMidEndSepPunct{\mcitedefaultmidpunct}
{\mcitedefaultendpunct}{\mcitedefaultseppunct}\relax
\EndOfBibitem
\bibitem{Hu:2020axt}
Q.-Y. Hu, X.-Q. Li, Y.-D. Yang, and D.-H. Zheng,
  \ifthenelse{\boolean{articletitles}}{\emph{{The measurable angular
  distribution of $ {\Lambda}_b^0\to {\Lambda}_c^{+}\left(\to
  {\Lambda}^0{\pi}^{+}\right){\tau}^{-}\left(\to
  {\pi}^{-}{v}_{\tau}\right){\overline{v}}_{\tau } $ decay}},
  }{}\href{https://doi.org/10.1007/JHEP02(2021)183}{JHEP \textbf{02} (2021)
  183}, \href{http://arxiv.org/abs/2011.05912}{{\normalfont\ttfamily
  arXiv:2011.05912}}\relax
\mciteBstWouldAddEndPuncttrue
\mciteSetBstMidEndSepPunct{\mcitedefaultmidpunct}
{\mcitedefaultendpunct}{\mcitedefaultseppunct}\relax
\EndOfBibitem
\bibitem{Lees:2013uzd}
BaBar collaboration, J.~P. Lees {\em et~al.},
  \ifthenelse{\boolean{articletitles}}{\emph{{Measurement of an excess of
  $\bar{B} \to D^{(*)}\tau^- \bar{\nu}_\tau$ decays and implications for
  charged Higgs bosons}},
  }{}\href{https://doi.org/10.1103/PhysRevD.88.072012}{Phys.\ Rev.\
  \textbf{D88} (2013) 072012},
  \href{http://arxiv.org/abs/1303.0571}{{\normalfont\ttfamily
  arXiv:1303.0571}}\relax
\mciteBstWouldAddEndPuncttrue
\mciteSetBstMidEndSepPunct{\mcitedefaultmidpunct}
{\mcitedefaultendpunct}{\mcitedefaultseppunct}\relax
\EndOfBibitem
\bibitem{Huschle:2015rga}
Belle collaboration, M.~Huschle {\em et~al.},
  \ifthenelse{\boolean{articletitles}}{\emph{{Measurement of the branching
  ratio of \mbox{$\bar{B} \to D^{(\ast)} \tau^- \bar{\nu}_\tau$} relative to
  $\bar{B} \to D^{(\ast)} \ell^- \bar{\nu}_\ell$ decays with hadronic tagging
  at Belle}}, }{}\href{https://doi.org/10.1103/PhysRevD.92.072014}{Phys.\ Rev.\
   \textbf{D92} (2015) 072014},
  \href{http://arxiv.org/abs/1507.03233}{{\normalfont\ttfamily
  arXiv:1507.03233}}\relax
\mciteBstWouldAddEndPuncttrue
\mciteSetBstMidEndSepPunct{\mcitedefaultmidpunct}
{\mcitedefaultendpunct}{\mcitedefaultseppunct}\relax
\EndOfBibitem
\bibitem{Sato:2016svk}
Belle collaboration, Y.~Sato {\em et~al.},
  \ifthenelse{\boolean{articletitles}}{\emph{{Measurement of the branching
  ratio of \mbox{$\bar{B}^0 \rightarrow D^{*+} \tau^- \bar{\nu}_{\tau}$}
  relative to $\bar{B}^0 \rightarrow D^{*+} \ell^- \bar{\nu}_{\ell}$ decays
  with a semileptonic tagging method}},
  }{}\href{https://doi.org/10.1103/PhysRevD.94.072007}{Phys.\ Rev.\
  \textbf{D94} (2016) 072007},
  \href{http://arxiv.org/abs/1607.07923}{{\normalfont\ttfamily
  arXiv:1607.07923}}\relax
\mciteBstWouldAddEndPuncttrue
\mciteSetBstMidEndSepPunct{\mcitedefaultmidpunct}
{\mcitedefaultendpunct}{\mcitedefaultseppunct}\relax
\EndOfBibitem
\bibitem{Hirose:2016wfn}
Belle collaboration, S.~Hirose {\em et~al.},
  \ifthenelse{\boolean{articletitles}}{\emph{{Measurement of the $\tau$ lepton
  polarization and $R(D^*)$ in the decay $\bar{B} \to D^* \tau^-
  \bar{\nu}_\tau$}},
  }{}\href{https://doi.org/10.1103/PhysRevLett.118.211801}{Phys.\ Rev.\ Lett.\
  \textbf{118} (2017) 211801},
  \href{http://arxiv.org/abs/1612.00529}{{\normalfont\ttfamily
  arXiv:1612.00529}}\relax
\mciteBstWouldAddEndPuncttrue
\mciteSetBstMidEndSepPunct{\mcitedefaultmidpunct}
{\mcitedefaultendpunct}{\mcitedefaultseppunct}\relax
\EndOfBibitem
\bibitem{LHCb-PAPER-2017-027}
LHCb collaboration, R.~Aaij {\em et~al.},
  \ifthenelse{\boolean{articletitles}}{\emph{{Test of lepton flavor
  universality by the measurement of the \mbox{\decay{\Bz}{D^{\ast-} \taup
  \nu_{\tau}}} branching fraction using three-prong $\tau$ decays}},
  }{}\href{https://doi.org/10.1103/PhysRevD.97.072013}{Phys.\ Rev.\
  \textbf{D97} (2018) 072013},
  \href{http://arxiv.org/abs/1711.02505}{{\normalfont\ttfamily
  arXiv:1711.02505}}\relax
\mciteBstWouldAddEndPuncttrue
\mciteSetBstMidEndSepPunct{\mcitedefaultmidpunct}
{\mcitedefaultendpunct}{\mcitedefaultseppunct}\relax
\EndOfBibitem
\bibitem{LHCb-PAPER-2015-025}
LHCb collaboration, R.~Aaij {\em et~al.},
  \ifthenelse{\boolean{articletitles}}{\emph{{Measurement of the ratio of
  branching fractions
  \mbox{$\BF(\decay{\Bzb}{\Dstarp\taum\neutb})/\BF(\decay{\Bzb}{\Dstarp\mun\neumb})$}}},
  }{}\href{https://doi.org/10.1103/PhysRevLett.115.111803}{Phys.\ Rev.\ Lett.\
  \textbf{115} (2015) 111803}, Publisher's Note
  \href{https://doi.org/10.1103/PhysRevLett.115.159901}{ibid.\   \textbf{115}
  (2015) 159901}, \href{http://arxiv.org/abs/1506.08614}{{\normalfont\ttfamily
  arXiv:1506.08614}}\relax
\mciteBstWouldAddEndPuncttrue
\mciteSetBstMidEndSepPunct{\mcitedefaultmidpunct}
{\mcitedefaultendpunct}{\mcitedefaultseppunct}\relax
\EndOfBibitem
\bibitem{PDG2020}
Particle Data Group, P.~A. Zyla {\em et~al.},
  \ifthenelse{\boolean{articletitles}}{\emph{{\href{http://pdg.lbl.gov/}{Review
  of particle physics}}}, }{}\href{https://doi.org/10.1093/ptep/ptaa104}{Prog.\
  Theor.\ Exp.\ Phys.\  \textbf{2020} (2020) 083C01}\relax
\mciteBstWouldAddEndPuncttrue
\mciteSetBstMidEndSepPunct{\mcitedefaultmidpunct}
{\mcitedefaultendpunct}{\mcitedefaultseppunct}\relax
\EndOfBibitem
\bibitem{Konig:1993wz}
B.~K\"{o}nig, J.~G. K\"orner, and M.~Kr\"amer,
  \ifthenelse{\boolean{articletitles}}{\emph{{On the determination of the $b
  \to c$ handedness using nonleptonic $\Lambda_c$ decays}},
  }{}\href{https://doi.org/10.1103/PhysRevD.49.2363}{Phys.\ Rev.\  \textbf{D49}
  (1994) 2363},
  \href{http://arxiv.org/abs/hep-ph/9310263}{{\normalfont\ttfamily
  arXiv:hep-ph/9310263}}\relax
\mciteBstWouldAddEndPuncttrue
\mciteSetBstMidEndSepPunct{\mcitedefaultmidpunct}
{\mcitedefaultendpunct}{\mcitedefaultseppunct}\relax
\EndOfBibitem
\bibitem{Mu:2019bin}
X.-L. Mu, Y.~Li, Z.-T. Zou, and B.~Zhu,
  \ifthenelse{\boolean{articletitles}}{\emph{{Investigation of effects of new
  physics in $\Lambda_b\to\Lambda_c \tau\bar\nu_\tau$ decay}},
  }{}\href{https://doi.org/10.1103/PhysRevD.100.113004}{Phys.\ Rev.\
  \textbf{D100} (2019) 113004},
  \href{http://arxiv.org/abs/1909.10769}{{\normalfont\ttfamily
  arXiv:1909.10769}}\relax
\mciteBstWouldAddEndPuncttrue
\mciteSetBstMidEndSepPunct{\mcitedefaultmidpunct}
{\mcitedefaultendpunct}{\mcitedefaultseppunct}\relax
\EndOfBibitem
\bibitem{Falk:1993rf}
A.~F. Falk and M.~E. Peskin,
  \ifthenelse{\boolean{articletitles}}{\emph{{Production, decay, and
  polarization of excited heavy hadrons}},
  }{}\href{https://doi.org/10.1103/PhysRevD.49.3320}{Phys.\ Rev.\  \textbf{D49}
  (1994) 3320},
  \href{http://arxiv.org/abs/hep-ph/9308241}{{\normalfont\ttfamily
  arXiv:hep-ph/9308241}}\relax
\mciteBstWouldAddEndPuncttrue
\mciteSetBstMidEndSepPunct{\mcitedefaultmidpunct}
{\mcitedefaultendpunct}{\mcitedefaultseppunct}\relax
\EndOfBibitem
\bibitem{Galanti:2015pqa}
M.~Galanti {\em et~al.}, \ifthenelse{\boolean{articletitles}}{\emph{{Heavy
  baryons as polarimeters at colliders}},
  }{}\href{https://doi.org/10.1007/JHEP11(2015)067}{JHEP \textbf{11} (2015)
  067}, \href{http://arxiv.org/abs/1505.02771}{{\normalfont\ttfamily
  arXiv:1505.02771}}\relax
\mciteBstWouldAddEndPuncttrue
\mciteSetBstMidEndSepPunct{\mcitedefaultmidpunct}
{\mcitedefaultendpunct}{\mcitedefaultseppunct}\relax
\EndOfBibitem
\bibitem{Goldstein:1999jr}
G.~R. Goldstein, \ifthenelse{\boolean{articletitles}}{\emph{{Polarization of
  inclusively produced {$\Lambda^+_c$} in a QCD based hybrid model}},
  }{}\href{http://arxiv.org/abs/hep-ph/9907573}{{\normalfont\ttfamily
  arXiv:hep-ph/9907573}}\relax
\mciteBstWouldAddEndPuncttrue
\mciteSetBstMidEndSepPunct{\mcitedefaultmidpunct}
{\mcitedefaultendpunct}{\mcitedefaultseppunct}\relax
\EndOfBibitem
\bibitem{Goldstein:2015aqa}
G.~R. Goldstein and S.~Liuti,
  \ifthenelse{\boolean{articletitles}}{\emph{{Angular momentum and polarization
  in hadron collisions up to LHC energies}},
  }{}\href{https://doi.org/10.1142/S2010194515600381}{Int.\ J.\ Mod.\ Phys.\
  Conf.\ Ser.\  \textbf{37} (2015) 1560038}\relax
\mciteBstWouldAddEndPuncttrue
\mciteSetBstMidEndSepPunct{\mcitedefaultmidpunct}
{\mcitedefaultendpunct}{\mcitedefaultseppunct}\relax
\EndOfBibitem
\bibitem{Mannel:1991bs}
T.~Mannel and G.~A. Schuler,
  \ifthenelse{\boolean{articletitles}}{\emph{{Semileptonic decays of bottom
  baryons at LEP}},
  }{}\href{https://doi.org/10.1016/0370-2693(92)91864-6}{Phys.\ Lett.\
  \textbf{B279} (1992) 194}\relax
\mciteBstWouldAddEndPuncttrue
\mciteSetBstMidEndSepPunct{\mcitedefaultmidpunct}
{\mcitedefaultendpunct}{\mcitedefaultseppunct}\relax
\EndOfBibitem
\bibitem{Botella:2016ksl}
F.~J. Botella {\em et~al.}, \ifthenelse{\boolean{articletitles}}{\emph{{On the
  search for the electric dipole moment of strange and charm baryons at LHC}},
  }{}\href{https://doi.org/10.1140/epjc/s10052-017-4679-y}{Eur.\ Phys.\ J.\
  \textbf{C77} (2017) 181},
  \href{http://arxiv.org/abs/1612.06769}{{\normalfont\ttfamily
  arXiv:1612.06769}}\relax
\mciteBstWouldAddEndPuncttrue
\mciteSetBstMidEndSepPunct{\mcitedefaultmidpunct}
{\mcitedefaultendpunct}{\mcitedefaultseppunct}\relax
\EndOfBibitem
\bibitem{Bagli:2017foe}
E.~Bagli {\em et~al.},
  \ifthenelse{\boolean{articletitles}}{\emph{{Electromagnetic dipole moments of
  charged baryons with bent crystals at the LHC}},
  }{}\href{https://doi.org/10.1140/epjc/s10052-017-5400-x}{Eur.\ Phys.\ J.\
  \textbf{C77} (2017) 828},
  \href{http://arxiv.org/abs/1708.08483}{{\normalfont\ttfamily
  arXiv:1708.08483}}\relax
\mciteBstWouldAddEndPuncttrue
\mciteSetBstMidEndSepPunct{\mcitedefaultmidpunct}
{\mcitedefaultendpunct}{\mcitedefaultseppunct}\relax
\EndOfBibitem
\bibitem{Marangotto:2713231}
D.~Marangotto, \ifthenelse{\boolean{articletitles}}{\emph{{Amplitude analysis
  and polarisation measurement of the $\Lambda^+_c$ baryon in $pK^-\pi^+$ final
  state for electromagnetic dipole moment experiment}}, }{}
\newblock PhD thesis, Universit\`a degli studi di Milano, Presented 16 Mar
  2020, \url{https://cds.cern.ch/record/2713231}\relax
\mciteBstWouldAddEndPuncttrue
\mciteSetBstMidEndSepPunct{\mcitedefaultmidpunct}
{\mcitedefaultendpunct}{\mcitedefaultseppunct}\relax
\EndOfBibitem
\bibitem{Aiola:2020yam}
S.~Aiola {\em et~al.}, \ifthenelse{\boolean{articletitles}}{\emph{{Progress
  towards the first measurement of charm baryon dipole moments}},
  }{}\href{https://doi.org/10.1103/PhysRevD.103.072003}{Phys.\ Rev.\
  \textbf{D103} (2020) 072003},
  \href{http://arxiv.org/abs/2010.11902}{{\normalfont\ttfamily
  arXiv:2010.11902}}\relax
\mciteBstWouldAddEndPuncttrue
\mciteSetBstMidEndSepPunct{\mcitedefaultmidpunct}
{\mcitedefaultendpunct}{\mcitedefaultseppunct}\relax
\EndOfBibitem
\bibitem{Aitala:1999uq}
E791 collaboration, E.~M. Aitala {\em et~al.},
  \ifthenelse{\boolean{articletitles}}{\emph{Multidimensional resonance
  analysis of {$\Lambda^+_c\to pK^-\pi^+$}},
  }{}\href{https://doi.org/10.1016/S0370-2693(99)01397-0}{Phys.\ Lett.\
  \textbf{B471} (2000) 449},
  \href{http://arxiv.org/abs/hep-ex/9912003}{{\normalfont\ttfamily
  arXiv:hep-ex/9912003}}\relax
\mciteBstWouldAddEndPuncttrue
\mciteSetBstMidEndSepPunct{\mcitedefaultmidpunct}
{\mcitedefaultendpunct}{\mcitedefaultseppunct}\relax
\EndOfBibitem
\bibitem{Marangotto:2019ucc}
D.~Marangotto, \ifthenelse{\boolean{articletitles}}{\emph{{Helicity amplitudes
  for generic multibody particle decays featuring multiple decay chains}},
  }{}\href{https://doi.org/10.1155/2020/6674595}{Adv.\ High Energy Phys.\
  \textbf{2020} (2020) 6674595},
  \href{http://arxiv.org/abs/1911.10025}{{\normalfont\ttfamily
  arXiv:1911.10025}}\relax
\mciteBstWouldAddEndPuncttrue
\mciteSetBstMidEndSepPunct{\mcitedefaultmidpunct}
{\mcitedefaultendpunct}{\mcitedefaultseppunct}\relax
\EndOfBibitem
\bibitem{LHCb-DP-2008-001}
LHCb collaboration, A.~A. Alves~Jr.\ {\em et~al.},
  \ifthenelse{\boolean{articletitles}}{\emph{{The \lhcb detector at the LHC}},
  }{}\href{https://doi.org/10.1088/1748-0221/3/08/S08005}{JINST \textbf{3}
  (2008) S08005}\relax
\mciteBstWouldAddEndPuncttrue
\mciteSetBstMidEndSepPunct{\mcitedefaultmidpunct}
{\mcitedefaultendpunct}{\mcitedefaultseppunct}\relax
\EndOfBibitem
\bibitem{LHCb-DP-2014-002}
LHCb collaboration, R.~Aaij {\em et~al.},
  \ifthenelse{\boolean{articletitles}}{\emph{{LHCb detector performance}},
  }{}\href{https://doi.org/10.1142/S0217751X15300227}{Int.\ J.\ Mod.\ Phys.\
  \textbf{A30} (2015) 1530022},
  \href{http://arxiv.org/abs/1412.6352}{{\normalfont\ttfamily
  arXiv:1412.6352}}\relax
\mciteBstWouldAddEndPuncttrue
\mciteSetBstMidEndSepPunct{\mcitedefaultmidpunct}
{\mcitedefaultendpunct}{\mcitedefaultseppunct}\relax
\EndOfBibitem
\bibitem{BBDT}
V.~V. Gligorov and M.~Williams,
  \ifthenelse{\boolean{articletitles}}{\emph{{Efficient, reliable and fast
  high-level triggering using a bonsai boosted decision tree}},
  }{}\href{https://doi.org/10.1088/1748-0221/8/02/P02013}{JINST \textbf{8}
  (2013) P02013}, \href{http://arxiv.org/abs/1210.6861}{{\normalfont\ttfamily
  arXiv:1210.6861}}\relax
\mciteBstWouldAddEndPuncttrue
\mciteSetBstMidEndSepPunct{\mcitedefaultmidpunct}
{\mcitedefaultendpunct}{\mcitedefaultseppunct}\relax
\EndOfBibitem
\bibitem{LHCb-PROC-2015-018}
T.~Likhomanenko {\em et~al.}, \ifthenelse{\boolean{articletitles}}{\emph{{LHCb
  topological trigger reoptimization}},
  }{}\href{https://doi.org/10.1088/1742-6596/664/8/082025}{J.\ Phys.\ Conf.\
  Ser.\  \textbf{664} (2015) 082025}\relax
\mciteBstWouldAddEndPuncttrue
\mciteSetBstMidEndSepPunct{\mcitedefaultmidpunct}
{\mcitedefaultendpunct}{\mcitedefaultseppunct}\relax
\EndOfBibitem
\bibitem{LHCb-PAPER-2012-048}
LHCb collaboration, R.~Aaij {\em et~al.},
  \ifthenelse{\boolean{articletitles}}{\emph{{Measurements of the \Lb, \Xibm,
  and \Omegab baryon masses}},
  }{}\href{https://doi.org/10.1103/PhysRevLett.110.182001}{Phys.\ Rev.\ Lett.\
  \textbf{110} (2013) 182001},
  \href{http://arxiv.org/abs/1302.1072}{{\normalfont\ttfamily
  arXiv:1302.1072}}\relax
\mciteBstWouldAddEndPuncttrue
\mciteSetBstMidEndSepPunct{\mcitedefaultmidpunct}
{\mcitedefaultendpunct}{\mcitedefaultseppunct}\relax
\EndOfBibitem
\bibitem{LHCb-PAPER-2013-011}
LHCb collaboration, R.~Aaij {\em et~al.},
  \ifthenelse{\boolean{articletitles}}{\emph{{Precision measurement of \D meson
  mass differences}}, }{}\href{https://doi.org/10.1007/JHEP06(2013)065}{JHEP
  \textbf{06} (2013) 065},
  \href{http://arxiv.org/abs/1304.6865}{{\normalfont\ttfamily
  arXiv:1304.6865}}\relax
\mciteBstWouldAddEndPuncttrue
\mciteSetBstMidEndSepPunct{\mcitedefaultmidpunct}
{\mcitedefaultendpunct}{\mcitedefaultseppunct}\relax
\EndOfBibitem
\bibitem{Sjostrand:2007gs}
T.~Sj\"{o}strand, S.~Mrenna, and P.~Skands,
  \ifthenelse{\boolean{articletitles}}{\emph{{A brief introduction to PYTHIA
  8.1}}, }{}\href{https://doi.org/10.1016/j.cpc.2008.01.036}{Comput.\ Phys.\
  Commun.\  \textbf{178} (2008) 852},
  \href{http://arxiv.org/abs/0710.3820}{{\normalfont\ttfamily
  arXiv:0710.3820}}\relax
\mciteBstWouldAddEndPuncttrue
\mciteSetBstMidEndSepPunct{\mcitedefaultmidpunct}
{\mcitedefaultendpunct}{\mcitedefaultseppunct}\relax
\EndOfBibitem
\bibitem{Sjostrand:2006za}
T.~Sj\"{o}strand, S.~Mrenna, and P.~Skands,
  \ifthenelse{\boolean{articletitles}}{\emph{{PYTHIA 6.4 physics and manual}},
  }{}\href{https://doi.org/10.1088/1126-6708/2006/05/026}{JHEP \textbf{05}
  (2006) 026}, \href{http://arxiv.org/abs/hep-ph/0603175}{{\normalfont\ttfamily
  arXiv:hep-ph/0603175}}\relax
\mciteBstWouldAddEndPuncttrue
\mciteSetBstMidEndSepPunct{\mcitedefaultmidpunct}
{\mcitedefaultendpunct}{\mcitedefaultseppunct}\relax
\EndOfBibitem
\bibitem{LHCb-PROC-2010-056}
I.~Belyaev {\em et~al.}, \ifthenelse{\boolean{articletitles}}{\emph{{Handling
  of the generation of primary events in Gauss, the LHCb simulation
  framework}}, }{}\href{https://doi.org/10.1088/1742-6596/331/3/032047}{J.\
  Phys.\ Conf.\ Ser.\  \textbf{331} (2011) 032047}\relax
\mciteBstWouldAddEndPuncttrue
\mciteSetBstMidEndSepPunct{\mcitedefaultmidpunct}
{\mcitedefaultendpunct}{\mcitedefaultseppunct}\relax
\EndOfBibitem
\bibitem{Lange:2001uf}
D.~J. Lange, \ifthenelse{\boolean{articletitles}}{\emph{{The EvtGen particle
  decay simulation package}},
  }{}\href{https://doi.org/10.1016/S0168-9002(01)00089-4}{Nucl.\ Instrum.\
  Meth.\  \textbf{A462} (2001) 152}\relax
\mciteBstWouldAddEndPuncttrue
\mciteSetBstMidEndSepPunct{\mcitedefaultmidpunct}
{\mcitedefaultendpunct}{\mcitedefaultseppunct}\relax
\EndOfBibitem
\bibitem{davidson2015photos}
N.~Davidson, T.~Przedzinski, and Z.~Was,
  \ifthenelse{\boolean{articletitles}}{\emph{{PHOTOS interface in C++:
  Technical and physics documentation}},
  }{}\href{https://doi.org/https://doi.org/10.1016/j.cpc.2015.09.013}{Comp.\
  Phys.\ Comm.\  \textbf{199} (2016) 86},
  \href{http://arxiv.org/abs/1011.0937}{{\normalfont\ttfamily
  arXiv:1011.0937}}\relax
\mciteBstWouldAddEndPuncttrue
\mciteSetBstMidEndSepPunct{\mcitedefaultmidpunct}
{\mcitedefaultendpunct}{\mcitedefaultseppunct}\relax
\EndOfBibitem
\bibitem{Allison:2006ve}
Geant4 collaboration, J.~Allison {\em et~al.},
  \ifthenelse{\boolean{articletitles}}{\emph{{Geant4 developments and
  applications}}, }{}\href{https://doi.org/10.1109/TNS.2006.869826}{IEEE
  Trans.\ Nucl.\ Sci.\  \textbf{53} (2006) 270}\relax
\mciteBstWouldAddEndPuncttrue
\mciteSetBstMidEndSepPunct{\mcitedefaultmidpunct}
{\mcitedefaultendpunct}{\mcitedefaultseppunct}\relax
\EndOfBibitem
\bibitem{Agostinelli:2002hh}
Geant4 collaboration, S.~Agostinelli {\em et~al.},
  \ifthenelse{\boolean{articletitles}}{\emph{{Geant4: A simulation toolkit}},
  }{}\href{https://doi.org/10.1016/S0168-9002(03)01368-8}{Nucl.\ Instrum.\
  Meth.\  \textbf{A506} (2003) 250}\relax
\mciteBstWouldAddEndPuncttrue
\mciteSetBstMidEndSepPunct{\mcitedefaultmidpunct}
{\mcitedefaultendpunct}{\mcitedefaultseppunct}\relax
\EndOfBibitem
\bibitem{LHCb-PROC-2011-006}
M.~Clemencic {\em et~al.}, \ifthenelse{\boolean{articletitles}}{\emph{{The
  \lhcb simulation application, Gauss: Design, evolution and experience}},
  }{}\href{https://doi.org/10.1088/1742-6596/331/3/032023}{J.\ Phys.\ Conf.\
  Ser.\  \textbf{331} (2011) 032023}\relax
\mciteBstWouldAddEndPuncttrue
\mciteSetBstMidEndSepPunct{\mcitedefaultmidpunct}
{\mcitedefaultendpunct}{\mcitedefaultseppunct}\relax
\EndOfBibitem
\bibitem{LHCb-DP-2018-004}
D.~M{\"u}ller, M.~Clemencic, G.~Corti, and M.~Gersabeck,
  \ifthenelse{\boolean{articletitles}}{\emph{{ReDecay: A novel approach to
  speed up the simulation at LHCb}},
  }{}\href{https://doi.org/10.1140/epjc/s10052-018-6469-6}{Eur.\ Phys.\ J.\
  \textbf{C78} (2018) 1009},
  \href{http://arxiv.org/abs/1810.10362}{{\normalfont\ttfamily
  arXiv:1810.10362}}\relax
\mciteBstWouldAddEndPuncttrue
\mciteSetBstMidEndSepPunct{\mcitedefaultmidpunct}
{\mcitedefaultendpunct}{\mcitedefaultseppunct}\relax
\EndOfBibitem
\bibitem{Poluektov:2014rxa}
A.~Poluektov, \ifthenelse{\boolean{articletitles}}{\emph{{Kernel density
  estimation of a multidimensional efficiency profile}},
  }{}\href{https://doi.org/10.1088/1748-0221/10/02/P02011}{JINST \textbf{10}
  (2015) P02011}, \href{http://arxiv.org/abs/1411.5528}{{\normalfont\ttfamily
  arXiv:1411.5528}}\relax
\mciteBstWouldAddEndPuncttrue
\mciteSetBstMidEndSepPunct{\mcitedefaultmidpunct}
{\mcitedefaultendpunct}{\mcitedefaultseppunct}\relax
\EndOfBibitem
\bibitem{Rogozhnikov:2016bdp}
A.~Rogozhnikov, \ifthenelse{\boolean{articletitles}}{\emph{{Reweighting with
  Boosted Decision Trees}},
  }{}\href{https://doi.org/10.1088/1742-6596/762/1/012036}{J.\ Phys.\ Conf.\
  Ser.\  \textbf{762} (2016) 012036},
  \href{http://arxiv.org/abs/1608.05806}{{\normalfont\ttfamily
  arXiv:1608.05806}}, \url{https://github.com/arogozhnikov/hep_ml}\relax
\mciteBstWouldAddEndPuncttrue
\mciteSetBstMidEndSepPunct{\mcitedefaultmidpunct}
{\mcitedefaultendpunct}{\mcitedefaultseppunct}\relax
\EndOfBibitem
\bibitem{Skwarnicki:1986xj}
T.~Skwarnicki, {\em {A study of the radiative cascade transitions between the
  Upsilon-prime and Upsilon resonances}}, PhD thesis, Institute of Nuclear
  Physics, Krakow, 1986,
  {\href{http://inspirehep.net/record/230779/}{DESY-F31-86-02}}\relax
\mciteBstWouldAddEndPuncttrue
\mciteSetBstMidEndSepPunct{\mcitedefaultmidpunct}
{\mcitedefaultendpunct}{\mcitedefaultseppunct}\relax
\EndOfBibitem
\bibitem{JacobWick}
M.~Jacob and G.~C. Wick, \ifthenelse{\boolean{articletitles}}{\emph{On the
  general theory of collisions for particles with spin},
  }{}\href{https://doi.org/https://doi.org/10.1016/0003-4916(59)90051-X}{Annals
  of Physics \textbf{7} (1959) 404 }\relax
\mciteBstWouldAddEndPuncttrue
\mciteSetBstMidEndSepPunct{\mcitedefaultmidpunct}
{\mcitedefaultendpunct}{\mcitedefaultseppunct}\relax
\EndOfBibitem
\bibitem{TFA}
\ifthenelse{\boolean{articletitles}}{\emph{{TensorFlowAnalysis}: A collection
  of useful functions and example scripts for performing amplitude fits using
  {TensorFlow}}, }{}
\newblock \url{https://gitlab.cern.ch/poluekt/TensorFlowAnalysis}\relax
\mciteBstWouldAddEndPuncttrue
\mciteSetBstMidEndSepPunct{\mcitedefaultmidpunct}
{\mcitedefaultendpunct}{\mcitedefaultseppunct}\relax
\EndOfBibitem
\bibitem{tensorflow2015-whitepaper}
M.~Abadi {\em et~al.}, \ifthenelse{\boolean{articletitles}}{\emph{{TensorFlow}:
  Large-scale machine learning on heterogeneous systems}, }{} 2015.
\newblock \url{https://www.tensorflow.org}\relax
\mciteBstWouldAddEndPuncttrue
\mciteSetBstMidEndSepPunct{\mcitedefaultmidpunct}
{\mcitedefaultendpunct}{\mcitedefaultseppunct}\relax
\EndOfBibitem
\bibitem{James:1975dr}
F.~James and M.~Roos, \ifthenelse{\boolean{articletitles}}{\emph{{Minuit: A
  system for function minimization and analysis of the parameter errors and
  correlations}},
  }{}\href{https://doi.org/10.1016/0010-4655(75)90039-9}{Comput.\ Phys.\
  Commun.\  \textbf{10} (1975) 343}\relax
\mciteBstWouldAddEndPuncttrue
\mciteSetBstMidEndSepPunct{\mcitedefaultmidpunct}
{\mcitedefaultendpunct}{\mcitedefaultseppunct}\relax
\EndOfBibitem
\bibitem{Brun:1997pa}
R.~Brun and F.~Rademakers, \ifthenelse{\boolean{articletitles}}{\emph{{ROOT: An
  object oriented data analysis framework}},
  }{}\href{https://doi.org/10.1016/S0168-9002(97)00048-X}{Nucl.\ Instrum.\
  Meth.\  \textbf{A389} (1997) 81}\relax
\mciteBstWouldAddEndPuncttrue
\mciteSetBstMidEndSepPunct{\mcitedefaultmidpunct}
{\mcitedefaultendpunct}{\mcitedefaultseppunct}\relax
\EndOfBibitem
\bibitem{Bugg:2005xx}
D.~V. Bugg, \ifthenelse{\boolean{articletitles}}{\emph{{The Kappa in E791 data
  for $D \to K \pi \pi$}},
  }{}\href{https://doi.org/10.1016/j.physletb.2005.11.019}{Phys.\ Lett.\
  \textbf{B632} (2006) 471},
  \href{http://arxiv.org/abs/hep-ex/0510019}{{\normalfont\ttfamily
  arXiv:hep-ex/0510019}}\relax
\mciteBstWouldAddEndPuncttrue
\mciteSetBstMidEndSepPunct{\mcitedefaultmidpunct}
{\mcitedefaultendpunct}{\mcitedefaultseppunct}\relax
\EndOfBibitem
\bibitem{Flatte:1976xu}
S.~M. Flatte, \ifthenelse{\boolean{articletitles}}{\emph{{Coupled - Channel
  Analysis of the $\pi \eta$ and $K \bar{K}$ Systems Near $K \bar{K}$
  Threshold}}, }{}\href{https://doi.org/10.1016/0370-2693(76)90654-7}{Phys.\
  Lett.\  \textbf{B63} (1976) 224}\relax
\mciteBstWouldAddEndPuncttrue
\mciteSetBstMidEndSepPunct{\mcitedefaultmidpunct}
{\mcitedefaultendpunct}{\mcitedefaultseppunct}\relax
\EndOfBibitem
\bibitem{Zhang:2013sva}
H.~Zhang, J.~Tulpan, M.~Shrestha, and D.~M. Manley,
  \ifthenelse{\boolean{articletitles}}{\emph{{Multichannel parametrization of
  $\bar K N$ scattering amplitudes and extraction of resonance parameters}},
  }{}\href{https://doi.org/10.1103/PhysRevC.88.035205}{Phys.\ Rev.\
  \textbf{C88} (2013) 035205},
  \href{http://arxiv.org/abs/1305.4575}{{\normalfont\ttfamily
  arXiv:1305.4575}}\relax
\mciteBstWouldAddEndPuncttrue
\mciteSetBstMidEndSepPunct{\mcitedefaultmidpunct}
{\mcitedefaultendpunct}{\mcitedefaultseppunct}\relax
\EndOfBibitem
\bibitem{Cameron:1978qi}
Rutherford-London collaboration, W.~Cameron {\em et~al.},
  \ifthenelse{\boolean{articletitles}}{\emph{{Partial wave analysis of $\bar{K}
  N \to \bar{K}^* N$ between 1830 MeV and 2170 MeV center-of-mass energy
  including new data below 1960 MeV}},
  }{}\href{https://doi.org/10.1016/0550-3213(78)90071-8}{Nucl.\ Phys.\
  \textbf{B146} (1978) 327}\relax
\mciteBstWouldAddEndPuncttrue
\mciteSetBstMidEndSepPunct{\mcitedefaultmidpunct}
{\mcitedefaultendpunct}{\mcitedefaultseppunct}\relax
\EndOfBibitem
\bibitem{Sozzi:1087897}
M.~S. Sozzi, {\em {Discrete symmetries and \CP violation: from experiment to
  theory}}, Oxford Graduate Texts, Oxford Univ. Press, New York, NY, 2008\relax
\mciteBstWouldAddEndPuncttrue
\mciteSetBstMidEndSepPunct{\mcitedefaultmidpunct}
{\mcitedefaultendpunct}{\mcitedefaultseppunct}\relax
\EndOfBibitem
\bibitem{Mikhasenko:2019rjf}
JPAC collaboration, M.~Mikhasenko {\em et~al.},
  \ifthenelse{\boolean{articletitles}}{\emph{{Dalitz-plot decomposition for
  three-body decays}},
  }{}\href{https://doi.org/10.1103/PhysRevD.101.034033}{Phys.\ Rev.\
  \textbf{D101} (2020) 034033},
  \href{http://arxiv.org/abs/1910.04566}{{\normalfont\ttfamily
  arXiv:1910.04566}}\relax
\mciteBstWouldAddEndPuncttrue
\mciteSetBstMidEndSepPunct{\mcitedefaultmidpunct}
{\mcitedefaultendpunct}{\mcitedefaultseppunct}\relax
\EndOfBibitem
\bibitem{Gourgoulhon}
E.~Gourgoulhon, {\em {Special Relativity in General Frames}}, Springer,
  2013\relax
\mciteBstWouldAddEndPuncttrue
\mciteSetBstMidEndSepPunct{\mcitedefaultmidpunct}
{\mcitedefaultendpunct}{\mcitedefaultseppunct}\relax
\EndOfBibitem
\bibitem{Blatt:1952ije}
J.~M. Blatt and V.~F. Weisskopf, {\em {Theoretical nuclear physics}},
  \href{https://doi.org/10.1007/978-1-4612-9959-2}{ Springer, New York,
  1952}\relax
\mciteBstWouldAddEndPuncttrue
\mciteSetBstMidEndSepPunct{\mcitedefaultmidpunct}
{\mcitedefaultendpunct}{\mcitedefaultseppunct}\relax
\EndOfBibitem
\bibitem{VonHippel:1972fg}
F.~Von~Hippel and C.~Quigg,
  \ifthenelse{\boolean{articletitles}}{\emph{{Centrifugal-barrier effects in
  resonance partial decay widths, shapes, and production amplitudes}},
  }{}\href{https://doi.org/10.1103/PhysRevD.5.624}{Phys.\ Rev.\  \textbf{D5}
  (1972) 624}\relax
\mciteBstWouldAddEndPuncttrue
\mciteSetBstMidEndSepPunct{\mcitedefaultmidpunct}
{\mcitedefaultendpunct}{\mcitedefaultseppunct}\relax
\EndOfBibitem
\end{mcitethebibliography}

\newpage
\centerline
{\large\bf LHCb collaboration}
\begin
{flushleft}
\small
R.~Aaij$^{32}$\lhcborcid{0000-0003-0533-1952},
A.S.W.~Abdelmotteleb$^{50}$\lhcborcid{0000-0001-7905-0542},
C.~Abellan~Beteta$^{44}$,
F.~Abudin{\'e}n$^{50}$\lhcborcid{0000-0002-6737-3528},
T.~Ackernley$^{54}$\lhcborcid{0000-0002-5951-3498},
B.~Adeva$^{40}$\lhcborcid{0000-0001-9756-3712},
M.~Adinolfi$^{48}$\lhcborcid{0000-0002-1326-1264},
H.~Afsharnia$^{9}$,
C.~Agapopoulou$^{13}$\lhcborcid{0000-0002-2368-0147},
C.A.~Aidala$^{76}$\lhcborcid{0000-0001-9540-4988}
% ...

{\footnotesize \it

% ...
$^{9}$Universit{\'e} Clermont Auvergne, CNRS/IN2P3, LPC, Clermont-Ferrand, France\\
% ...
$^{13}$LPNHE, Sorbonne Universit{\'e}, Paris Diderot Sorbonne Paris Cit{\'e}, CNRS/IN2P3, Paris, France\\
% ...
$^{32}$Nikhef National Institute for Subatomic Physics, Amsterdam, Netherlands\\
% ...
$^{40}$Instituto Galego de F{\'\i}sica de Altas Enerx{\'\i}as (IGFAE), Universidade de Santiago de Compostela, Santiago de Compostela, Spain\\
% ...
$^{44}$Physik-Institut, Universit{\"a}t Z{\"u}rich, Z{\"u}rich, Switzerland\\
% ...
$^{48}$H.H. Wills Physics Laboratory, University of Bristol, Bristol, United Kingdom\\
% ...
$^{50}$Department of Physics, University of Warwick, Coventry, United Kingdom\\
% ...
$^{54}$Oliver Lodge Laboratory, University of Liverpool, Liverpool, United Kingdom\\
% ...
$^{62}$Syracuse University, Syracuse, NY, United States\\
% ...
$^{76}$University of Michigan, Ann Arbor, MI, United States, associated to $^{62}$\\
\bigskip
% ...
}
\end{flushleft}

\end{document}